\documentclass[twocolumn]{aastex701}

\begin{document}

\title{Mock Observations: Morphological Analysis of Galactic Bars in IllustrisTNG Simulations}
\correspondingauthor{Weipeng Lin}
\email{linweip5@mail.sysu.edu.cn}

\correspondingauthor{Jun-Yu Gong}
\email{gongjy27@mail.sysu.edu.cn}

\author{Jun-Yu Gong}
\affiliation{School of Physics and Astronomy, Sun Yat-sen University, DaXue Road 2, 519082, Zhuhai, China}
\affiliation{CSST Science Center for the Guangdong-Hongkong-Macau Greater Bay Area, DaXue Road 2, 519082, Zhuhai, China}
\email{gongjy27@mail.sysu.edu.cn}  

\author{Weipeng Lin}
\affiliation{School of Physics and Astronomy, Sun Yat-sen University, DaXue Road 2, 519082, Zhuhai, China}
\affiliation{CSST Science Center for the Guangdong-Hongkong-Macau Greater Bay Area, DaXue Road 2, 519082, Zhuhai, China}
\email{linweip5@mail.sysu.edu.cn}

\author{Weishan Zhu}
\affiliation{School of Physics and Astronomy, Sun Yat-sen University, DaXue Road 2, 519082, Zhuhai, China}
\affiliation{CSST Science Center for the Guangdong-Hongkong-Macau Greater Bay Area, DaXue Road 2, 519082, Zhuhai, China}
\email{zhuwshan5@mail.sysu.edu.cn}

\author{Lin Tang}
\affiliation{School of Physics and Astronomy, China West Normal University, ShiDa Road 1, 637002, Nanchong, China}
\affiliation{CSST Science Center for the Guangdong-Hongkong-Macau Greater Bay Area, DaXue Road 2, 519082, Zhuhai, China}
\email{tanglin23@cwnu.edu.cn}

\begin{abstract}

Galactic bars are important tracers of disk-galaxy evolution, yet reported bar fractions vary among observational studies and between observations and cosmological simulations. We compare bar statistics and properties in the IllustrisTNG simulations (TNG100 and TNG50 at $z\simeq0.1$) with Sloan Digital Sky Survey (SDSS) galaxies at $0.1 \le z \le 0.1264$. We generate synthetic $r$-band TNG images, match SDSS observational conditions, and analyze mock TNG and real SDSS images using a common image-based framework of source segmentation, morphological classification, and isophotal ellipse fitting.

For face-on late-type galaxies with $\log(M_*/M_{\odot})>10$, the global bar fractions are $26.68_{-0.23}^{+0.23}\%$, $24.96_{-0.76}^{+0.78}\%$, and $13.42_{-1.07}^{+1.15}\%$ for SDSS, TNG100, and TNG50, respectively. TNG100 broadly agrees with SDSS over $10^{10} \lesssim M_* \lesssim 10^{11.2}\,M_{\odot}$ but shows a higher $f_{\rm bar}$ at higher masses, whereas TNG50 remains lower than SDSS over most of $10^{10} \lesssim M_* \lesssim 10^{11.7}\,M_{\odot}$. Bulge--disk decompositions show fewer high-$n$ bulges in TNG100 than in SDSS, particularly at high masses, but more in TNG50. These differences in bulge concentration may contribute to the different bar fractions.

Bar length generally increases with stellar mass, and this trend depends little on the adopted bar-length definition. However, the outer ellipticity-decline estimator ($R_{\rm 0.85max}$) yields systematically larger bar lengths than the ellipticity-peak estimator ($R_{\epsilon,\max}^{\rm bar}$). Massive TNG100 galaxies have more extended stellar disks than SDSS, which may contribute to their smaller relative bar lengths ($R_{\rm bar}/R_{\rm e}$).

Our results indicate that observational conditions, sample selection, and bar-identification methodology can affect the inferred bar fraction, emphasizing comparable observational conditions and analysis procedures in simulation--observation comparisons. Our SDSS measurements may underestimate the bar fraction relative to observations with higher spatial resolution.

\end{abstract}


\section{Introduction}

Stellar bars are fundamental non-axisymmetric structures residing in the central regions of a significant fraction of disk galaxies. While observational surveys often suggest that approximately two-thirds of disk galaxies host a bar \citep[e.g.,][]{Eskridge2000, Menendez2007, Marinova2007, Saha2018}, this figure includes weak and intermediate features. Restricting the census to strong bars decreases the observed fraction to approximately 30\% \citep[e.g.,][]{Nair2010, Barazza2009, Masters2011}. This discrepancy arises primarily because visual classifications typically adopt stricter criteria, while automated methods can detect smaller bars but may also misidentify oval distortions as bars \citep{Lee2012, Masters2011}. Moreover, the bar fraction strongly correlates with stellar mass, with higher bar fractions found in more massive systems \citep{Masters2012, Oh2012, Melvin2014, Gavazzi2015}. Recent observations further indicate that bars were already present at high redshifts \citep{Fragkoudi2020, Costantin2023, Guo2023, Le2024, LeConte2026}.

Several mechanisms have been proposed for the formation and evolution of galactic bars. Bars form via disk instabilities when a disk becomes sufficiently self-gravitating and dynamically cool \citep{Toomre1964, Sellwood1993, Rosas2020, Romeo2023}, while tidal interactions and mergers can also trigger bar formation \citep{Noguchi1987, Gerin1990, Barnes1991}. The properties of the dark matter halo, including its angular momentum and non-axisymmetric structure, may further affect bar formation and evolution \citep{Saha2013, Heller2007}, while gas accretion can replenish the disk and lead to the formation or re-formation of bars \citep{Bournaud2002}. Bars subsequently grow stronger and longer by transferring angular momentum outward to the dark matter halo \citep{Athanassoula1986, Athanassoula2002, Saha2013}, whereas central mass concentrations and gas inflows may weaken or, under some conditions, destroy them \citep{Athanassoula2005, Bournaud2005, Berentzen2007, Du2017, Guo2020}. Bars play a critical role in redistributing angular momentum among stars, gas, and dark matter \citep{Lynden-Bell1972, Weinberg1985, Debattista2000, Athanassoula2003, Kormendy2013, Collier2021}, facilitating gas inflow and the formation of various central structures, including nuclear disks \citep{Sanders1980, Hawarden1986, Athanassoula1992, Knapen1995, Gadotti2020}, disky pseudo-bulges \citep{Kormendy1993, Kormendy2004, Jogee2005, Liang2025}, and boxy/peanut-shaped bulges \citep{Combes1981, Kuijken1995, Bureau2005, Shen2010}. The connection between bar-driven gas inflow and Active Galactic Nucleus (AGN) activity also remains debated \citep{Ho1997, Hao2009, Lee2012, Cisternas2015}. Despite extensive theoretical and observational studies, the relative importance of these processes in bar formation and evolution remains uncertain.

Numerical simulations are essential for studying bar formation and evolution, and substantial progress has been made beyond the TNG framework. For example, using the EAGLE simulations, \citet{Cavanagh2022} revealed that bar destruction is primarily driven by major merging events. In the NewHorizon simulation, \citet{Reddish2022} demonstrated that strong dark matter dominance suppresses bar formation in low-mass galaxies, while massive bulges prevent bar persistence in high-mass systems. 

Alongside these efforts, numerous studies have used the IllustrisTNG project \citep[hereafter TNG;][]{Springel2018, Pillepich2018_firstresult, Nelson2018, Marinacci2018, Naiman2018, Pillepich2019, Nelson2019} to investigate galactic bars. For TNG100, \citet{Zhou2020} and \citet{Zhao2020} found that the global bar fraction and its dependence on stellar mass are broadly consistent with observations. However, \citet{Zhao2020} also reported an excess of bars in massive galaxies and an overproduction of relatively short bars. In contrast, \citet{Roshan2021} found that both the simulated bar fraction and its stellar-mass dependence differ substantially from observations, together with a significant discrepancy in the ratio of corotation radius to bar length. For TNG50, \citet{Rosas-Guevara2022} showed that the inferred evolution of the bar fraction depends strongly on the adopted limits on bar size and spatial resolution. More recently, \citet{Lu2025} found an excess of short bars and related the presence or absence of bars to both mergers and internal galaxy structure. Taken together, these studies show that the inferred bar fractions, stellar-mass dependences, and bar sizes are not fully consistent among different TNG analyses, while the degree of agreement between TNG and observations also remains uncertain. Recent mock-observation studies further demonstrate that observational effects can significantly affect the measured properties and detectability of bars: \citet{Gonccalves2025} found systematic wavelength dependence in the measured ellipticity and length of TNG50 bars, while \citet{Goncalves2026} showed that realistic Euclid-like observing conditions can substantially reduce the inferred bar fraction in TNG50.

These studies highlight the importance of accounting for both observational effects and differences in analysis methods when comparing simulations with observations. Mock observations provide a direct approach to such comparisons by placing simulated galaxies under realistic observing conditions and allowing the same analysis procedures to be applied to both simulated and observed data. \citet{Rodriguez-Gomez2019} created Panoramic Survey Telescope and Rapid Response System (Pan-STARRS) synthetic images, and found general morphological alignment in TNG despite lingering discrepancies in color and size relations. For spectroscopy, \citet{Nanni2022} demonstrated that mock Mapping Nearby Galaxies at Apache Point Observatory (MaNGA) cubes allow for the accurate recovery of intrinsic age and metallicity gradients. \citet{Zhou2025} introduced GalaxyGenius, a multi-wavelength framework that serves as a testbed for observational analysis. Crucially, \citet{Tang2021} established a pipeline using observational surface brightness limits rather than gravitational binding energy, revealing that the reported mismatch in the luminosity function was largely an artifact of simplified galaxy sample definitions. Building on this, \citet{Tang2023} confirmed the stripped-satellite origin of intracluster light, while \citet{Lan2024} successfully reproduced the intrinsic alignment trends of massive galaxies. \citet{Gong2025} showed that these brightness-based definitions recover morphological distributions matching Sloan Digital Sky Survey (SDSS) data significantly better than traditional gravity-based methods.

In this work, we perform a fair and robust comparison between TNG simulations and SDSS observations. To ensure consistency, we first apply a mock-observation pipeline to the TNG data to mimic the SDSS data quality, and then process both datasets using a closely matched analysis framework. The paper is organized as follows: Section~\ref{sec:data} introduces the TNG and SDSS samples, Section~\ref{sec:methods} presents the mock-observation and measurement pipeline, Section~\ref{sec:results} reports the bar statistics, Section~\ref{sec:discussion} interprets the physical implications, and Section~\ref{sec:summary} summarizes our conclusions.

\section{Data and Sample Selection}\label{sec:data}

\subsection{Simulations}\label{sec:tng_data}
In this study, we utilize data from the TNG100-1 and TNG50-1 runs (hereafter TNG100 and TNG50) of the IllustrisTNG project. These large-scale magnetohydrodynamical simulations are evolved using the moving-mesh code \texttt{Arepo} \citep{Springel2010}. Building on the original Illustris model, TNG incorporates updated magnetohydrodynamics and improved subgrid prescriptions for galactic winds and black hole feedback \citep{Pillepich2018, Terrazas2020, Zinger2020}, and it reproduces multiple global galaxy observables (e.g., stellar mass content, colors, and size trends) reasonably well \citep{Genel2018, Donnari2019, Huertas-Company2019, Rodriguez-Gomez2019, Tacchella2019}. The simulations adopt a standard $\Lambda$CDM cosmology with parameters consistent with \citet{Planck2016}: $\Omega_m = 0.3089$, $\Omega_\Lambda = 0.6911$, $\Omega_b = 0.0486$, $\sigma_8 = 0.8159$, $n_s = 0.9667$, and $h = 0.6774$. 

TNG100 simulates a cubic volume with a side length of $75\,h^{-1}\,\mathrm{Mpc}$ ($\approx 110.7$\,Mpc) containing $2 \times 1820^3$ resolution elements. This run achieves a baryonic mass resolution of $1.4 \times 10^6 M_{\odot}$ and a gravitational softening length for stellar particles of $\varepsilon = 0.74$\,kpc for $z \le 1$. To capture finer structural details, we also employ the TNG50 simulation, which evolves a smaller volume of $35\,h^{-1}\,\mathrm{Mpc}$ ($\approx 51.7$\,Mpc). TNG50 provides significantly higher resolution, with a baryon mass resolution of $8.5 \times 10^4 M_{\odot}$ and a stellar softening length of approximately 0.29\,kpc. Dark matter halos are identified using the friends-of-friends (FoF) algorithm \citep{Davis1985}, which serves as the basis for our subsequent mock observation pipeline.

\subsection{Observations}\label{sec:sdss_data}

Our observational parent sample is drawn from the catalog presented by \citet{Meert2015} (hereafter \hyperlink{cite.Meert2015}{M15}), which comprises approximately $7 \times 10^5$ spectroscopically selected galaxies from the SDSS Data Release 7 (DR7). \hyperlink{cite.Meert2015}{M15} provides robust 2D photometric decompositions in the $r$-band, utilizing the \texttt{GALFIT} \citep{Peng2002} algorithm and the \texttt{Pymorph} \citep{Vikram2010} pipeline to fit varying models (e.g., Sérsic, de Vaucouleurs) to galaxy images. Included is a comprehensive, physically motivated flagging system, which indicates that over 90\% of two-component fits are reliable for analysis, ensuring a high degree of sample completeness.

To complement the photometric data with physical parameters, we cross-match our sample with the GALEX–SDSS–WISE Legacy Catalog \citep[hereafter GSWLC;][]{Salim2016}. This catalog provides our primary parameter of interest, stellar mass, alongside dust attenuations and star formation rates for galaxies with $z < 0.3$, covering 90\% of the SDSS footprint. The properties are derived via Bayesian spectral energy distribution (SED) fitting to UV/optical photometry, incorporating corrections for blending and emission lines. To address gaps in the GSWLC coverage, we incorporate the Max Planck Institute for Astrophysics--Johns Hopkins University (MPA-JHU) DR7 catalog\footnote{\url{https://wwwmpa.mpa-garching.mpg.de/SDSS/DR7/}} as a supplementary resource. Specifically, we utilize their stellar mass estimates, which are similarly derived via SED fitting to the SDSS broad-band photometry.

Finally, we utilize the deep-learning-based morphological catalog of \citet{Dominguez2018} to assign continuous T-types to our sample. While this catalog provides various classifications derived from models trained on visual data from Galaxy Zoo 2 and \citet{Nair2010}, our primary purpose in adopting it is to reliably distinguish between early-type galaxies (ETGs) and late-type galaxies (LTGs). Therefore, we strictly use these T-types to filter out early-type systems and restrict our final sample exclusively to LTGs.

\subsection{Sample Selection}\label{sec:sample_selection}

The selection strategy for our analysis is primarily driven by the physical resolution limits of the simulations. The gravitational softening length of TNG100 is $\varepsilon \approx 0.74$\,kpc. For mock observations to be robust, the projected angular size of $\varepsilon$ must not significantly exceed the pixel scale of the target instrument. At redshifts $z < 0.1$, the projected softening length becomes larger than the SDSS pixel scale ($0.396^{\prime\prime}$), potentially introducing numerical artifacts into the synthetic images. At $z=0.1$, $\varepsilon$ projects to $\approx 0.39''$, which is comparable to the pixel scale. Consequently, for the simulated sample, we strictly utilize Snapshot 91, which corresponds to $z=0.1$. This choice is applied to both TNG100 and TNG50 to maintain consistency, although TNG50 possesses higher intrinsic resolution. We note that while the raw snapshots contain all galaxy types, our final analysis is restricted to LTGs. This morphological selection is performed a posteriori following the mock observation pipeline detailed in Section~\ref{sec:morph_class}.

For the observational comparison, we draw from the \hyperlink{cite.Meert2015}{M15} parent sample and restrict the analysis to the redshift interval $0.1 \le z \le 0.1264$. The lower limit of $z=0.1$ is chosen to match the redshift of the simulation snapshot and to keep the physical spatial resolution of the SDSS images comparable to that of the mock observations. The upper redshift limit is calculated to define a comoving radial depth of $\approx 110.7$\,Mpc, matching the TNG100 box side ($75\,h^{-1}\,\mathrm{Mpc}$). This ensures the observational slice spans a line-of-sight distance comparable to the simulation box. To select LTGs consistent with the simulation analysis, we utilize the morphological classifications from \citet{Dominguez2018}, requiring a T-type $> 0$. Applying these criteria to the \hyperlink{cite.Meert2015}{M15} catalog yields an observational sample of 52,674 galaxies. Stellar masses are assigned hierarchically: the majority are from GSWLC; for the 7,401 galaxies missing from GSWLC, 7,110 are supplemented by the MPA-JHU catalog. For the remaining 291 galaxies, we estimate their stellar masses using the color-dependent mass-to-light ratio relation from \citet{Zibetti2009}: $\log (M_*/L_i) = -0.963 + 1.032(g - i)$, adopting a solar absolute magnitude of $M_{\odot, i} = 4.58$ \citep{Blanton2001}.

\section{Methodology}\label{sec:methods}

\subsection{Synthetic Image Generation}
To perform a robust morphological comparison between simulations and observations, it is essential to bridge the gap between the intrinsic physical properties of simulated galaxies and observable quantities. To this end, we adopt the \texttt{GalaxyGenius} framework \citep{Zhou2025} to generate synthetic images using the Monte Carlo radiative transfer code \texttt{SKIRT} \citep{Baes2011, Camps2020}. This framework allows us to rigorously model the emission of starlight and its subsequent propagation through the interstellar medium (ISM), accounting for absorption and scattering by dust grains. As detailed in Section~\ref{sec:sample_selection}, we specifically utilize Snapshot 91 ($z=0.1$) from the TNG simulations to strictly match the lower bound of our observational SDSS sample.

A key modification we implement in the standard input generation pipeline is the selection of particle data. Previous studies generally identify individual galaxies with \textsc{Subfind} and analyze the particles associated with each subhalo \citep[e.g.,][]{Rodriguez-Gomez2019, Huertas-Company2019, Zhou2025}. In contrast, we do not render individual \textsc{Subfind} galaxies. Instead, we render the entire parent FoF group, including all of its stellar and gas particles. Before image generation, we require the FoF group to contain stellar particles and to satisfy $M_{200c}>10^8 M_\odot/h$, where $M_{200c}$ is the total mass enclosed within a sphere whose mean density is 200 times the critical density of the Universe at the corresponding snapshot. Individual galaxies are then identified from the resulting mock images through the source-segmentation procedure described in Section~\ref{sec:segmentation}, rather than being pre-selected from the \textsc{Subfind} catalog. This approach preserves intracluster light, satellites, and neighboring structures in the synthetic images, allowing source detection and deblending to operate under conditions more closely analogous to real observations. Consequently, no explicit minimum stellar-particle number or intrinsic subhalo stellar-mass cut is imposed prior to image generation; the effective galaxy selection is instead performed subsequently in observational space.

Radiation sources are stellar particles in each FoF group. We assign SEDs using a two-age scheme: particles older than 10\,Myr use Bruzual \& Charlot SSP templates with a Chabrier IMF \citep{Bruzual2003, Chabrier2003}, while younger particles ($\le 10$\,Myr) use MAPPINGS III templates \citep{Groves2008}. For the young component, we adopt $P = 1.38 \times 10^{-12}$\,Pa and a Gaussian compactness distribution centered at $\log C = 5$.

Dust is assigned from FoF gas cells with non-zero star formation or $T < 8000$\,K, excluding hot halo gas. We adopt a constant dust-to-metal ratio $f_{\text{dust}} = 0.3$ and the Zubko et al. grain mixture (silicate, graphite, and PAH) \citep{Zubko2004}. The dust field is mapped to an octree grid with refinement levels 6--10.

Each FoF group is rendered from three orthogonal views ($xy$, $yz$, $zx$), with a field of view (FoV) that encloses the full stellar extent. Unless otherwise stated, all three projections are included together in the following analysis. We find no significant systematic dependence of the population-level bar statistics on projection, although some differences among the three views are present; a detailed comparison is presented in Appendix~\ref{app:projection}. For all FoF groups, the \texttt{SKIRT} instrument distance is fixed to that corresponding to the fiducial redshift $z=0.1$, giving a sampling of $0.396^{\prime\prime}$ ($\approx0.7542$\,kpc) per pixel. Individual galaxy cutouts identified through source segmentation are extracted directly from these parent FoF images and therefore inherit the same pixel scale. For the SDSS galaxies, the physical pixel scale is calculated individually from the angular-diameter distance at the redshift of each galaxy, adopting the SDSS pixel scale of $0.396^{\prime\prime}$ per pixel. We run \texttt{SKIRT} in `ExtinctionOnly' mode using 300 wavelength bins from $0.1$ to $1.2\,\mu\mathrm{m}$ and $10^8$ photon packets. The output cubes are sampled to the SDSS pixel scale ($0.396^{\prime\prime}$; $\approx 0.7542$\,kpc at $z=0.1$), and the spatially resolved SEDs are integrated through the SDSS $r$-band response using photon counting, including the SDSS 2.5\,m effective collecting area and $t_{\text{exp}}=54$\,s. The final products therefore correspond to expected photo-electron counts at SDSS depth.

\subsection{Mock Observations}\label{sec:mock_obs}

\begin{figure*}[thpb]
    \centering
    \includegraphics[width=\textwidth]{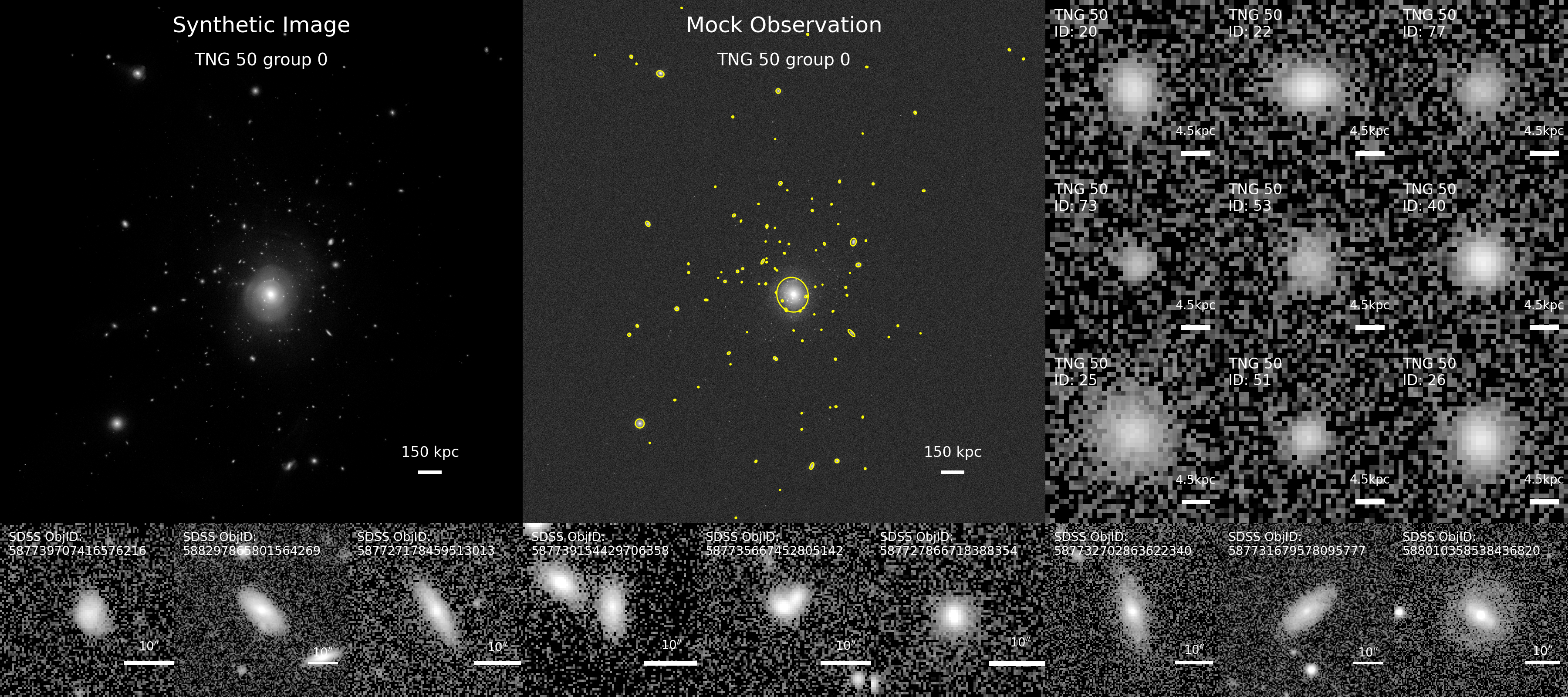}
    \caption{Morphological comparison between TNG50 synthetic galaxies and SDSS observations. The upper section displays TNG50 Group 0 at redshift $z=0.1$. The large panels show the original synthetic image (left) and the corresponding SDSS $r$-band mock observation (middle), both annotated with a 150 kpc scale bar. Yellow elliptical apertures indicate the member galaxies. The top-right $3\times3$ grid presents a closer look at 9 randomly selected individual galaxies from this galaxy group. The bottom row displays a control sample of 9 randomly selected LTGs from the SDSS sample for direct visual comparison. All panels share the same contrast level.}
    \label{fig:mock_comparison}
\end{figure*}

To bridge the gap between the intrinsic properties of simulated galaxies and ground-based observations, we process the ideal broadband images through a mock observation pipeline. We adopt a methodology that integrates the frameworks of \citet{Zhou2025} and \citet{deGraaff2022}, with specific adaptations to strictly match our observational setup. This procedure is applied to the SDSS $r$ band, sequentially incorporating sky background, Point Spread Function (PSF) convolution, and instrumental noise.

The image degradation begins with adding a uniform sky background ($\mu_{\text{sky}} = 20.9$\,mag\,arcsec$^{-2}$). To simulate atmospheric turbulence and instrumental scattering, we convolve the images with a double-Gaussian PSF \citep{Abdurrouf2021}, comprising a narrow ``core'' and a broad ``wing''. As indicated by \citet{Michard2002} and \citet{Wu2005}, the broad wing primarily represents stable instrumental scattering. To determine the wing's profile, we directly adopt the composite empirical PSF (ePSF) data constructed by \citet{Gong2025} through stacking isolated SDSS stars to capture the telescope's average optical response. Fitting this ePSF fixes the wing's full width at half maximum (FWHM) to approximately $2.52^{\prime\prime}$.

The narrow core accounts for variable atmospheric seeing. Rather than defining it directly, we determine it dynamically based on the final desired resolution. For each galaxy group, we randomly draw a target total FWHM from a log-normal distribution representing typical SDSS conditions (median $1.43^{\prime\prime}$, lower/upper quartiles $1.30^{\prime\prime}$ and $1.58^{\prime\prime}$)\footnote{Statistics derived from the SDSS DR7 seeing distribution: \url{https://classic.sdss.org/dr7/products/general/seeing.php}}. We then derive the exact width of the core component so that the combined PSF mathematically matches this targeted total seeing.

Following the convolution, we model the detector noise by first converting the images from physical flux units to electron counts. The total noise budget incorporates a mean dark current derived from a dark variance of $1.32$\,analog-to-digital units (ADU)$^2$ with a gain of $4.73$\,$e^-$\,ADU$^{-1}$, Poissonian shot noise applied to the total accumulated signal (source plus sky and dark current), and a Gaussian readout noise (RON) with a standard deviation of $\sigma_{\text{RON}} = 4.3$\,$e^-$. The mock observations are finalized by subtracting the background level, which is estimated by masking sources detected above a $3\sigma$ threshold. The images are then returned to units of Jy\,pixel$^{-1}$. A visual comparison between the initial synthetic images and the final mock observations is presented in Figure~\ref{fig:mock_comparison}.

\subsection{Source Segmentation and Catalog Generation}
\label{sec:segmentation}

Since our mock observation images are constructed from the full FoF groups, the FoV may contain multiple galaxy members interacting with intracluster light. To effectively isolate individual galaxies, we address the challenge of balancing detection sensitivity against spurious fragmentation by employing \texttt{SExtractor} \citep{Bertin1996} with a dual-configuration strategy. Adapted from the Galaxy Evolution from Morphologies and SEDs (GEMS) survey methodology \citep{Rix2004}, this approach involves running two distinct detection passes—a ``cold'' mode and a ``hot'' mode—followed by a merging step.

The ``cold'' mode is designed to define the footprints of bright, extended galaxies without over-deblending their substructures. It utilizes a conservative configuration with a detection threshold of $2.3 \sigma_{\text{bkg}}$, a minimum detected area of 100 pixels, and deblending parameters set to $\text{MINCONT}=0.065$ and $\text{NTHRESH}=64$. Conversely, the ``hot'' mode is tuned to recover faint objects missed by the conservative pass. This configuration employs a more sensitive threshold of $1.65\sigma_{\text{bkg}}$, a smaller minimum area of 45 pixels, and slightly more aggressive deblending settings ($\text{MINCONT}=0.06$, $\text{NTHRESH}=32$).

The final catalog is built by keeping all cold-run sources and adding only hot-run detections that do not overlap any cold-run source in the segmentation map. Based on this consolidated segmentation map, we generate image cutouts and mask maps for each target to block other segmented regions. This process ensures the target's light profile is effectively isolated for subsequent analysis, as demonstrated by the isolated galaxy images in the top-right $3\times3$ stamps of Figure~\ref{fig:mock_comparison}.

This identical pipeline is applied to the SDSS sample to generate segmentation and mask maps. Target galaxies are then identified via RA and Dec cross-matching.

\subsection{Morphological Classification of TNG Samples}\label{sec:morph_class}

To ensure a rigorous comparison between simulations and observations, the morphological selection criteria applied to the simulated galaxies must be methodologically consistent with those used for the observational dataset. The SDSS LTG sample is defined based on the T-type classifications from \citet{Dominguez2018}, which were derived using deep learning algorithms. In alignment with this approach, we employ a Convolutional Neural Network (CNN) to classify the post-processed synthetic images of the TNG galaxies.

Specifically, we adopt the binary classification architecture presented in \citet{Gong2025}, which is explicitly designed to distinguish ETGs (T-type $\le 0$) from LTGs (T-type $> 0$). For details of the network architecture and training procedure, we refer interested readers to that study. Here, we apply this pre-trained model to the source cutouts generated in Section~\ref{sec:segmentation}, classifying a simulated galaxy as an LTG if the model output probability satisfies $P_{\text{LTG}} > 0.5$.

Although \citet{Dominguez2018} utilized a regression-based CNN to predict continuous T-types, the binary classifier adopted here corresponds to Model 1 of \citet{Gong2025}, designed specifically to separate ETGs ($T\leq0$) from LTGs ($T>0$). This model was previously validated using both internal test samples and independent Galaxy Zoo classifications, achieving an accuracy of $\approx95\%$. \citet{Gong2025} further showed that, when applied to mock-observed TNG galaxies defined through image-based source segmentation, the resulting morphological populations reproduce the observed SDSS structural distributions substantially better than traditional gravity-based galaxy definitions. We therefore adopt the same validated classifier here for the ETG/LTG separation.

\subsection{Bar Measurement and Quantification}\label{sec:bar_measurement}

A fundamental advantage of our comprehensive processing workflow—spanning synthetic image generation, mock observation effects, and source segmentation—is the ability to apply a closely matched analysis framework to both the simulated TNG galaxies and the SDSS observations. Although residual differences between mock and real observations are unavoidable, the mock-observed TNG images are processed to match the SDSS data as closely as possible in pixel scale, PSF distribution, sky background, and noise properties. We therefore apply the same code and logical definitions to both samples for bar characterization, thereby minimizing methodological differences introduced by data processing. Nevertheless, some methodological differences remain, including those in morphological classification and stellar-mass estimation.

\begin{figure*}[thpb]
    \centering
    \includegraphics[width=\textwidth]{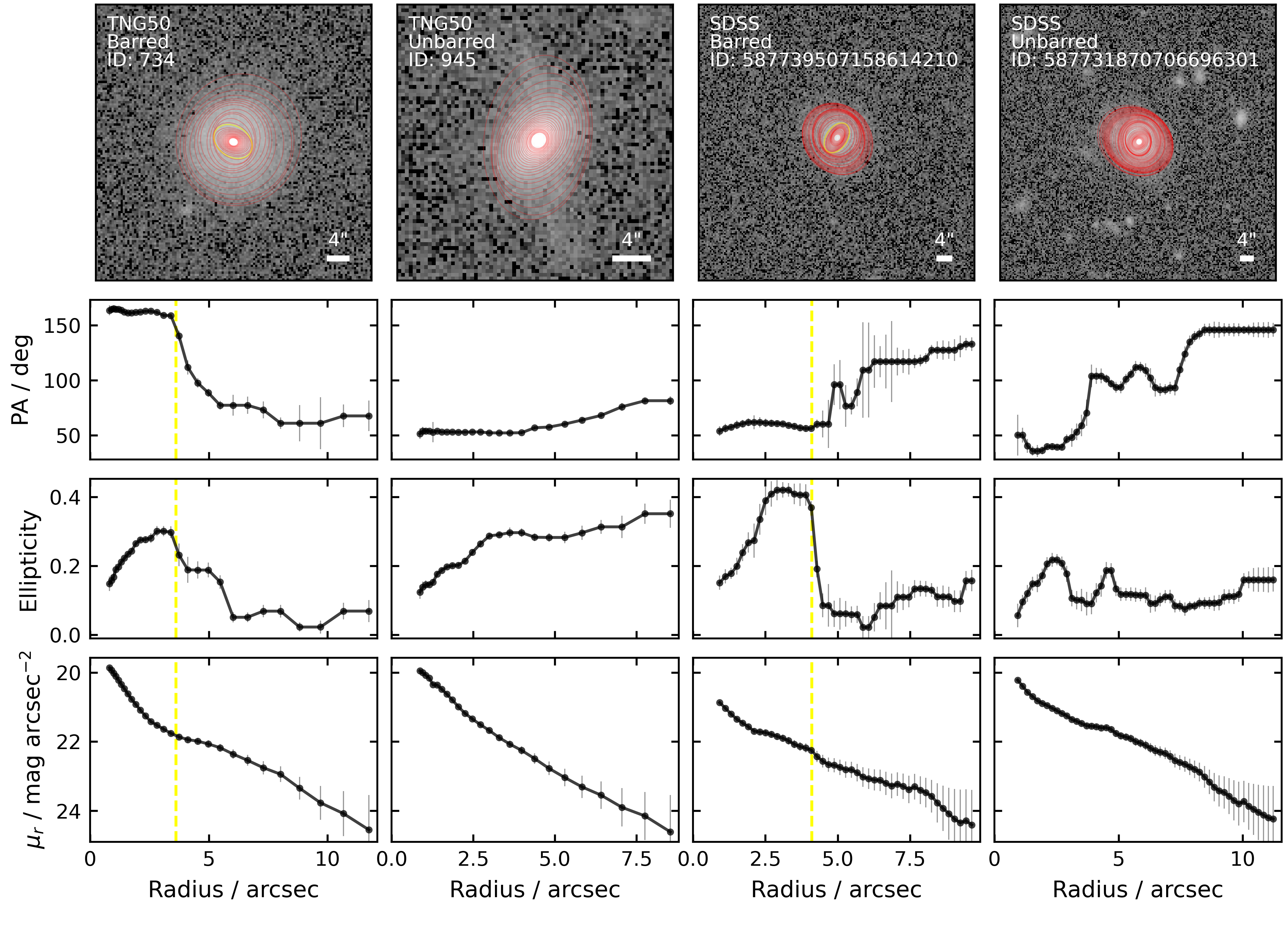}
    \caption{Isophotal fitting results for representative galaxies classified as barred and unbarred. From left to right, the columns show a barred TNG50 galaxy (ID: 734), an unbarred TNG50 galaxy (ID: 945), a barred SDSS galaxy (ObjID: 587739507158614210), and an unbarred SDSS galaxy (ObjID: 587731870706696301). The top row shows the $r$-band images with fitted isophotes (thin red ellipses), using the same contrast scale for all galaxies; a $4^{\prime\prime}$ scale bar is shown in the bottom right corner. For barred galaxies, the yellow ellipse marks the isophote corresponding to the measured bar. The remaining rows show the radial profiles of PA, ellipticity, and $r$-band surface brightness ($\mu_r$), respectively. Error bars indicate the isophotal-fitting uncertainties; for $\mu_r$, the fitting uncertainty is combined in quadrature with the single-pixel background standard deviation. The vertical yellow dashed lines mark the measured bar length ($R_{\rm bar}$).}

    \label{fig:bar_examples}
\end{figure*}

Following the methodology of \citet{Zhao2020}, we quantify the bar properties by fitting ellipses to the isophote contours of the galaxy images. This procedure is performed using the \texttt{photutils} library \citep{Bradley2020}, which extracts the radial profiles of surface brightness, ellipticity ($\epsilon$), and position angle (PA), as illustrated across the different morphological classes in Figure~\ref{fig:bar_examples}. We adopt the bar identification criteria from \citet{Marinova2007}, which are similarly applied in \citet{Zhao2020}. A galaxy is classified as barred only if it satisfies the following conditions: (1) the maximum ellipticity ($\epsilon_{\max}$) must exceed $0.25$, while the PA simultaneously varies by less than $10^\circ$; and (2) the ellipticity must exhibit a distinct drop of $\Delta \epsilon > 0.1$ from the maximum outward to the disk region (see the contrast between the barred and unbarred columns in Figure~\ref{fig:bar_examples}).

For galaxies classified as barred, following \citet{Zhao2020}, we adopt the $R_{\rm 0.85max}$ metric as our bar length estimator. It is defined as the radius where the ellipticity decreases to $85\%$ of $\epsilon_{\max}$ (the vertical yellow dashed lines in Figure~\ref{fig:bar_examples}). Hereafter, $R_{\rm 0.85max}$ is denoted as $R_{\rm bar}$. Because $R_{\rm bar}$ can be sensitive to gradual outer ellipticity declines or contamination by non-bar structures, we additionally measure $R_{\epsilon,\max}^{\rm bar}$, defined as the radius of maximum ellipticity within the identified bar region, both to assess the dependence of our results on the adopted bar length definition and to facilitate direct comparison with previous observational studies that commonly adopt the ellipticity-peak radius as a bar length estimator. Our ellipse fitting starts at a minimum semimajor axis of 2 pixels ($\simeq1.5$ kpc), chosen to avoid the innermost regions where anomalously enhanced ellipticities\footnote{In addition, the PSF convolution applied in our mock-observation procedure further suppresses this effect \citep{Goncalves2024}.} have been reported in simulated barred galaxies \citep{Zhao2020,Lu2025}. A comparison of the results obtained with $R_{\rm bar}$ and $R_{\epsilon,\max}^{\rm bar}$ is presented in Section~\ref{sec:bar_length_definition}.

We characterize the background noise by $\sigma_{\rm n}$, the standard deviation of pixel intensities in regions free of the target galaxy and other identifiable sources, identified using the source-segmentation masks. In the TNG sample, $\sigma_{\rm n}$ is measured from each parent FoF image and shared by all galaxies extracted from that group; the SDSS value is measured from the full SDSS field image. Using the full images provides a sufficiently large area of source-free pixels for estimating the background fluctuations. This quantity measures the typical pixel-to-pixel fluctuations about the mean background. Following the isophotes outward, we define the galaxy aperture using the last isophote before the mean background-subtracted intensity first falls to or below $\sigma_{\rm n}$. This aperture defines the extent of each galaxy and is used consistently for its global measurements. The axis ratio $a/b$ measured from this aperture is used to exclude highly inclined systems, with moderately inclined or face-on disk galaxies ($i \lesssim 60^\circ$) selected by requiring $\log(a/b) < 0.3$. The same aperture is used to measure the apparent $r$-band magnitude. For the TNG galaxies, rather than adopting stellar masses from the \textsc{Subfind} catalogue, we apply this aperture to the projected stellar-mass maps and obtain the galaxy stellar mass by summing the stellar-particle masses within the aperture. Although this differs from the SED-based stellar-mass estimates adopted for the SDSS sample, the TNG masses are likewise measured within an observationally defined galaxy extent.

Because our apertures are determined by the low-surface-brightness outskirts, we tested whether stochastic Monte Carlo (MC) noise from SKIRT could significantly affect their definition. We compared finite-photon-sampling fluctuations with the observational noise introduced by our mock-observation procedure, including source and sky photon-counting noise and detector readout noise. For a limited sample of galaxy groups, we also reran SKIRT with different photon counts and random seeds and remeasured the apertures. These tests indicate that MC fluctuations are generally smaller than observational noise and produce only modest aperture changes for most tested objects. Although nearly circular apertures can exhibit larger position-angle changes, these rotations have little effect on spatial coverage. Since modest aperture perturbations mainly affect faint outer regions, their impact on integrated luminosity and stellar mass is expected to be small; MC sampling noise is therefore unlikely to drive our main statistical conclusions.

\section{Results} \label{sec:results}

In this section, we present the statistical properties of galactic bars derived from our processed datasets, beginning with a summary of the final sample statistics. Given the limited numerical resolution of TNG100, we restrict all subsequent analyses to galaxies with $\log(M_*/M_\odot)>10$ to ensure more robust structural measurements and a consistent comparison between the simulated and observational samples. A detailed breakdown of the sample selection process for both observations and simulations is provided in Table~\ref{tab:sample_stats}.

For the SDSS observational sample, starting from the LTGs selected in Section~\ref{sec:sample_selection} and applying the stellar-mass cut described above, we performed isophote fitting to extract isophotal parameters. During this process, we excluded 209 galaxies for which the fitting algorithm failed to converge, primarily due to low signal-to-noise ratios or severe contamination from nearby objects. From the remaining 51,826 systems, we restricted our analysis to face-on disk galaxies using an inclination criterion based on $\log(a/b) < 0.3$. This selection yielded a final observational sample of 38,286 face-on LTGs, within which 10,213 were classified as barred.

For the TNG simulations, we adopted the same apparent-magnitude interval as that of the SDSS parent sample of \hyperlink{cite.Meert2015}{M15}, $14 < r < 17.77$, with the TNG $r$-band magnitudes measured from our isophotal aperture photometry. Based on the morphological classification detailed in Section~\ref{sec:morph_class}, this selection yielded 4,046 LTGs in TNG100 and 1,348 in TNG50. Isophote fitting was then performed for these sources, with 9 TNG100 galaxies and 1 TNG50 galaxy excluded due to fitting failures. After applying the identical inclination threshold used for the observations, the final face-on samples consist of 3,169 TNG100 galaxies and 946 TNG50 galaxies, among which our automated pipeline classified 791 and 127 galaxies as barred, respectively.

\begin{deluxetable*}{lccccc}[t!]
\tablecaption{Statistics of the Sample Selection Process \label{tab:sample_stats}}
\tablecolumns{6}
\tablewidth{0pt}
\tablehead{
\colhead{Sample} &
\colhead{Initial LTGs} &
\colhead{Successful Fits\tablenotemark{a}} &
\colhead{Face-on Sample} &
\colhead{Barred Galaxies} &
\colhead{Bar Fraction} \\
\colhead{} &
\colhead{($14 < r < 17.77$)} &
\colhead{(Isophote)} &
\colhead{($\log(a/b) < 0.3$)} &
\colhead{} &
\colhead{(\%)}
}
\startdata
SDSS ($0.1 \le z \le 0.1264$) & 52,035 & 51,826 & 38,286 & 10,213 & $26.68_{-0.23}^{+0.23}$ \\
\hline
TNG100 ($z\simeq0.1$, Snapshot 91) & 4,046 & 4,037 & 3,169 & 791 & $24.96_{-0.76}^{+0.78}$ \\
TNG50 ($z\simeq0.1$, Snapshot 91) & 1,348 & 1,347 & 946 & 127 & $13.42_{-1.07}^{+1.15}$ \\
\enddata
\tablenotetext{a}{All galaxy counts reported in this table already include the stellar-mass cut $\log(M_*/M_\odot)>10$. For SDSS, this cut reduces the LTG sample from 52,674 to 52,035 galaxies. Number of galaxies remaining after excluding objects where ellipse fitting failed due to low S/N or contamination (209 for SDSS, 9 for TNG100, and 1 for TNG50).}
\end{deluxetable*}

\subsection{Bar Fraction}\label{sec:bar_fraction}
\begin{figure}[thpb]
    \centering
    \includegraphics[width=\columnwidth]{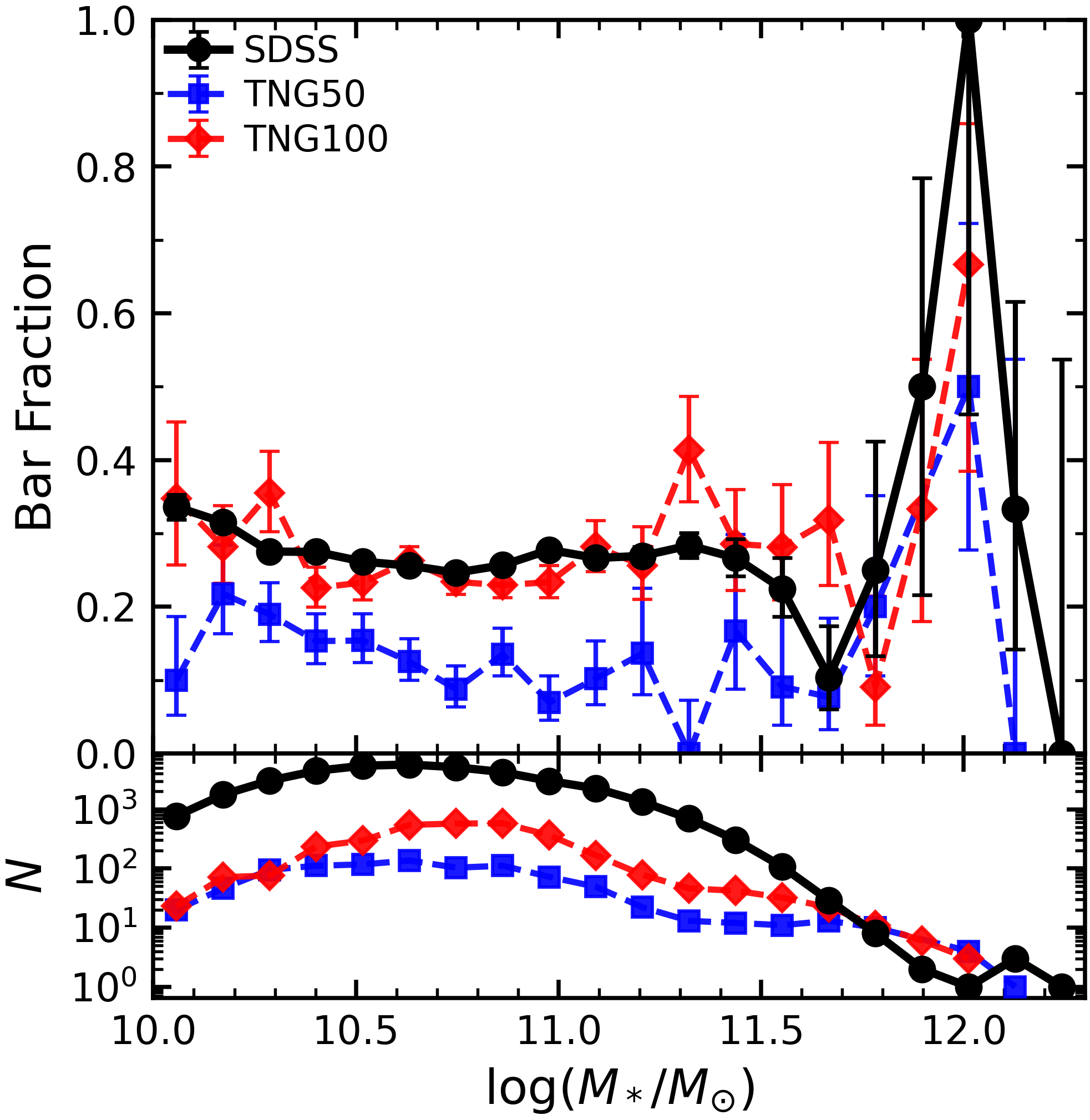}
    \caption{ Bar fraction and sample size as a function of stellar mass. Top panel: The fraction of barred galaxies as a function of stellar mass, $\log(M_*/M_{\odot})$. The observational results from SDSS are shown as black circles connected by a solid line, while TNG100 and TNG50 are shown as red diamonds and blue squares, respectively, connected by dashed lines. The vertical error bars represent the $1\sigma$ confidence intervals derived from the Beta distribution, which remains applicable even for bins with small sample sizes where the fractions approach 0 or 1. Bottom panel: The total number of galaxies in each stellar-mass bin for the corresponding samples, displayed on a logarithmic scale.} \label{fig:fbar_mass}
\end{figure}

The bar fraction ($f_{\rm bar}$) is defined as the fraction of face-on LTGs classified as barred according to the photometric criteria described in Section~\ref{sec:bar_measurement}. In this subsection, $1\sigma$ uncertainties in $f_{\rm bar}$ are estimated using Bayesian binomial confidence intervals with a Jeffreys prior \citep{Cameron2011}. The SDSS observations have a global $f_{\rm bar}$ of $26.68_{-0.23}^{+0.23}\%$. TNG100 shows a comparable global $f_{\rm bar}$ of $24.96_{-0.76}^{+0.78}\%$, whereas TNG50 presents a substantially lower value of $13.42_{-1.07}^{+1.15}\%$.

To probe the dependence of $f_{\rm bar}$ on stellar mass, we measure the bar fraction in bins of $\log(M_*/M_{\odot})$. Figure~\ref{fig:fbar_mass} compares the trends for SDSS, TNG100, and TNG50, with the bottom panel showing the galaxy counts in each stellar-mass bin on a logarithmic scale. The number of galaxies decreases rapidly toward the high-mass end, particularly for the TNG samples, resulting in increasingly large statistical uncertainties in the individual-bin bar fractions. Over $10.0 \lesssim \log(M_*/M_{\odot}) \lesssim 11.3$, the SDSS $f_{\rm bar}$ remains relatively flat, varying between approximately $24\%$ and $33\%$. Above $\log(M_*/M_{\odot}) \approx 11.3$, $f_{\rm bar}$ begins to decline, reaching approximately $10\%$ at $\log(M_*/M_{\odot}) \approx 11.7$.

TNG100 (red line) shows close agreement with SDSS over $10^{10} \lesssim M_* \lesssim 10^{11.2}\,M_{\odot}$, with many stellar-mass bins overlapping within the uncertainties. Over this range, $f_{\rm bar}$ remains relatively flat at approximately $22\%$--$35\%$. Above $M_* \approx 10^{11.2}\,M_{\odot}$, however, TNG100 generally predicts higher $f_{\rm bar}$ values than SDSS. At $M_* \approx 10^{11.3}\,M_{\odot}$, the TNG100 bar fraction reaches $\sim 41\%$, compared with $\sim 27\%$ for SDSS, and the discrepancy persists toward the high-mass end, where TNG100 still has $f_{\rm bar}\sim31\%$ at $M_* \approx 10^{11.7}\,M_{\odot}$, while the SDSS value decreases to only $\sim10\%$. In contrast, TNG50 (blue line) generally shows lower $f_{\rm bar}$ than SDSS over $10^{10} \lesssim M_* \lesssim 10^{11.7}\,M_{\odot}$, together with stronger bin-to-bin fluctuations because of its smaller sample size.

The matched observational treatment and common bar-identification method reduce methodological differences between the samples, although sample-dependent detectability may still contribute to the measured differences. In brief, TNG100 agrees well with SDSS at intermediate stellar masses but shows a higher $f_{\rm bar}$ at the high-mass end, whereas TNG50 remains systematically lower over most of its well-sampled mass range. These differences may reflect variations in the central structural properties of the galaxies produced by the two simulations, potentially including effects associated with numerical resolution. We explore the possible structural and resolution-related origins of these trends further in Section~\ref{sec:discussion_structure}.

\subsection{Bar Length}\label{sec:bar_length}

\begin{figure}[thpb]
    \centering
    \includegraphics[width=\columnwidth]{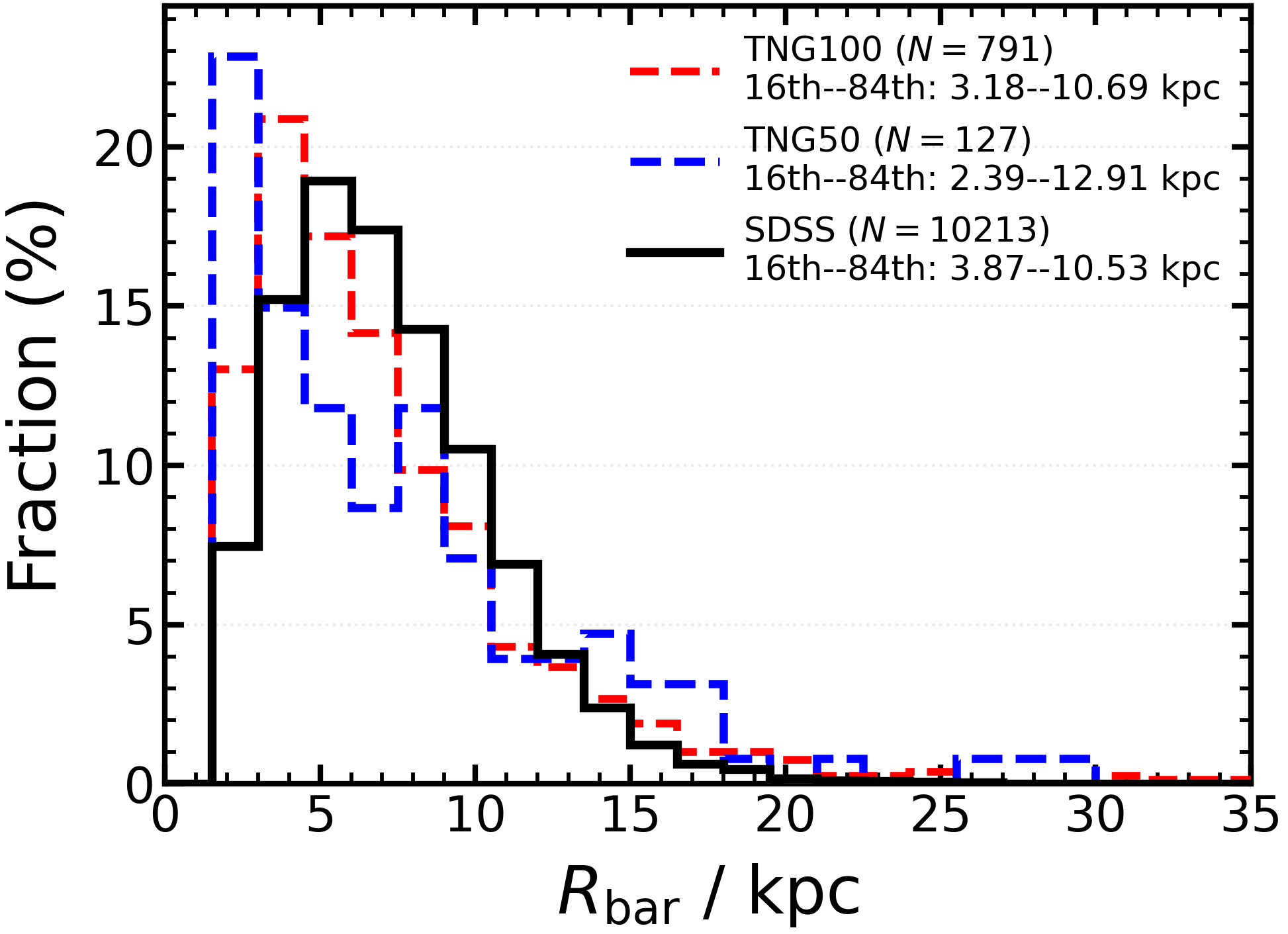}
    \caption{
    Fractional histograms of the absolute bar length, $R_{\mathrm{bar}}$, for SDSS (black), TNG100 (red dashed), and TNG50 (blue dashed).
    The vertical axis gives the fraction of barred galaxies in each bin.
    Sample sizes and 16th--84th percentile ranges are given in the legend.
    All three samples use common, uniform bins of width $1.5\,\mathrm{kpc}$, starting at zero.
    }
    \label{fig:Rbar_hist}
\end{figure}

\begin{figure*}[thpb]
    \centering
    \includegraphics[width=\textwidth]{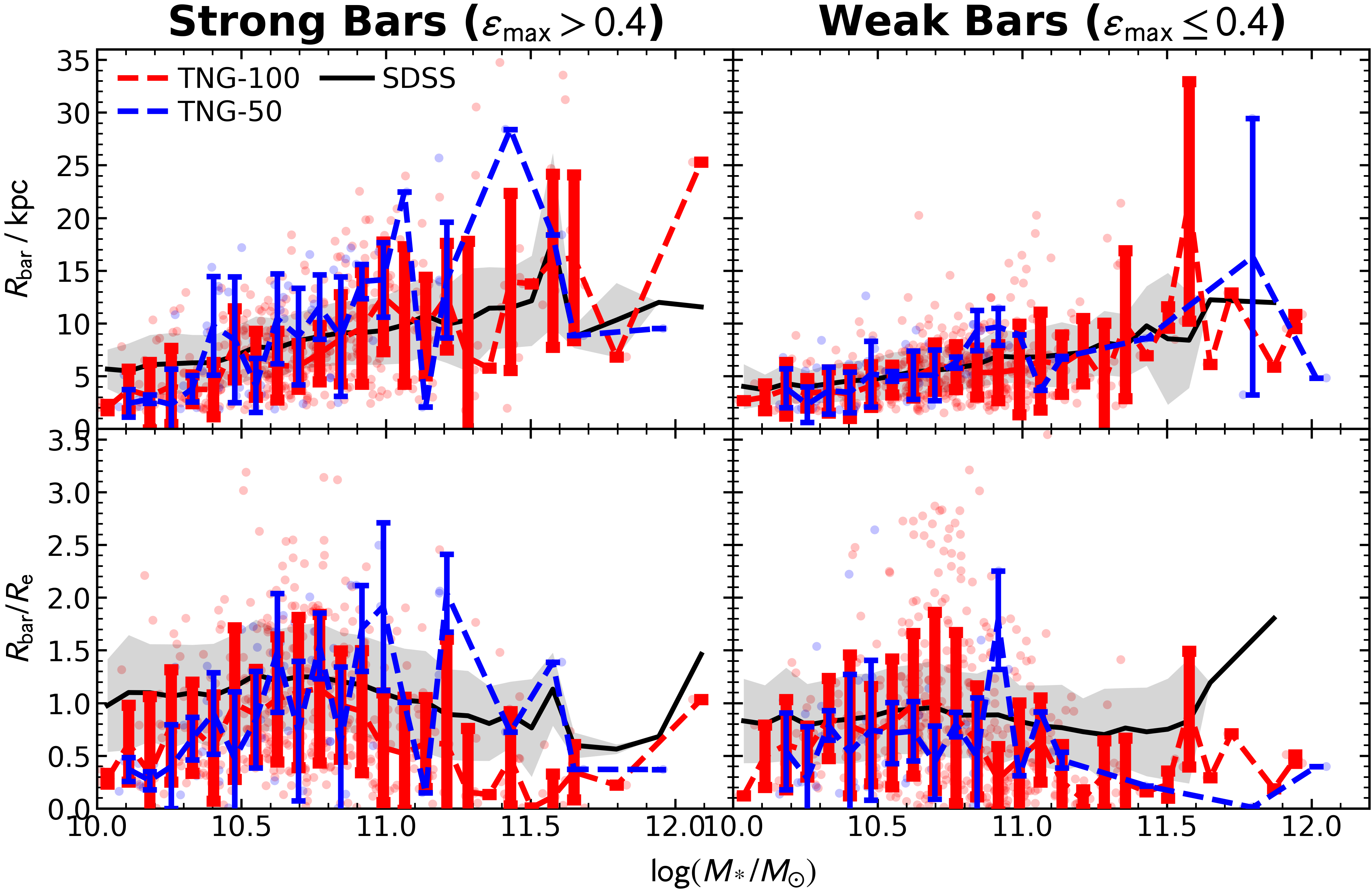}
    \caption{
    The dependence of bar properties on stellar mass for TNG and SDSS galaxies.
    The top row shows the absolute bar length ($R_{\mathrm{bar}}$), and the bottom row shows the relative bar length ($R_{\mathrm{bar}}/R_{\mathrm{e}}$).
    The semi-transparent red and blue circles in the background show the individual measurements for TNG100 and TNG50 galaxies, respectively.
    The sample is divided into strong bars ($\epsilon_{\mathrm{max}} > 0.4$, left column) and weak bars ($\epsilon_{\mathrm{max}} \leq 0.4$, right column).
    The black solid line represents the median values for SDSS, while the red and blue dashed lines correspond to TNG100 and TNG50, respectively. The shaded regions for SDSS and the error bars for TNG100 and TNG50 indicate the standard deviation within each stellar-mass bin.
    Bins containing only a single data point do not display errors, as the dispersion cannot be estimated in these cases.
    }
    \label{fig:bar_length}
\end{figure*}

The physical dimensions of bars provide valuable insights into the evolutionary history of their host galaxies. Bar size is characterized using two metrics: the absolute physical bar length ($R_{\mathrm{bar}}$) in kpc and the relative bar length normalized by the galaxy effective radius ($R_{\mathrm{bar}}/R_{\mathrm{e}}$). Based on the maximum isophotal ellipticity ($\epsilon_{\mathrm{max}}$), bars are classified into ``strong bars'' ($\epsilon_{\mathrm{max}} > 0.4$) and ``weak bars'' ($\epsilon_{\mathrm{max}} \leq 0.4$), consistent with widely adopted observational criteria \citep[e.g.,][]{Jogee2004, Barazza2008}.

The effective radius ($R_{\mathrm{e}}$) for each galaxy is derived via 2D surface brightness profile fitting using \texttt{GALFIT}. The light distribution of each galaxy is modeled with a single-component Sérsic profile, defined as:
\begin{equation}
    I(r) = I_e \exp \left\{ -\kappa \left[ \left( \frac{r}{R_{\rm e}} \right)^{1/n} - 1 \right] \right\},
\end{equation}
where $I(r)$ signifies the intensity at radius $r$, $I_e$ is the intensity at the effective radius $R_{\mathrm{e}}$, $n$ is the Sérsic index, and $\kappa$ is a coefficient associated with $n$. For the SDSS observational sample, a custom PSF is constructed for each galaxy using field stars selected from the surrounding wide-field frame to accurately model the seeing conditions. For the TNG synthetic images, the identical double-Gaussian PSF kernel used during the image convolution step (see Section~\ref{sec:mock_obs}) is employed, ensuring that seeing effects are consistently treated in the structural fitting of the simulated and observational samples.

Figure~\ref{fig:Rbar_hist} shows the overall distributions of the absolute bar length, $R_{\mathrm{bar}}$. The SDSS distribution has a median bar length of $6.67\,\mathrm{kpc}$ and a 16th--84th percentile range of $3.87$--$10.53\,\mathrm{kpc}$. The TNG100 distribution substantially overlaps with that of SDSS, with a median of $5.91\,\mathrm{kpc}$ and a 16th--84th percentile range of $3.18$--$10.69\,\mathrm{kpc}$. TNG50 exhibits a broader distribution, with a median of $6.02\,\mathrm{kpc}$ and a 16th--84th percentile range of $2.39$--$12.91\,\mathrm{kpc}$, although this broader range should be interpreted with caution because of its smaller sample size.

Figure~\ref{fig:bar_length} presents the dependence of bar length on stellar mass for strong (left column) and weak (right column) bars. In terms of absolute size (top panels), the median $R_{\mathrm{bar}}$ relations in TNG100 and TNG50 broadly agree with SDSS over most of the stellar-mass range, although larger fluctuations appear toward the mass extremes. For example, at $M_* \approx 10^{10.5}\,M_{\odot}$, the median $R_{\mathrm{bar}}$ values in all three samples are approximately $7.5\,\mathrm{kpc}$ for strong bars and $5\,\mathrm{kpc}$ for weak bars. In contrast, the relative bar lengths (bottom panels) show clearer differences, with both simulations tending toward smaller $R_{\mathrm{bar}}/R_{\mathrm{e}}$ at the low- and high-mass ends, particularly for TNG100.

For strong bars, the median $R_{\mathrm{bar}}$ in SDSS increases from $\sim 5.5\,\mathrm{kpc}$ at $M_* \approx 10^{10}\,M_{\odot}$ to $\sim 12\,\mathrm{kpc}$ at $M_* \approx 10^{11.5}\,M_{\odot}$. TNG100 shows a similar increasing trend, and its plotted within-bin dispersion ranges overlap those of SDSS in several stellar-mass bins, with the median $R_{\mathrm{bar}}$ increasing from $\sim 2\,\mathrm{kpc}$ at $M_* \approx 10^{10}\,M_{\odot}$ to $\sim 13.5\,\mathrm{kpc}$ at $M_* \approx 10^{11.5}\,M_{\odot}$. TNG50 also shows an overall increase with stellar mass but exhibits more pronounced bin-to-bin fluctuations, with the median $R_{\mathrm{bar}}$ increasing from $\sim 2\,\mathrm{kpc}$ at $M_* \approx 10^{10.1}\,M_{\odot}$ to $\sim 14\,\mathrm{kpc}$ at $M_* \approx 10^{11}\,M_{\odot}$. At higher stellar masses, the small number of TNG50 galaxies prevents a robust characterization of the trend. Among weak bars, the intervals traced by the within-bin standard deviations intersect for the three samples in some stellar-mass bins. The median $R_{\mathrm{bar}}$ in both SDSS and TNG100 increases from approximately $2$--$3\,\mathrm{kpc}$ at $M_* \approx 10^{10}\,M_{\odot}$ to $\sim 8\,\mathrm{kpc}$ at $M_* \approx 10^{11.3}\,M_{\odot}$, while TNG50 follows a similar trend, increasing to $\sim 8\,\mathrm{kpc}$ by $M_* \approx 10^{11}\,M_{\odot}$.

The relative bar length ($R_{\mathrm{bar}}/R_{\mathrm{e}}$; Figure~\ref{fig:bar_length}, bottom panels) shows larger differences between the simulations and observations. Strong bars in SDSS exhibit a mild rise-and-fall trend: the median $R_{\mathrm{bar}}/R_{\mathrm{e}}$ increases from $\sim 0.98$ at $M_* \approx 10^{10}\,M_{\odot}$ to a peak of $\sim 1.25$ at $M_* \approx 10^{10.7}\,M_{\odot}$, before declining to $\sim 0.74$ at $M_* \approx 10^{11.5}\,M_{\odot}$. TNG100 shows a similar overall trend but with more pronounced variation, rising from $\sim 0.24$ at $M_* \approx 10^{10}\,M_{\odot}$ to $\sim 1.2$ at $M_* \approx 10^{10.7}\,M_{\odot}$ and subsequently declining toward higher stellar masses. TNG50 broadly follows this behavior, with relative bar length increasing toward intermediate stellar masses but showing larger bin-to-bin fluctuations. For weak bars, the SDSS median $R_{\mathrm{bar}}/R_{\mathrm{e}}$ remains relatively stable at approximately $0.8$--$0.9$ over $10^{10} \lesssim M_* \lesssim 10^{11.5}\,M_{\odot}$. Although the SDSS standard-deviation bands intersect the TNG error bars in some stellar-mass bins, the median trends differ among the three samples. TNG100 shows a pronounced rise-and-fall trend, increasing from $\sim 0.1$ at $M_* \approx 10^{10}\,M_{\odot}$ to a peak of $\sim 0.98$ at $M_* \approx 10^{10.7}\,M_{\odot}$, before declining to $\sim 0.2$ at $M_* \approx 10^{11.5}\,M_{\odot}$. In contrast, TNG50 remains relatively stable at approximately $0.3$--$0.7$ over $10^{10.2} \lesssim M_* \lesssim 10^{11.1}\,M_{\odot}$.

At the low stellar-mass end, strong bars in both TNG100 and TNG50 are shorter than those in SDSS in both absolute and relative terms, indicating that their lower $R_{\mathrm{bar}}/R_{\mathrm{e}}$ values primarily reflect smaller absolute bar sizes. Toward higher stellar masses, the median absolute bar lengths in TNG100 become broadly comparable to those measured in our SDSS sample, while the median relative bar lengths remain systematically smaller, particularly at the high-mass end. This contrast suggests that differences in the size scaling of the host galaxies may contribute to the discrepancy in $R_{\mathrm{bar}}/R_{\mathrm{e}}$, in addition to any differences in the physical lengths of the bars themselves. For weak bars, the within-bin standard-deviation intervals in absolute bar length share a common range across the three samples in some stellar-mass bins, despite their different median trends in $R_{\mathrm{bar}}/R_{\mathrm{e}}$, again suggesting an important contribution from differences in galaxy size. TNG50 remains more difficult to characterize at high stellar masses because of its smaller sample size. We examine these structural differences further in Section~\ref{sec:discussion_structure}.

\section{Discussion} \label{sec:discussion}

\subsection{Structural Differences Associated with Bar Statistics and Scaling Relations} \label{sec:discussion_structure}

\begin{figure*}[thpb]
    \centering
    \includegraphics[width=\textwidth]{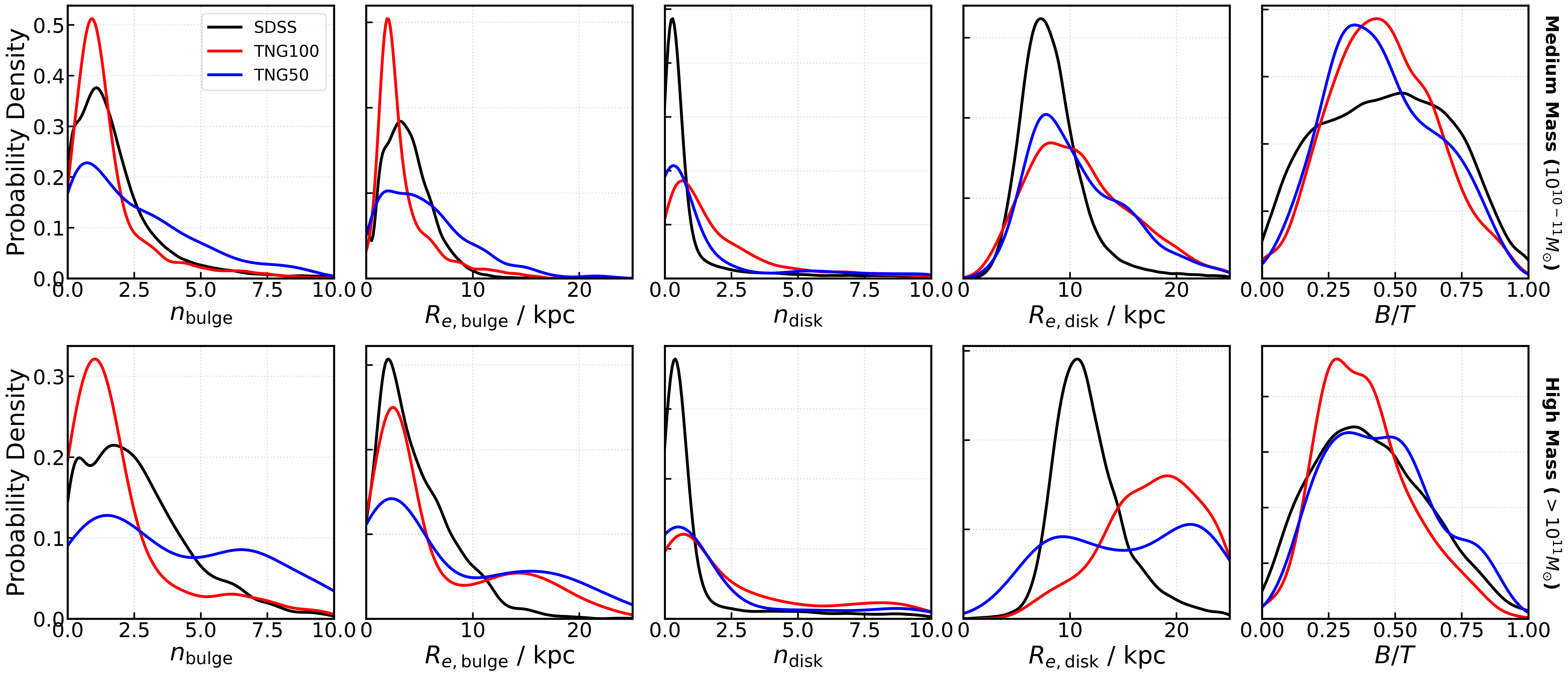}
    \caption{The distribution of structural parameters derived from the double-Sérsic \texttt{GALFIT} decomposition. The top row displays galaxies in the intermediate stellar mass range ($10^{10} < M_* < 10^{11} M_{\odot}$), while the bottom row shows the high stellar mass regime ($M_* > 10^{11} M_{\odot}$). From left to right, the panels represent the bulge Sérsic index ($n_{\rm bulge}$), bulge effective radius ($R_{\rm{e, bulge}}$), disk Sérsic index ($n_{\rm disk}$), disk effective radius ($R_{\rm{e, disk}}$), and the Bulge-to-Total flux ratio ($B/T$). Colors indicate the different datasets: SDSS (black), TNG100 (red), and TNG50 (blue).}
    \label{fig:decomposition}
\end{figure*}

\begin{figure*}[thpb]
    \centering
    \includegraphics[width=\textwidth]{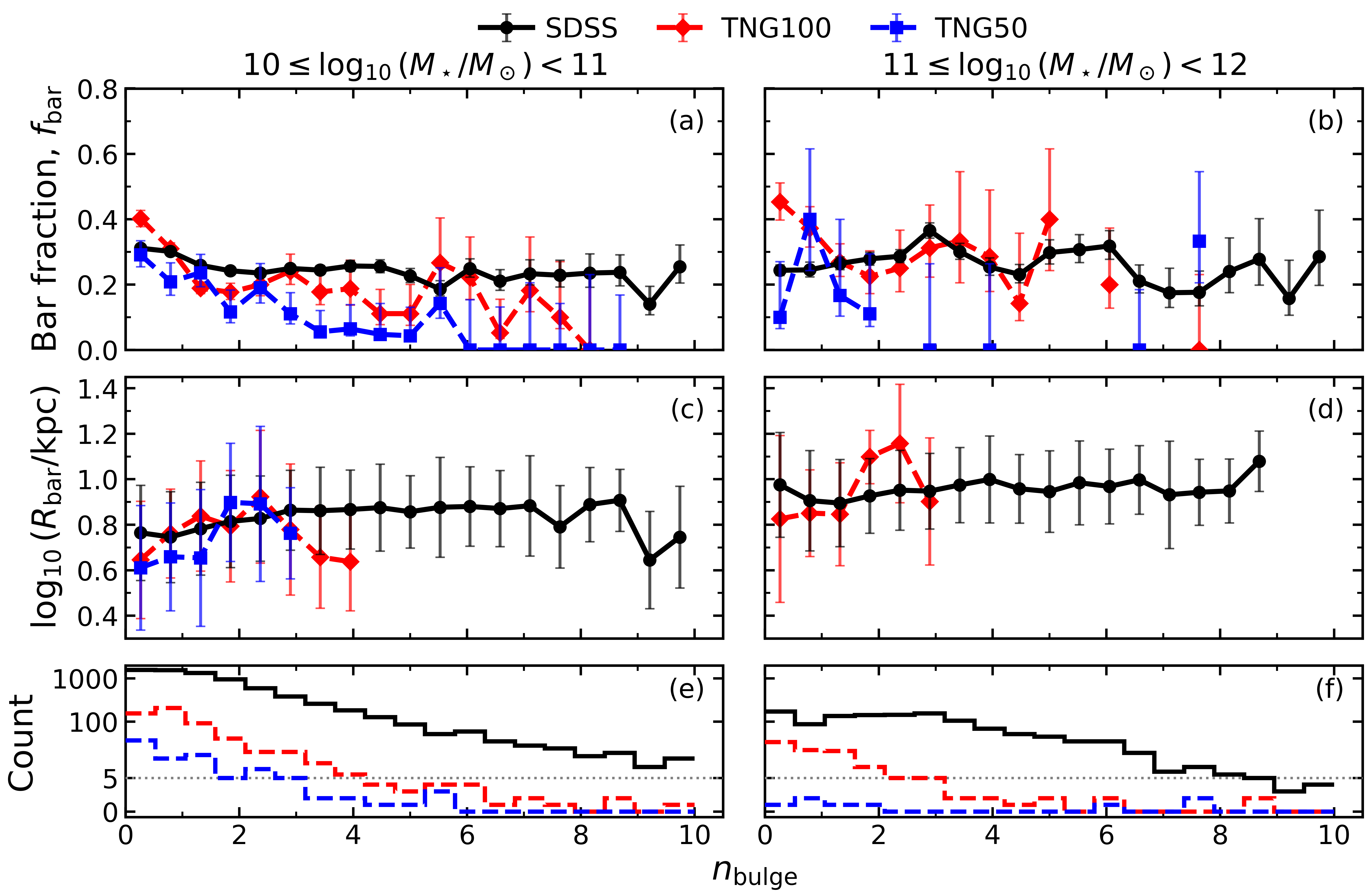}
    \caption{Bar fraction ($f_{\rm bar}$, top), median logarithmic bar length ($\log_{10}(R_{\rm bar}/\mathrm{kpc})$, middle), and barred-galaxy counts (bottom) as functions of the bulge Sérsic index ($n_{\rm bulge}$). The left and right columns show intermediate-mass ($10.0\leq\log_{10}(M_*/M_\odot)<11.0$) and high-mass ($11.0\leq\log_{10}(M_*/M_\odot)<12.0$) galaxies, respectively. Error bars show beta-distribution intervals for $f_{\rm bar}$ and within-bin standard deviations for logarithmic bar length. The upper panels include bins containing at least five galaxies in the parent sample, while the middle panels additionally require at least five barred galaxies per bin. The bottom row includes all bins, with the dotted line marking a count of five barred galaxies. All samples use 19 equal-width bins over $0\leq n_{\rm bulge}\leq10$. The count axis is linear up to 5 and logarithmic above 5. SDSS, TNG100, and TNG50 are shown as black circles, red diamonds, and blue squares, respectively.}
    \label{fig:nbulge}
\end{figure*}

As shown in Section~\ref{sec:results}, TNG100 shows good agreement with the SDSS $f_{\rm bar}$ over the intermediate stellar-mass range but remains systematically higher at the high-mass end, where the SDSS bar fraction declines. TNG50, in contrast, exhibits a substantially lower $f_{\rm bar}$ over most of its well-sampled mass range, although the SDSS measurements approach the TNG50 values at the highest masses, where the limited TNG50 sample prevents a detailed comparison. In addition, massive TNG100 galaxies have absolute bar lengths comparable to those in SDSS but systematically smaller relative bar lengths ($R_{\rm bar}/R_{\rm e}$).

To investigate the physical drivers of these discrepancies, we perform a two-component structural decomposition using \texttt{GALFIT}, modeling the light distribution of each galaxy as a superposition of two Sérsic components. Systems where either component yields an unphysical Sérsic index ($n > 10$) are excluded exclusively in the analysis of this subsection. This cut removes 4,893 SDSS galaxies, 591 TNG100 images, and 270 TNG50 images with otherwise valid two-component fits; the TNG counts include the three projections. We adopt the component-switching criterion of \citet{Lange2016}, starting with component 1 as the nominal disk and component 2 as the nominal bulge. The components are exchanged if $R_{\rm e,bulge}>1.1R_{\rm e,disk}$, or if $R_{\rm e,disk}<R_{\rm e,bulge}\leq1.1R_{\rm e,disk}$ and $n_{\rm disk}>n_{\rm bulge}$; otherwise their labels are retained. Thus, the bulge is not required to have the higher S\'ersic index.

Figure~\ref{fig:decomposition} presents the derived structural parameters. In the intermediate stellar-mass regime ($10^{10} < M_* < 10^{11} M_{\odot}$, top row), the $n_{\rm bulge}$ distributions of all three samples peak at similarly low values near $n_{\rm bulge}\sim0.7$--$1.1$, but their overall distributions differ. TNG100 is more strongly concentrated toward low $n_{\rm bulge}$, while SDSS exhibits a somewhat broader distribution extending toward higher Sérsic indices. TNG50 shows an even broader high-$n_{\rm bulge}$ tail, indicating a larger population of more concentrated bulges. The disk-size distributions also differ moderately: the SDSS $R_{\rm e,disk}$ distribution peaks at $\sim 7.2$~kpc, TNG50 at $\sim 7.7$~kpc, and TNG100 at  $\sim 8.2$~kpc, with the simulated samples showing broader distributions toward larger disk sizes. In the high stellar-mass regime ($M_* > 10^{11} M_{\odot}$, bottom row), the differences in bulge structure become substantially stronger. TNG100 remains dominated by low-$n_{\rm bulge}$ systems, SDSS contains a broader population extending toward higher $n_{\rm bulge}$, and TNG50 exhibits the strongest high-$n_{\rm bulge}$ component, including a broad secondary enhancement at high Sérsic index. The high-mass $R_{\rm e,disk}$ distributions peak at approximately $10.7$~kpc for SDSS and $\sim 19$~kpc for TNG100, while TNG50 shows a broad distribution with enhancements near $9$ and $21$~kpc.

The differences in the $n_{\rm bulge}$ distributions are associated with the measured $f_{\rm bar}$. At intermediate stellar masses, the relatively similar $n_{\rm bulge}$ distributions of TNG100 and SDSS are accompanied by their close agreement in $f_{\rm bar}$. In contrast, TNG50 contains a larger fraction of high-$n_{\rm bulge}$ systems than SDSS over both stellar-mass regimes and shows a systematically lower $f_{\rm bar}$ over most of the well-sampled mass range. At high stellar masses, TNG100 shows the opposite behavior, with a clear deficit of high-$n_{\rm bulge}$ systems relative to SDSS while maintaining a higher $f_{\rm bar}$. These results suggest an association between the central structure of galaxies and their measured bar fractions.

Such a connection is consistent with previous theoretical studies showing that central mass concentrations can suppress or weaken bars by modifying the gravitational potential and disrupting bar-supporting orbital families \citep{Shen2004, Athanassoula2005}, while angular-momentum transfer through gravitational torques may further contribute to bar evolution \citep{Bournaud2005}. The larger population of concentrated bulges in TNG50 and the relative lack of such systems in massive TNG100 galaxies may therefore contribute to their lower and higher bar fractions, respectively. However, bulge concentration is unlikely to be the sole factor controlling $f_{\rm bar}$. In particular, TNG100 and SDSS show very similar bar fractions at intermediate stellar masses despite differences in their $n_{\rm bulge}$ distributions, suggesting that other structural and dynamical properties also play a role. To assess whether the different $f_{\rm bar}$ values in TNG100 and TNG50 could be primarily associated with their different simulation volumes, we further divide TNG100 into eight equal subvolumes (Appendix~\ref{app:subvolume}). The resulting global bar fractions have a median of $25.13\%$ and a standard deviation of $3.40$ percentage points, compared with $24.96\%$ for the full TNG100 sample, and range from $20.23\%$ to $29.24\%$. Even the lowest subvolume value remains substantially above the TNG50 value of $13.42\%$, suggesting that finite-volume effects alone are unlikely to account for the TNG50--TNG100 difference and that numerical resolution or other resolution-dependent galaxy properties may play a more important role.

The disk-size distributions provide a more direct connection to the relative bar length discrepancy. Massive TNG100 galaxies have substantially larger $R_{\rm e,disk}$ than SDSS, which may contribute to their smaller $R_{\rm bar}/R_{\rm e}$ even when their absolute bar lengths are comparable to those observed in SDSS. TNG50 shows a broader disk-size distribution, spanning both SDSS-like and more extended disks. The extended disks in TNG100 may be related in part to numerical resolution and associated effects involving the stellar-to-halo mass relation, two-body heating, and gravitational softening; \citet{Pillepich2019} showed that higher-resolution TNG runs generally produce more compact galaxy sizes.

To further constrain the connection between bars and bulges, Figure~\ref{fig:nbulge} plots $f_{\rm bar}$ and absolute bar length as a function of $n_{\rm bulge}$. To reduce small-number fluctuations, the $f_{\rm bar}$ panels include only bins containing at least five galaxies in the parent sample, while the bar-length panels additionally require at least five barred galaxies per bin. While $f_{\rm bar}$ exhibits substantial statistical fluctuations and shows no clear monotonic dependence on $n_{\rm bulge}$ across the simulated samples, the absolute bar length shows a more consistent dependence in both stellar-mass regimes. In SDSS, bar length generally increases at low \(n_{\rm bulge}\) and shows a weaker dependence at higher \(n_{\rm bulge}\). The simulated samples exhibit larger fluctuations, and no high-mass TNG50 bin satisfies $N_{\rm bar}\geq5$, preventing a bar-length comparison in this regime.

This behavior suggests an association between bar length and bulge concentration in the low-concentration regime ($n_{\rm bulge}<2$), whereas the dependence is weaker at higher $n_{\rm bulge}$ in SDSS. The absence of a clear decline in bar length at larger $n_{\rm bulge}$ provides no clear evidence that high-concentration systems systematically host shorter bars. Together with the non-monotonic behavior of $f_{\rm bar}$ with $n_{\rm bulge}$, these results suggest that central concentration is associated with bar properties but is unlikely to be the sole factor underlying the observed trends.

In summary, the structural decomposition shows clear differences in the bulge and disk properties of SDSS, TNG100, and TNG50 that are associated with their different bar statistics. In both stellar-mass regimes, TNG100 generally contains fewer high-$n_{\rm bulge}$ systems than SDSS, whereas TNG50 contains a larger fraction of concentrated bulges. This trend supports a possible role of central structure in regulating $f_{\rm bar}$, although the differences in $n_{\rm bulge}$ alone cannot fully account for the measured bar fractions, indicating that other structural and dynamical properties are also important. Moreover, we find no clear evidence that high-$n_{\rm bulge}$ systems systematically host shorter bars. Because our analysis is based on a single epoch, it cannot directly determine whether concentrated bulges suppress bar formation, limit subsequent bar growth, or destroy pre-existing bars. In addition, the substantially larger disk sizes of massive TNG100 galaxies may contribute to their smaller $R_{\rm bar}/R_{\rm e}$ despite having absolute bar lengths comparable to those in SDSS.

\subsection{Bar-Length Definition and Extreme Measurements}
\label{sec:bar_length_definition}

\begin{figure}[thpb]
    \centering
    \includegraphics[width=\columnwidth]{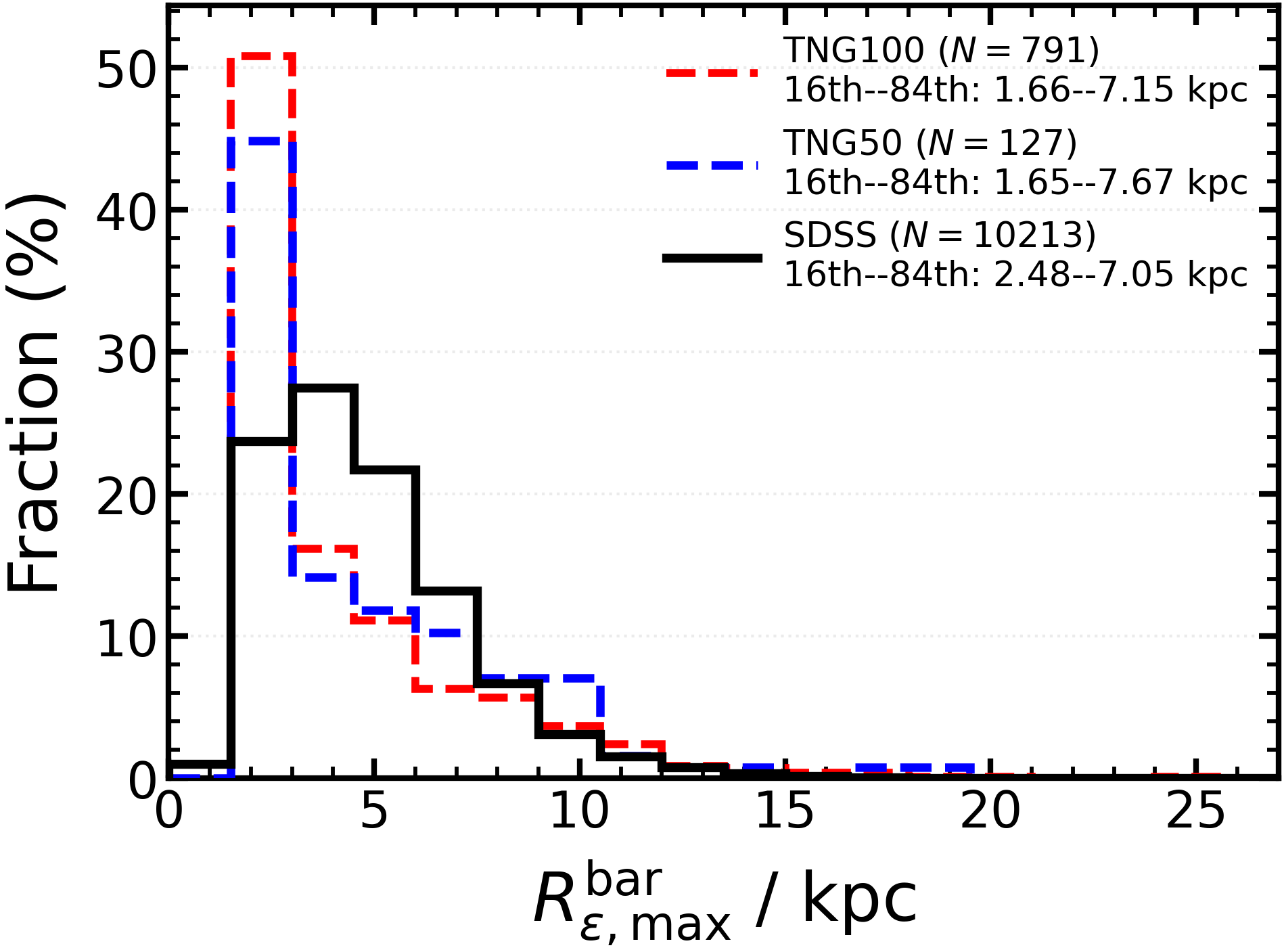}
    \caption{
        Same as Figure~\ref{fig:Rbar_hist}, but for $R_{\epsilon,\max}^{\rm bar}$, the radius of maximum ellipticity within the identified bar region. The same binning scheme is adopted, with a bin width of $1.5\,\mathrm{kpc}$.
    }
    \label{fig:Rmax_hist}
\end{figure}

\begin{figure*}[thpb]
    \centering
    \includegraphics[width=\textwidth]{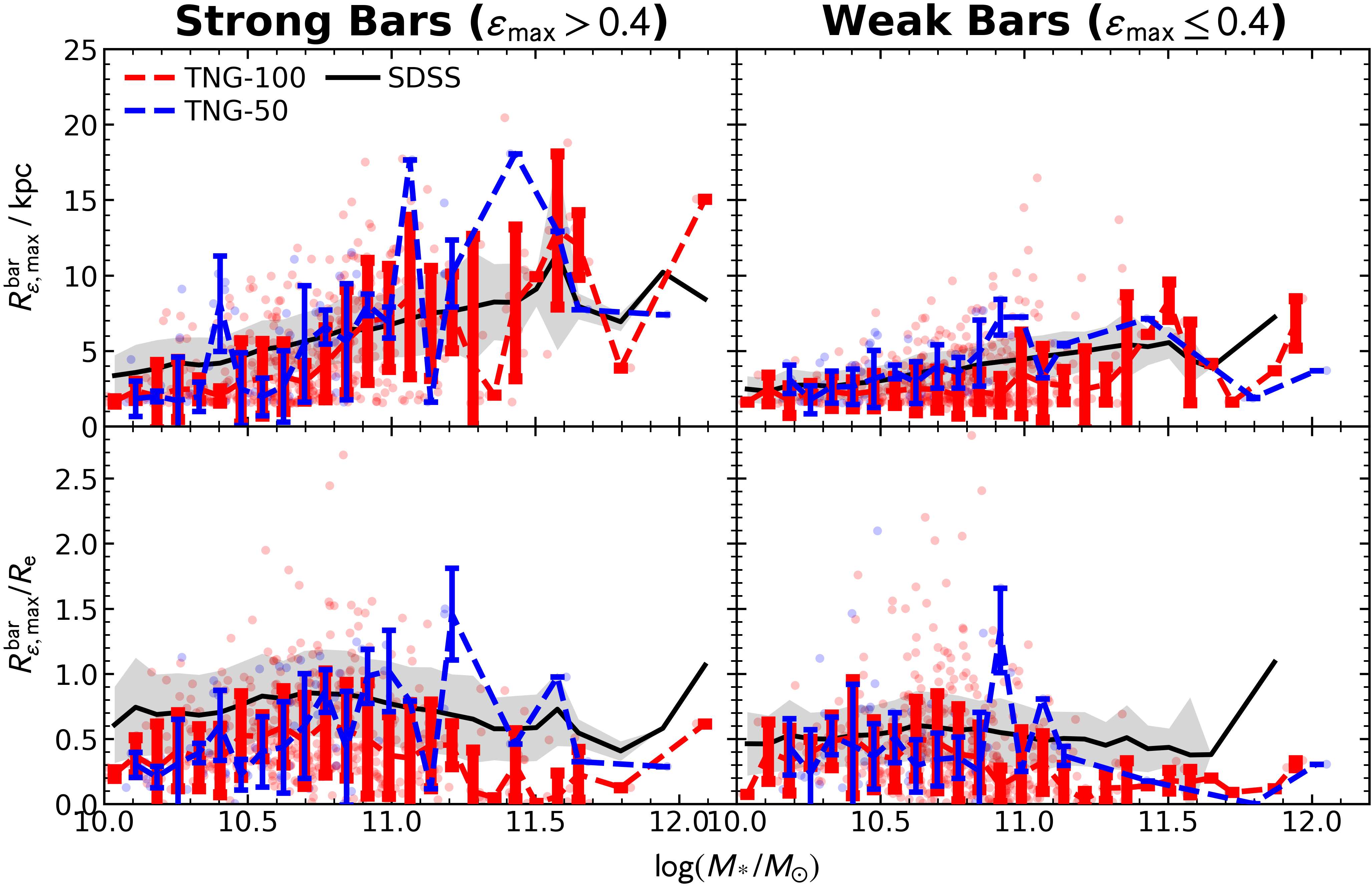}
    \caption{
        Same as Figure~\ref{fig:bar_length}, but using $R_{\epsilon,\max}^{\rm bar}$ instead of $R_{\mathrm{bar}}$.
        The top and bottom rows show $R_{\epsilon,\max}^{\rm bar}$ and $R_{\epsilon,\max}^{\rm bar}/R_{\mathrm{e}}$, respectively.
    }
    \label{fig:bar_length_Rmax}
\end{figure*}

Figure~\ref{fig:Rbar_hist} shows that the $R_{\mathrm{bar}}$ distributions of SDSS, TNG100, and TNG50 overlap substantially but are broader than those commonly reported for nearby barred galaxies. The SDSS sample has a median $R_{\mathrm{bar}}$ of $6.67\,\mathrm{kpc}$, with a 16th--84th percentile range of $3.87$--$10.53\,\mathrm{kpc}$. TNG100 has a median of $5.91\,\mathrm{kpc}$ and a corresponding range of $3.18$--$10.69\,\mathrm{kpc}$, while TNG50 has a median of $6.02\,\mathrm{kpc}$ and a broader range of $2.39$--$12.91\,\mathrm{kpc}$. The fractions above 10, 12, and $15\,\mathrm{kpc}$ are $19.85\%$, $12.51\%$, and $6.20\%$ in TNG100, $23.62\%$, $18.90\%$, and $10.24\%$ in TNG50, and $19.17\%$, $9.27\%$, and $2.80\%$ in SDSS, respectively. By comparison, nearby optical and near-infrared studies typically find median bar lengths of $\sim3$--$5\,\mathrm{kpc}$, with most bars below $\sim10$--$14\,\mathrm{kpc}$ \citep[e.g.,][]{Erwin2005, Marinova2007, Menendez2007, DiazGarcia2016, Erwin2019}. In SDSS-based studies, \citet{Aguerri2009} found bar lengths spanning approximately 1--$12\,\mathrm{kpc}$, with a median near $3.5\,\mathrm{kpc}$, while \citet{Gadotti2011} reported a median near $4.5\,\mathrm{kpc}$ and a full range of approximately 1.5--$13.5\,\mathrm{kpc}$. \citet{Neumann2024} reported a median bar length of $4.6\,\mathrm{kpc}$ for the MaNGA sample; although a few bars reach larger sizes, most remain below $\sim15\,\mathrm{kpc}$. The broader distributions in Figure~\ref{fig:Rbar_hist} therefore motivate an examination of the dependence of these measurements on the adopted bar length definition, particularly because the ellipticity-peak radius has been widely used as a bar length estimator in previous observational studies.

As described in Section~\ref{sec:bar_measurement}, we measure bar length using both $R_{\mathrm{bar}}=R_{\rm 0.85max}$ and the ellipticity-peak estimator $R_{\epsilon,\max}^{\rm bar}$. Figure~\ref{fig:Rmax_hist} shows that adopting $R_{\epsilon,\max}^{\rm bar}$ shifts the distributions to smaller radii. For TNG100, the median decreases from $5.91$ to $2.82\,\mathrm{kpc}$, and the 16th--84th percentile range changes from $3.18$--$10.69\,\mathrm{kpc}$ to $1.66$--$7.15\,\mathrm{kpc}$. For TNG50, the median decreases from $6.02$ to $3.42\,\mathrm{kpc}$, with the range changing from $2.39$--$12.91\,\mathrm{kpc}$ to $1.65$--$7.67\,\mathrm{kpc}$. For SDSS, the median decreases from $6.67$ to $4.34\,\mathrm{kpc}$, with the range changing from $3.87$--$10.53\,\mathrm{kpc}$ to $2.48$--$7.05\,\mathrm{kpc}$. The fractions above 10, 12, and $15\,\mathrm{kpc}$ decrease to $6.07\%$, $2.78\%$, and $1.14\%$ in TNG100, $7.87\%$, $3.15\%$, and $1.58\%$ in TNG50, and $3.78\%$, $1.45\%$, and $0.32\%$ in SDSS, respectively.

The corresponding stellar-mass relations are shown in Figure~\ref{fig:bar_length_Rmax}. Relative to the results obtained with $R_{\mathrm{bar}}$ in Figure~\ref{fig:bar_length}, replacing $R_{\mathrm{bar}}$ with $R_{\epsilon,\max}^{\rm bar}$ mainly lowers the absolute and relative bar lengths without substantially changing the shapes of the median relations. Bar length still increases with stellar mass, and the principal differences between TNG100, TNG50, and SDSS remain qualitatively similar for both strong and weak bars. Thus, the choice of bar length estimator mainly shifts the measured length distributions toward smaller radii, without qualitatively changing the overall stellar-mass dependence.

At the small-length end, Figures~\ref{fig:Rbar_hist} and \ref{fig:bar_length} show that both TNG100 and TNG50 exhibit a higher fraction of short bars than SDSS, particularly bars with $R_{\mathrm{bar}}\lesssim5\,\mathrm{kpc}$ at relatively low stellar masses ($10.0 \lesssim \log(M_*/M_\odot) \lesssim 10.5$). This short-bar excess is qualitatively consistent with previous TNG-based studies. \citet{Zhao2020} found that TNG100 overproduces short bars with radii of $\sim1.4$--$3\,\mathrm{kpc}$ relative to the near-infrared mass--bar-size relation. \citet{Frankel2022} found TNG50 bars to be shorter on average than bars in their MaNGA comparison sample, and showed that excluding simulated bars shorter than $2.5\,\mathrm{kpc}$ reduces the discrepancy. \citet{Habibi2024} likewise reported smaller TNG50 bars than in a local CALIFA+MaNGA sample and found no strong positive bar length--stellar-mass correlation comparable to observations. \citet{Lu2025} further separated short-bar galaxies in TNG100 and TNG50 and argued that such bars are either limited in growth or become shortened, with mergers, central mass density, and gravitational softening as possible contributors.

A limitation of the ellipse-based method is that morphologically ambiguous or projected systems can produce bar-like ellipticity profiles across the full range of measured bar lengths, with such ambiguity becoming more prominent toward the long-bar tail. Previous studies of IllustrisTNG have also identified elongated bar-like systems outside the conventional population of bars embedded in rotationally supported disks \citep{Lokas2021,Du2026}. Some of these structures may correspond to genuine elongated stellar features rather than being produced solely by projection effects. Our image-based analysis provides a photometric characterization of these systems and offers an additional view of their apparent bar-like structures. Representative examples for SDSS, TNG100, and TNG50 are shown in Appendix~\ref{app:extreme_cases}.

In summary, the broader $R_{\mathrm{bar}}$ distributions relative to much of the previous literature are partly associated with the use of the $R_{\rm 0.85max}$ definition; when $R_{\epsilon,\max}^{\rm bar}$ is used instead, the distributions shift toward smaller radii and become closer to previous measurements based on the ellipticity peak. Both TNG100 and TNG50 nevertheless retain an excess of short bars relative to SDSS, indicating a remaining difference between the simulated and observed bar-size distributions. Importantly, the positive dependence of median bar length on stellar mass remains present when $R_{\epsilon,\max}^{\rm bar}$ is adopted, indicating that the main bar length trend is not an artifact of the adopted outer bar length definition.

\subsection{Comparison with Previous Works
}\label{sec:comparison_previous_works}

\begin{figure*}[thpb]
    \centering
    \includegraphics[width=\textwidth]{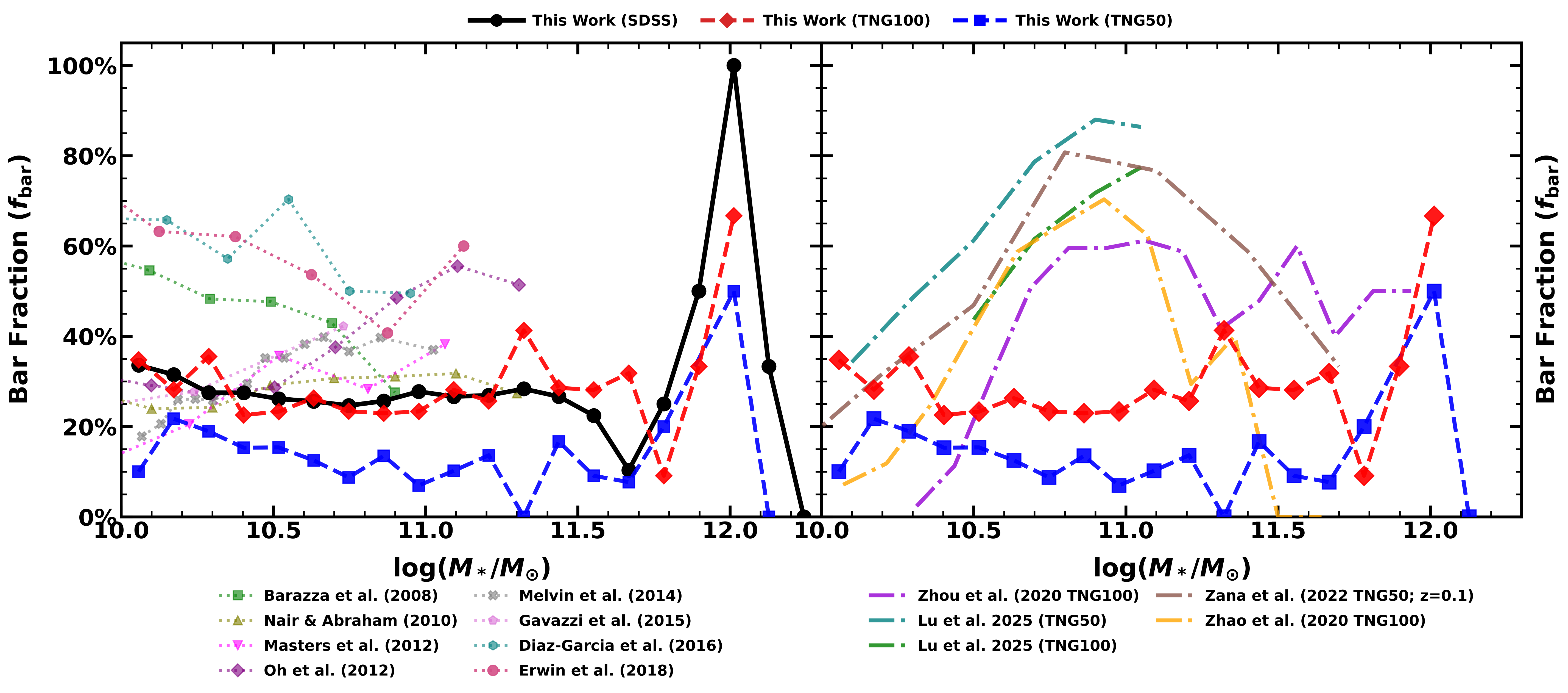}
    \caption{
    The bar fraction, $f_{\rm bar}$, as a function of stellar mass,
    $\log(M_*/M_\odot)$. Left panel: Our SDSS measurements are
    shown as black circles connected by a solid line, while TNG100 and TNG50 are shown as red diamonds and blue squares, respectively, connected by dashed lines. Previous observational measurements are shown with markers connected by dotted lines \citep{Barazza2008, Nair2010, Masters2012, Oh2012, Melvin2014, Gavazzi2015, DiazGarcia2016, Erwin2018}. Right panel: Our TNG100 and TNG50 measurements are shown using the same colors, markers, and line styles as in the left panel, while previous cosmological simulation results are shown as dash-dotted lines without markers \citep{Zhao2020, Zhou2020, Zana2022, Lu2025}.
    }
    \label{fig:bar_fraction_compare}
\end{figure*}

We compare the derived $f_{\rm bar}$ with previous observational and simulation studies. Our SDSS measurements remain relatively flat and decline only toward higher stellar masses. Among the previous observational studies shown in Figure~\ref{fig:bar_fraction_compare}, the results of \citet{Nair2010} show the closest overall agreement with our measurements, including the decline at the high-mass end, while \citet{Masters2012} also yields broadly comparable bar fractions over much of the overlapping stellar-mass range. In contrast, \citet{Oh2012}, \citet{Melvin2014}, and \citet{Gavazzi2015} generally report higher bar fractions and a stronger increase with stellar mass. The measurements of \citet{Barazza2008}, \citet{DiazGarcia2016}, and \citet{Erwin2018} are also systematically higher than ours. As suggested by \citet{Erwin2018}, the relatively high $f_{\rm bar}$ reported by \citet{Barazza2008} may be partly related to their lower-redshift sample, for which the imaging provides finer physical resolution and facilitates the detection of smaller bars. The higher fractions in \citet{DiazGarcia2016} and \citet{Erwin2018} are based on \textit{Spitzer} near-infrared imaging, which is less affected by dust obscuration than optical observations and thus facilitates bar detection.

Our SDSS data are broadly consistent with multiple previous studies over $10^{10} < M_* < 10^{10.5}\,M_\odot$, but tend to show lower $f_{\rm bar}$ at $M_* > 10^{10.5}\,M_\odot$. Our sample lies at slightly higher redshift than these previous samples, suggesting that the difference at the high-mass end may partly reflect the greater detectability of bars at lower redshift. Together with the higher bar fractions obtained from \textit{Spitzer} near-infrared imaging, these comparisons suggest that bar fractions measured from SDSS optical images may underestimate the underlying bar fraction because of limitations in spatial resolution and wavelength coverage.

Furthermore, the previous TNG-based studies shown in the right panel of Figure~\ref{fig:bar_fraction_compare} generally recover a broad intermediate-mass enhancement in $f_{\rm bar}$, with lower values toward both lower and higher stellar masses and the highest fractions concentrated around $M_* \sim 10^{10.5}$--$10^{11}\,M_\odot$ \citep{Zhao2020, Zhou2020, Zana2022, Lu2025}. In contrast, our TNG100 and TNG50 measurements show a substantially flatter stellar-mass dependence and more closely resemble the relation obtained from our SDSS sample, particularly for TNG100. For \citet{Zana2022}, we recompute $f_{\rm bar}$ from their public TNG50-1 catalogue at Snapshot 91 ($z\simeq0.1$) using their published disc and bar criteria.\footnote{We select disc galaxies satisfying $M_{\rm thin}+M_{\rm thick}+M_{\rm pseudo-bulge}>0.5M_*$, where the three terms denote the stellar masses assigned by MORDOR to the thin-disc, thick-disc, and pseudo-bulge components, respectively. We further require $\kappa_{\rm rot}\geq0.4$, where $\kappa_{\rm rot}$ measures the fraction of stellar kinetic energy in ordered rotation, and select objects with the catalogue bar flag and $A_{2,\max}(R)\geq0.2$, where $A_{2,\max}$ is the maximum amplitude of the $m=2$ Fourier mode of the stellar surface-density distribution and measures bar strength. This corresponds to their sample of proper bars.}

Because our comparison with \citet{Zana2022} is based on the same TNG50 snapshot, it provides a particularly useful case for examining the effects of different measurement frameworks. \citet{Zana2022} characterize galaxies from their intrinsic stellar kinematics and identify bars from the $m=2$ Fourier mode of the stellar surface-density distribution. Our analysis instead follows a more observation-oriented approach: we begin with radiative-transfer surface-brightness projections of entire galaxy groups, apply SDSS-like observational effects to the images, and then perform source segmentation, image-based morphological classification, and isophotal bar identification. These methodological differences can jointly affect sample membership, the LTG denominator, bar detectability, and the measured properties of bars, although the contribution of each individual step cannot be isolated here. A direct comparison of the bar-length measurements is presented in Appendix~\ref{app:zana_bar_length}.

This interpretation is supported by \citet{Goncalves2026}, who found that using radiative-transfer modeling together with realistic Euclid-like observational effects for TNG50 galaxies at $z=0.5$ reduced the inferred bar fraction from $44\%$ in idealized stellar-mass maps to $12$--$33\%$, with the exact value depending on the adopted detection method. Their results directly demonstrate that observational effects can reduce the measured bar fraction relative to that obtained from the intrinsic stellar distribution, while the adopted bar-identification method introduces an additional dependence.

\begin{deluxetable*}{llccc}[t!]
\tablecaption{
Galaxy Counts and Bar Fractions for This Work and Previous Studies
\label{tab:fbar_counts}
}
\tablecolumns{5}
\tablewidth{0pt}
\tablehead{
\colhead{Study} &
\colhead{Sample} &
\colhead{Morphology Counts} &
\colhead{$N_{\rm bar}/N_{\rm parent}$} &
\colhead{$f_{\rm bar}$ (\%)}
}
\startdata
\textbf{This Work}
& SDSS
& --
& $10{,}213/38{,}286$
& $26.68$
\\
\hline
\textbf{This Work}
& TNG50
& ETG/LTG: $111/1348$
& \begin{tabular}[t]{@{}c@{}}
$xy$: $48/318$ \\
$yz$: $32/307$ \\
$zx$: $47/321$ \\
Combined images: $127/946$
\end{tabular}
& \begin{tabular}[t]{@{}c@{}}
$15.09$ \\
$10.42$ \\
$14.64$ \\
$13.42$
\end{tabular}
\\
\citet{Lu2025}
& TNG50
& Non-disc/Disc: $201/615$
& $338/615$
& $54.96$
\\
\citet{Zana2022}
& TNG50
& Non-disc/Disc: $1486/2516$
& $441/2485$
& $17.75$
\\
\hline
\textbf{This Work}
& TNG100
& ETG/LTG: $778/4046$
& \begin{tabular}[t]{@{}c@{}}
$xy$: $239/1024$ \\
$yz$: $279/1043$ \\
$zx$: $273/1102$ \\
Combined images: $791/3169$
\end{tabular}
& \begin{tabular}[t]{@{}c@{}}
$23.34$ \\
$26.75$ \\
$24.77$ \\
$24.96$
\end{tabular}
\\
\citet{Zhou2020}
& TNG100
& Non-disc/Disc: $3849/2658$
& $871/2658$
& $32.77$
\\
\citet{Lu2025}
& TNG100
& Non-disc/Disc: $1383/1662$
& $911/1662$
& $54.81$
\\
\citet{Zhao2020}
& TNG100
& Non-disc/Disc: $2641/3866$
& $1182/3866$
& $30.57$
\\
\enddata
\tablecomments{
Counts for this work are restricted to
$\log_{10}(M_\star/M_\odot) > 10$.
The TNG100 sample of \citet{Zhou2020} is selected at
$M_\star > 10^{10}\,M_\odot$ using the total stellar mass, whereas
\citet{Zhao2020} adopt
$M_{\star,30,{\rm kpc}} \geq 10^{10}\,M_\odot$.
The public TNG50 catalogue of \citet{Zana2022} is restricted to galaxies
with $M_\star \gtrsim 10^{9}\,M_\odot$.
The stellar-mass ranges adopted by \citet{Lu2025} are
$10.0 \leq \log_{10}(M_{\star,30,{\rm kpc}}/M_\odot) \leq 11.1$
for TNG50 and
$10.4 \leq \log_{10}(M_{\star,30,{\rm kpc}}/M_\odot) \leq 11.1$
for TNG100.
For \citet{Zana2022}, the morphology counts are based on the MORDOR disc criterion, while the bar-fraction parent sample additionally requires $\kappa_{\rm rot}\geq0.4$.
}
\end{deluxetable*}

Table~\ref{tab:fbar_counts} further illustrates the differences in the bar fractions obtained in this work and previous TNG-based studies. Because our TNG samples include qualifying projected galaxy images from three orthogonal viewing directions, the combined parent and barred counts do not represent unique physical galaxies and are therefore not directly comparable to the galaxy counts reported in previous studies. We therefore also examine the three projections separately. For TNG100, the $xy$, $yz$, and $zx$ projections yield global bar fractions of $23.34\%$, $26.75\%$, and $24.77\%$, respectively, with a median of $24.77\%$, close to the combined-image value of $24.96\%$. For TNG50, the corresponding values are $15.09\%$, $10.42\%$, and $14.64\%$, with a median of $14.64\%$, compared with $13.42\%$ for the combined-image sample. The close agreement between the combined and individual-projection results suggests that the inferred bar fractions are unlikely to be strongly affected by any particular viewing direction or by combining the three projections.

The result of \citet{Zana2022} demonstrates that disc selection can substantially affect the denominator and hence the inferred bar fraction within a given analysis framework. However, cross-study differences in stellar-mass selection, morphological classification, projection treatment, observational conditions, and bar-identification methodology can affect both the parent sample and the identified bars. The projection-separated counts in Table~\ref{tab:fbar_counts} provide a breakdown of our results by viewing direction, although differences in stellar-mass limits and selection criteria prevent direct comparisons of absolute sample counts across studies.

For TNG100, the individual projections contain $1,024$--$1,102$ parent galaxies and $239$--$279$ identified bars. Both counts are numerically smaller than those reported by \citet{Zhao2020}, \citet{Zhou2020}, and \citet{Lu2025}, while all three projections also give lower $f_{\rm bar}$. Similarly, the TNG50 projections contain $307$--$321$ parent galaxies and $32$--$48$ identified bars, with $f_{\rm bar}=10.42\%$--$15.09\%$, compared with $17.75\%$ for the recomputed \citet{Zana2022} catalogue and $54.96\%$ in \citet{Lu2025}. Because these studies adopt different stellar-mass limits and selection criteria, the absolute parent-sample counts cannot determine the contribution of denominator effects. Both parent-sample selection and bar identification may therefore contribute to the differences in $f_{\rm bar}$.  In particular, \citet{Goncalves2026} independently showed that realistic observational effects can reduce bar detectability, providing an additional reason to consider observing conditions when interpreting cross-study differences in $f_{\rm bar}$.

\section{Summary and Conclusions} \label{sec:summary}

In this work, we compare galactic bars in the TNG100 and TNG50 mock-observation samples at $z\simeq0.1$ with those in SDSS galaxies at $0.1 \le z \le 0.1264$. We generate radiative-transfer images of the simulated galaxies, apply SDSS-like observational effects, and process the mock and real images using closely matched source-segmentation, morphological-classification, and isophotal bar-measurement procedures. This approach allows the bar statistics and structural properties of TNG100, TNG50, and SDSS to be compared using broadly consistent observational criteria and measurements. Our main conclusions are summarized as follows:

\begin{enumerate}

\item \textbf{Bar Fractions and Stellar-Mass Dependence:} For 38,286 face-on SDSS LTGs, we measure a global bar fraction of $26.68_{-0.23}^{+0.23}\%$ (10,213 galaxies classified as barred). TNG100 yields $24.96_{-0.76}^{+0.78}\%$ (791 of 3,169 galaxies), comparable to the global SDSS value, whereas TNG50 yields a lower fraction of $13.42_{-1.07}^{+1.15}\%$ (127 of 946 galaxies). TNG100 shows good agreement with the SDSS $f_{\rm bar}$ over $10^{10} \lesssim M_* \lesssim 10^{11.2}\,M_{\odot}$ but becomes higher at larger stellar masses as the SDSS bar fraction declines. TNG50 remains lower than SDSS over most of $10^{10} \lesssim M_* \lesssim 10^{11.7}\,M_{\odot}$ and exhibits stronger bin-to-bin fluctuations because of its smaller sample size; at the highest masses, the limited statistics prevent a robust comparison.

\item \textbf{Bar-Length Distributions and Scaling Relations:} The $R_{\rm bar}$ distributions of SDSS, TNG100, and TNG50 show considerable overlap, with median values of $6.67$, $5.91$, and $6.02$\,kpc and 16th--84th percentile ranges of $3.87$--$10.53$, $3.18$--$10.69$, and $2.39$--$12.91$\,kpc, respectively. Median bar length generally increases with stellar mass in all three samples. The outer ellipticity-decline estimator ($R_{\rm 0.85max}$) yields systematically larger bar lengths and a broader long-bar tail than the ellipticity-peak estimator ($R_{\epsilon,\max}^{\rm bar}$), with some extremely large bar-length measurements associated with morphologically ambiguous or projected systems. Adopting $R_{\epsilon,\max}^{\rm bar}$ shifts the distributions toward smaller radii while preserving the positive stellar-mass dependence. In relative terms, $R_{\rm bar}/R_{\rm e}$ shows clearer differences from SDSS, particularly toward the low- and high-mass ends and most prominently for TNG100.

\item \textbf{Structural Differences:} Double-Sérsic decompositions reveal systematic differences in the bulge and disk structures of the three samples. Across both intermediate- and high-mass regimes, TNG100 contains a smaller relative population of high-$n_{\rm bulge}$ systems than SDSS, whereas TNG50 contains a larger high-$n_{\rm bulge}$ population. These differences in bulge concentration may contribute to the different $f_{\rm bar}$ values, although they cannot fully account for the bar-fraction trends. Massive TNG100 galaxies also have substantially more extended stellar disks than SDSS, which may contribute to their smaller $R_{\rm bar}/R_{\rm e}$ at comparable absolute bar lengths.

\item \textbf{Observational Effects and Measurement Dependence:} Our SDSS bar fractions are lower than those reported by several previous observational studies, particularly those based on lower-redshift or near-infrared samples, suggesting that the SDSS measurements in this work may underestimate the bar fraction relative to observations with higher spatial resolution. Comparisons with previous TNG studies likewise show that the inferred $f_{\rm bar}$ depends on the adopted analysis framework. Observational conditions, stellar-mass selection, morphological classification, parent-sample definition, projection treatment, and bar-identification methodology may all affect the measured bar fraction. These results emphasize the importance of applying comparable observational conditions and analysis procedures when comparing simulations with observations.

\end{enumerate}

\begin{acknowledgments}
The authors express their gratitude to the Illustris and TNG projects for providing the simulation data. This work was supported by the National Natural Science Foundation of China (Nos. 12573007, 12073089) and the China Manned Space Program with grant No. MS-CSST-2025-A10. W.S.Z. is supported by the National Natural Science Foundation of China through grant No. 12595314 and the National SKA Program of China (grant Nos. 2025SKA0150104). L.T. is also supported by Natural Science Foundation of Sichuan Province (No. 2022NSFSC1842), the Fundamental Research Funds of China West Normal University (CWNU, No. 21E029), and the Sichuan Youth Science and Technology Innovation Research Team (21CXTD0038). Most computations were performed on the LOONG at the School of Physics and Astronomy, Sun Yat-sen University. This research has made use of data from SDSS-II Data Release 7 (DR7) and SDSS-III. Funding for these phases of the SDSS was provided by the Alfred P. Sloan Foundation, the Participating Institutions, the National Science Foundation, the U.S. Department of Energy, the National Aeronautics and Space Administration, the Japanese Monbukagakusho, the Max Planck Society, and the Higher Education Funding Council for England. This research has also made use of the SAO/NASA's Astrophysics Data System Bibliographic Services. This work is supported by China Manned Space Program through its Space Application System.
\end{acknowledgments}

\appendix

\section{Projection Dependence of the Bar Statistics}
\label{app:projection}

\begin{figure*}[thpb]
    \centering
    \includegraphics[width=\textwidth]{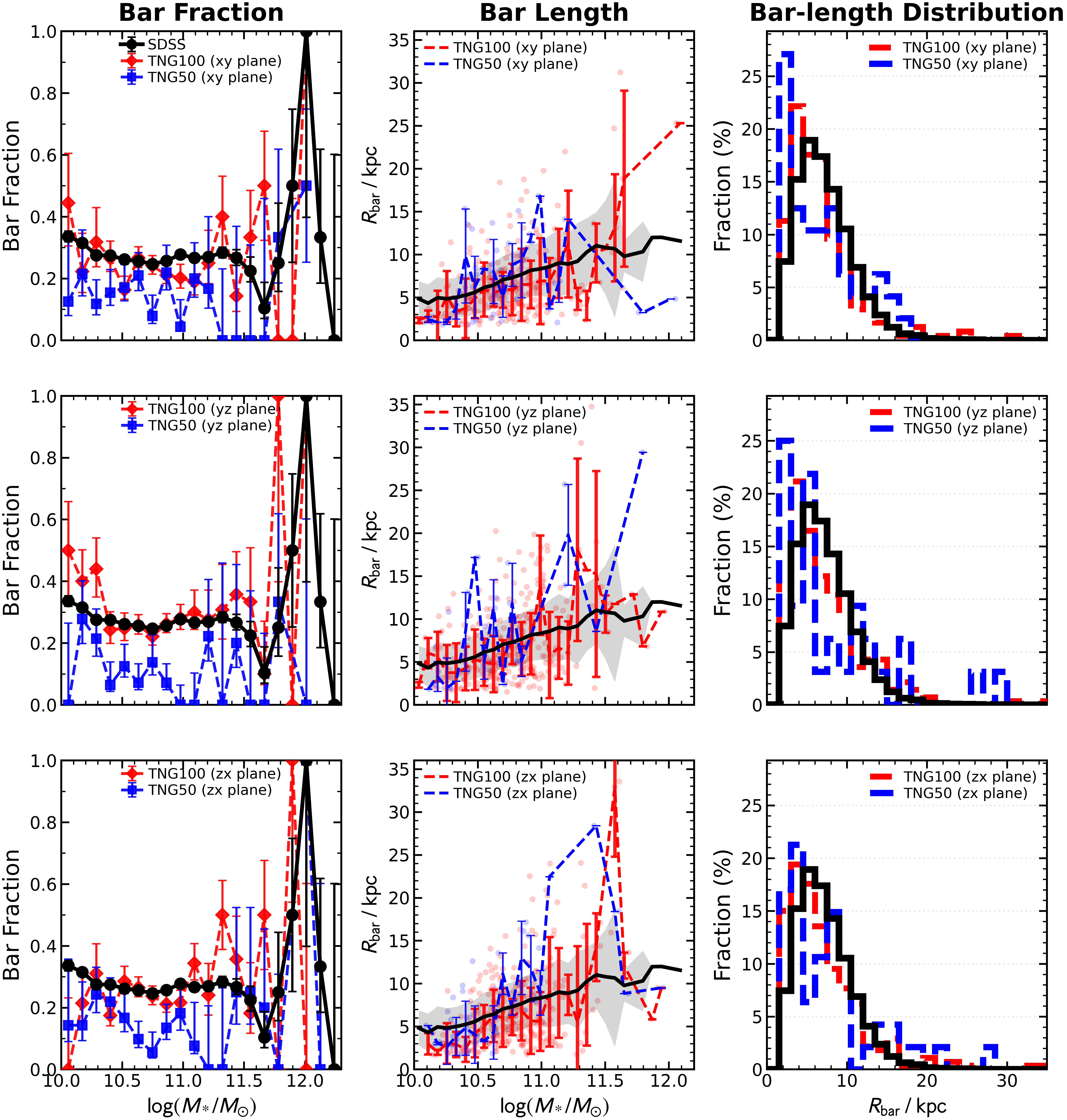}
    \caption{Comparison of the bar statistics obtained from the three orthogonal TNG projections. The top, middle, and bottom rows show the $xy$, $yz$, and $zx$ projections, respectively. From left to right, the panels present the bar fraction as a function of stellar mass, the median $R_{\rm bar}$ as a function of stellar mass, and the fractional $R_{\rm bar}$ distribution. TNG100 and TNG50 are shown in red and blue, respectively, while SDSS is shown in black as a common observational reference. The uncertainties are defined as in Figures~\ref{fig:fbar_mass} and~\ref{fig:bar_length}. }
    \label{fig:projection_comparison}
\end{figure*}

As described in Section~\ref{sec:methods}, each TNG FoF group is rendered along three orthogonal viewing directions ($xy$, $yz$, and $zx$), and all qualifying images from the three projections are retained in the final sample. To examine whether this treatment introduces a systematic projection dependence, Figure~\ref{fig:projection_comparison} compares the bar statistics obtained independently from the three viewing directions.

The three projections recover broadly consistent population-level results. Over the well-sampled stellar-mass range, the $xy$, $yz$, and $zx$ projections show similar $f_{\rm bar}$--$M_*$ relations and comparable $R_{\rm bar}$--$M_*$ trends. Their fractional bar length distributions also have similar overall shapes and characteristic scales. Larger fluctuations appear mainly in sparsely populated bins, particularly toward the high-mass end of TNG50. We therefore find no clear evidence for a systematic viewing-direction bias in either the bar fraction or the statistical distribution of bar lengths, and the main conclusions of this work are not driven by any particular projection.

\section{Dependence of Bar Fraction on Simulation Volume}
\label{app:subvolume}

\begin{figure*}[thpb]
    \centering
    \includegraphics[width=\textwidth]{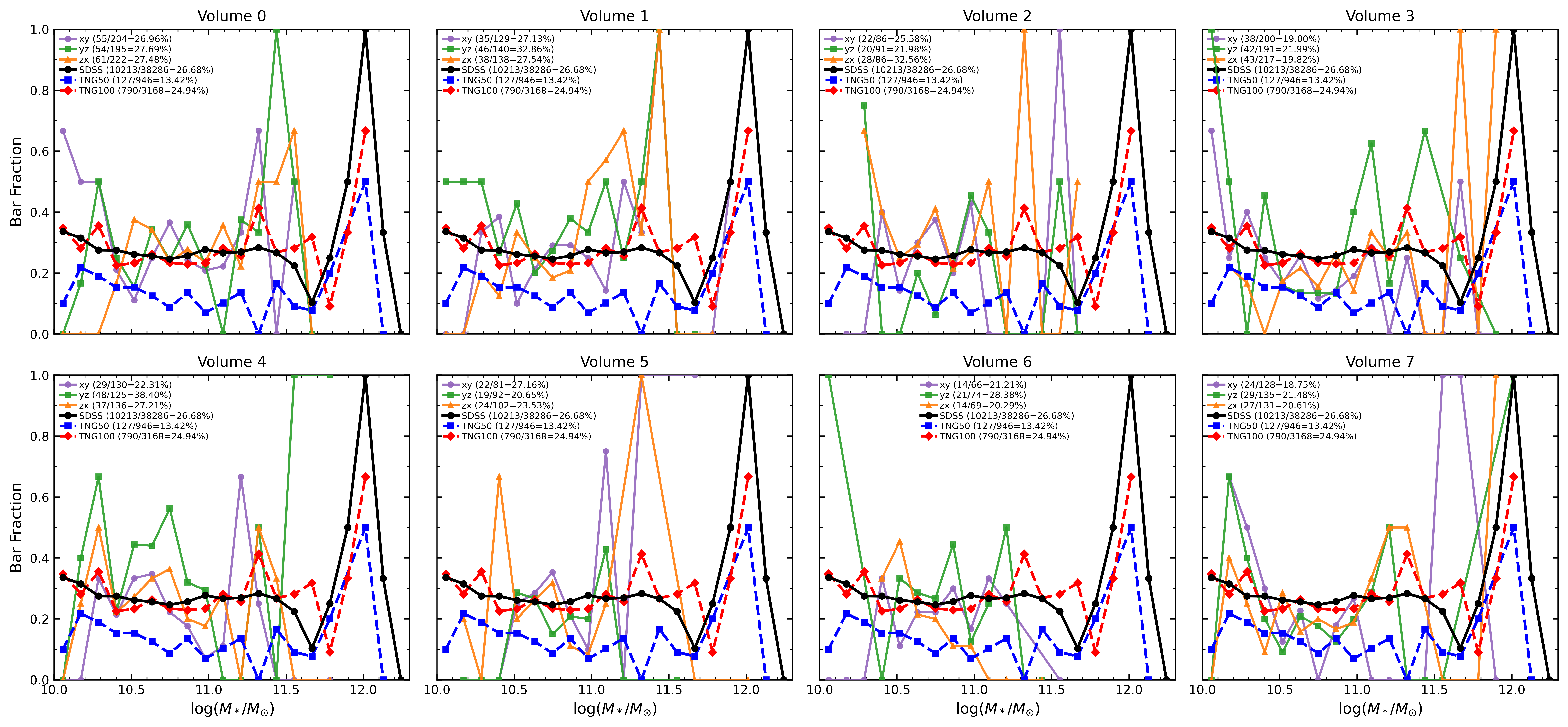}
    \caption{Bar fraction as a function of stellar mass for the eight equal subvolumes of TNG100. Each panel corresponds to one subvolume, with the $xy$, $yz$, and $zx$ projections shown separately. The TNG100 sample used in this test, together with the TNG50 and SDSS samples, is shown for reference. The numbers in the legend give the number of barred galaxies, the number of galaxies in the corresponding face-on LTG parent sample, and the resulting global bar fraction.}
    \label{fig:subvolume}
\end{figure*}

To examine whether the difference in $f_{\rm bar}$ between TNG100 and TNG50 may be affected by the smaller simulation volume of TNG50, we divide the TNG100 box into eight equal subvolumes. Each subvolume has a side length of $37.5\,h^{-1}\,\mathrm{Mpc}$, comparable to the $35\,h^{-1}\,\mathrm{Mpc}$ side length of TNG50. For each galaxy identified from the segmented mock images, we use the three-dimensional position of the most massive stellar particle associated with the line of sight through the galaxy-center pixel to assign the galaxy to one of the eight subvolumes. This procedure is used only to determine the spatial location of the galaxy defined by the segmentation map and does not require an object-by-object match with the \textsc{Subfind} catalogue\footnote{One TNG100 galaxy image classified as barred could not be assigned to a subvolume using this auxiliary procedure. It remains included in the main analysis but is omitted from this test, which therefore gives $f_{\rm bar}=790/3168=24.94\%$.}.

Figure~\ref{fig:subvolume} shows that the TNG100 subvolumes exhibit moderate variations in $f_{\rm bar}$, together with larger bin-to-bin fluctuations in bins with small galaxy samples. Combining the three projections within each subvolume, the global bar fractions range from $20.23\%$ to $29.24\%$, with a median of $25.13\%$ and a standard deviation of $3.40$ percentage points. The median is close to the overall TNG100 bar fraction of $24.96\%$, while even the lowest subvolume value remains substantially above the TNG50 value of $13.42\%$. The larger fluctuations in the mass-binned measurements are mainly associated with the reduced number of galaxies in each subvolume. These results suggest that finite-volume effects alone are unlikely to account for the TNG50--TNG100 difference, which may instead be more strongly associated with numerical resolution or other resolution-dependent galaxy properties.

\section{Extreme Bar-Length Examples}
\label{app:extreme_cases}

\begin{figure*}[thpb]
\centering
\includegraphics[width=\textwidth]{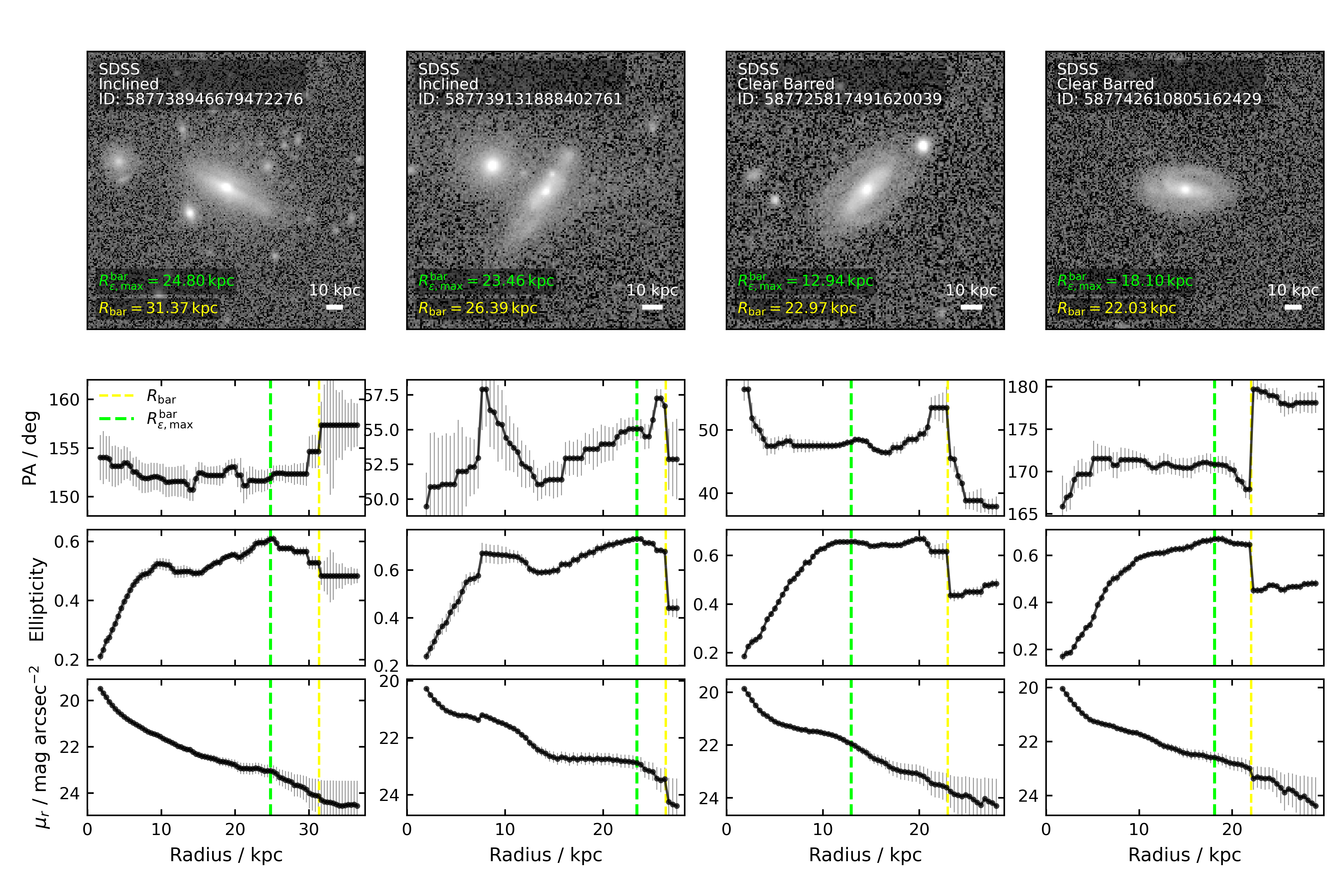}
\caption{
Extreme SDSS examples selected from the high-$R_{\mathrm{bar}}$ tail. The general layout follows Figure~\ref{fig:bar_examples}, but the radii are shown in kpc and the image panels list both $R_{\mathrm{bar}}$ and $R_{\epsilon,\max}^{\rm bar}$. The vertical yellow and green dashed lines in the profiles mark $R_{\mathrm{bar}}$ and $R_{\epsilon,\max}^{\rm bar}$, respectively. The two left columns show the two largest-$R_{\mathrm{bar}}$ objects among visually identified inclined or morphologically ambiguous systems, while the two right columns show the two largest-$R_{\mathrm{bar}}$ clear barred galaxies found by visual inspection.
}
\label{fig:app_sdss_extreme}
\end{figure*}

\begin{figure*}[thpb]
\centering
\includegraphics[width=\textwidth]{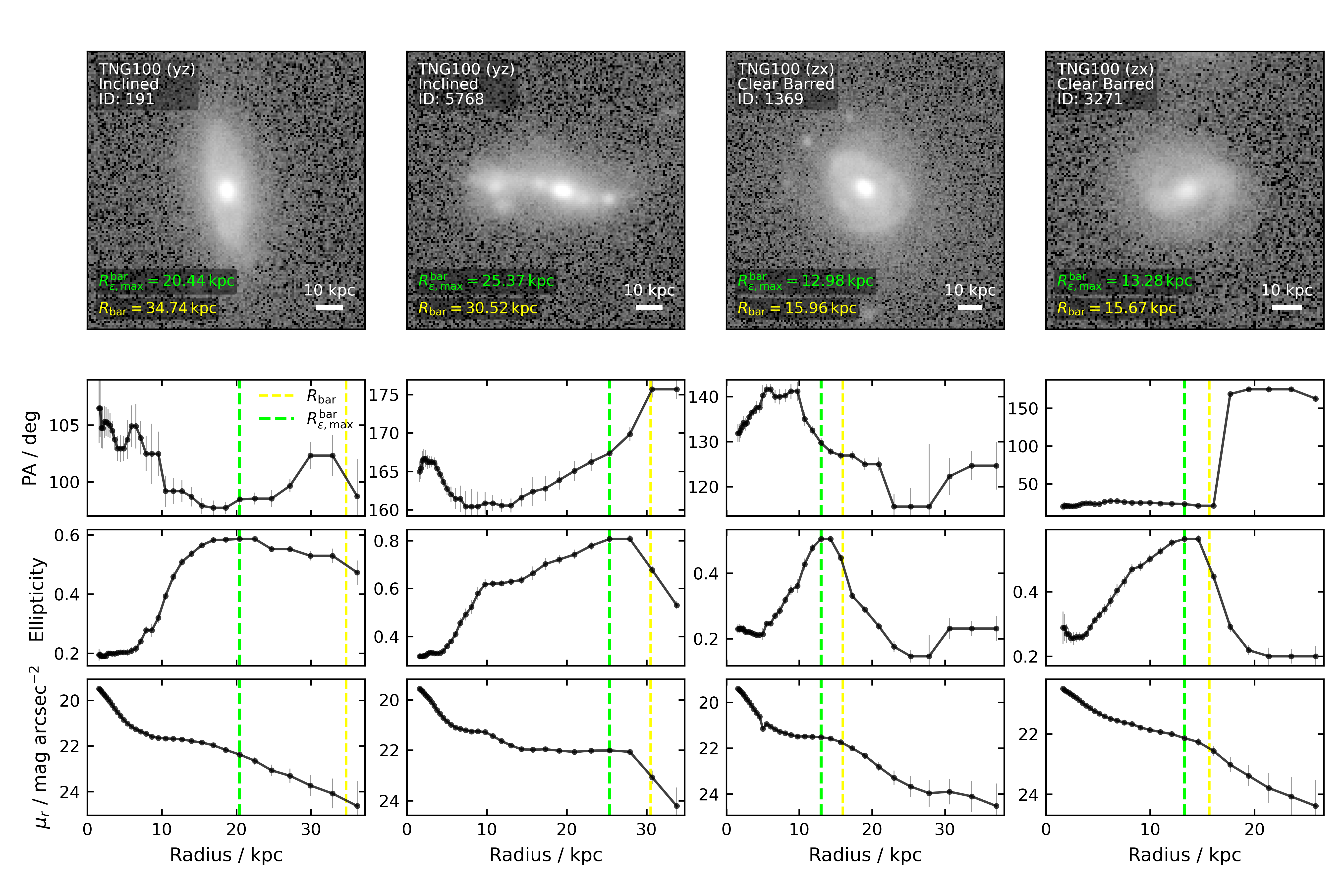}
\caption{
Same as Figure~\ref{fig:app_sdss_extreme}, but for TNG100. 
}
\label{fig:app_tng100_extreme}
\end{figure*}

\begin{figure*}[thpb]
\centering
\includegraphics[width=\textwidth]{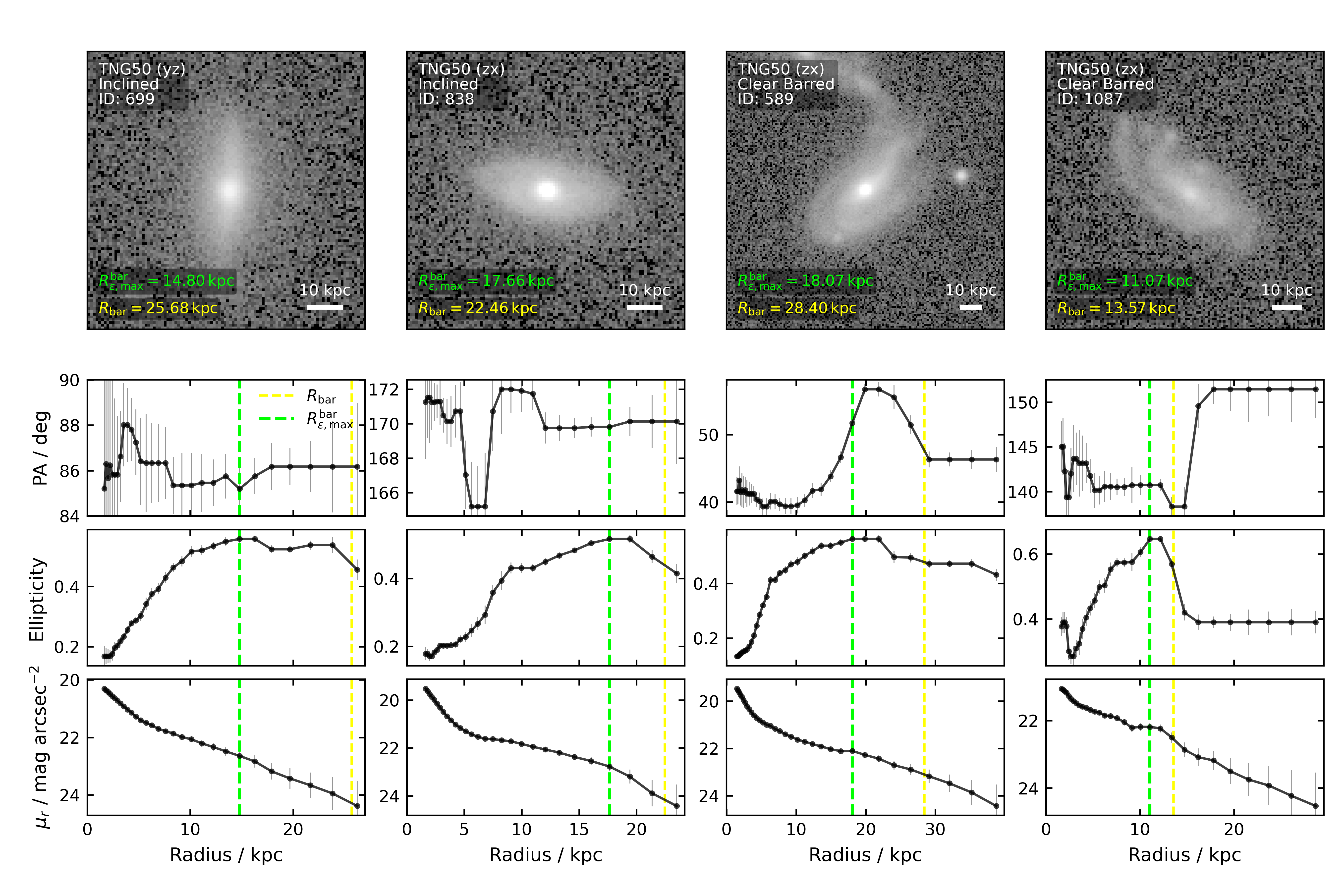}
\caption{
Same as Figure~\ref{fig:app_sdss_extreme}, but for TNG50. 
}
\label{fig:app_tng50_extreme}
\end{figure*}

Figures~\ref{fig:app_sdss_extreme}, \ref{fig:app_tng100_extreme}, and \ref{fig:app_tng50_extreme} show examples of systems with some of the largest measured bar lengths in SDSS, TNG100, and TNG50, respectively. For each sample, a limited set of candidates at the extreme end of the $R_{\mathrm{bar}}$ distribution was visually inspected. The two left columns show the two largest-$R_{\mathrm{bar}}$ cases with inclined or morphologically ambiguous appearances among these candidates, whereas the two right columns show the two largest cases containing clear stellar bars. Despite their substantially different image morphologies, the selected examples in both categories exhibit similar isophotal signatures: a broad radial region of high ellipticity with PA variations smaller than $10^\circ$, followed by an apparent ellipticity drop at larger radii. Hereafter, we refer to systems that satisfy our photometric bar-identification criteria but do not show a clear stellar bar embedded in a well-defined disk as ``bar-like systems.''

For these bar-like systems, the characteristic radius identified from either $R_{\mathrm{bar}}$ or $R_{\epsilon,\max}^{\rm bar}$ may trace the radial extent of the elongated system as a whole rather than the size of a stellar bar embedded in a galactic disk, which is the structure of primary interest in this work. The projected light distribution may therefore be dominated by an extended elongated component even when the underlying physical nature of the system is different from that of a conventional barred disk galaxy. Our image-based analysis cannot distinguish the physical origin of individual bar-like systems, and we therefore do not assign them a specific dynamical interpretation. These examples instead indicate that additional structural or kinematic information would be required to determine the physical nature of such systems, particularly for objects in the extreme bar length tail.

There are at least two possible physical origins for these bar-like systems. First, some may be intrinsically elongated or triaxial stellar systems rather than conventional bars embedded in rotationally supported disks. Observational studies have shown that galaxies can possess triaxial or prolate intrinsic stellar distributions, particularly among massive early-type systems \citep{Kimm2007,Padilla2008,Krajnovic2018}. Similar elongated bar-like systems have also been identified in IllustrisTNG, including prolate stellar systems in TNG100 and slowly rotating non-axisymmetric structures in dispersion-dominated TNG50 galaxies \citep{Lokas2021,Du2026}. Second, some bar-like systems may contain genuine stellar bars embedded in relatively low-surface-brightness disks. Previous observational and simulation studies have suggested that strong and extended bars can be associated with reduced stellar surface brightness in the surrounding disk \citep{Gadotti2003,Gadotti2008,Kim2016}.

\section{Comparison with the Zana et al. (2022) Bar-Length Catalogue}
\label{app:zana_bar_length}

\begin{figure*}[thpb]
    \centering
    \includegraphics[width=\textwidth]{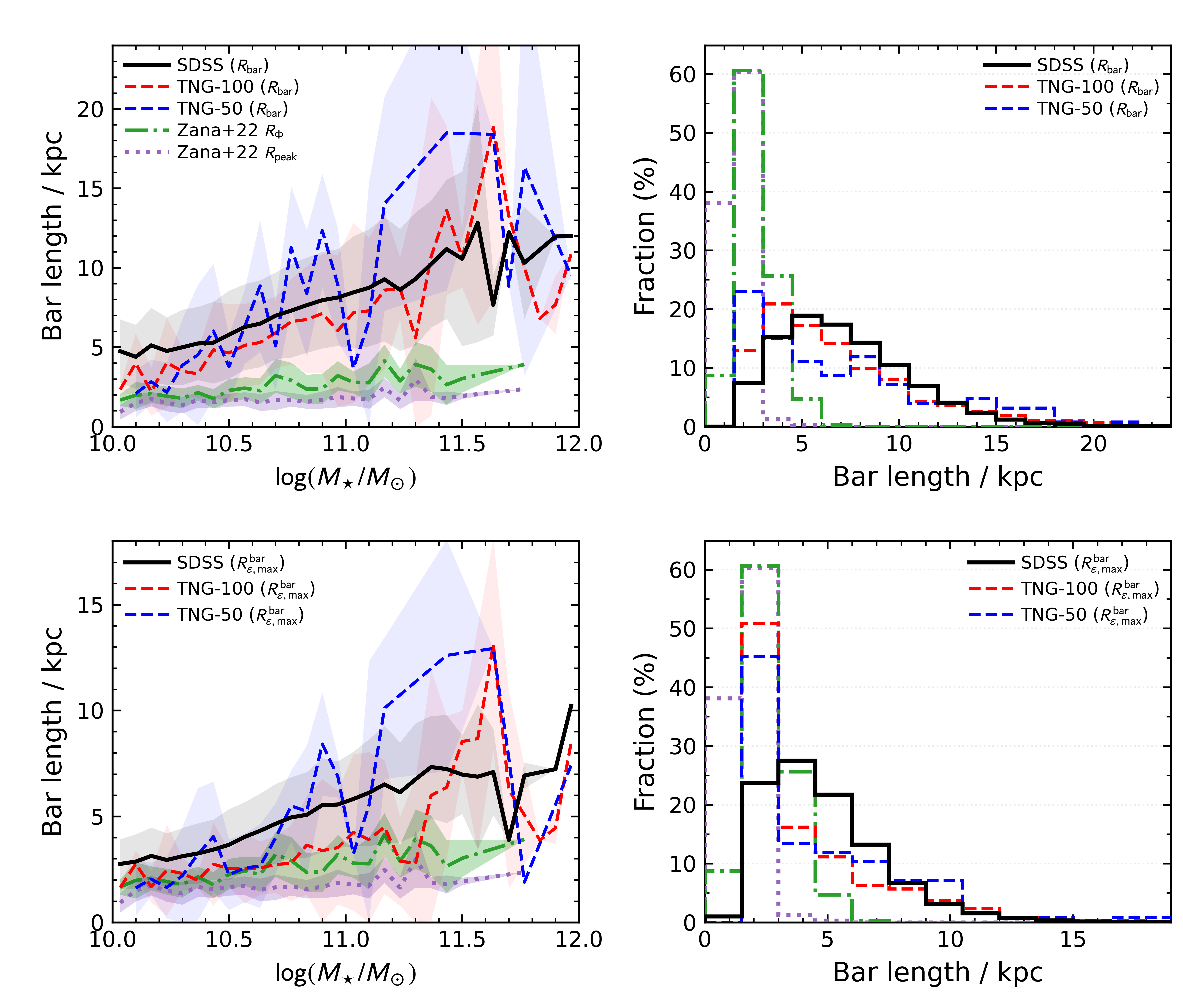}
    \caption{Comparison of our TNG50 bar length measurements with the public TNG50 catalogue of \citet{Zana2022}, with TNG100 and SDSS shown as additional references. The left panels show the median bar length as a function of stellar mass, with shaded regions indicating the standard deviation within each stellar-mass bin; the right panels show the corresponding fractional distributions. The top panels present $R_{\rm bar}$, while the bottom panels present $R_{\epsilon,\max}^{\rm bar}$. Our SDSS, TNG100, and TNG50 samples are shown in black, red, and blue, respectively, while the $R_{\Phi}$ and $R_{\rm peak}$ measurements of \citet{Zana2022} are shown as comparison curves.}
    \label{fig:zana_bar_length}
\end{figure*}

For an independent comparison, Figure~\ref{fig:zana_bar_length} presents our bar length measurements together with the public TNG50-1 results of \citet{Zana2022}, obtained from the same Snapshot 91. Our TNG100 and SDSS measurements are also shown as additional references. Their catalogue provides $R_{\Phi}$, the outer edge of the radial interval over which the phase of the $m=2$ Fourier mode remains approximately constant and their adopted bar half-length, together with $R_{\rm peak}$, the radius at which the cumulative Fourier amplitude $A_2(<R)$ reaches its maximum. The top panels compare these measurements with our $R_{\rm bar}$, while the bottom panels use $R_{\epsilon,\max}^{\rm bar}$, as defined in Section~\ref{sec:bar_length_definition}.

Our $R_{\rm bar}$ measurements are generally larger and have broader distributions than both \citet{Zana2022} estimators. The numerical differences relative to the \citet{Zana2022} estimators become substantially smaller when $R_{\epsilon,\max}^{\rm bar}$ is used. In particular, our TNG50 measurements are broadly consistent with $R_{\Phi}$ over $10.0 \lesssim \log(M_*/M_\odot) \lesssim 10.5$, while our TNG100 measurements, shown for reference, have similar values over most of their well-sampled stellar-mass range. At higher masses, however, our TNG50 bar lengths remain systematically larger. The \citet{Zana2022} $R_{\rm peak}$ values are generally smaller than $R_{\Phi}$ because $R_{\rm peak}$ traces an internal maximum of the cumulative Fourier amplitude, whereas $R_{\Phi}$ traces the outer boundary of the phase-coherent bar region.

The remaining high-mass difference may partly reflect the different sample selections and measurement frameworks. Compact bars embedded in bright central components may be preferentially missed in the mock images, biasing the recovered barred sample toward larger structures. In addition, our image-based LTG selection and isophotal bar identification differ from the intrinsic Fourier analysis and the additional rotational, phase-coherence, velocity-dispersion, and density-symmetry criteria adopted by \citet{Zana2022}. The two procedures may therefore retain different galaxy and bar populations, particularly at the high-mass end. The limited number of massive TNG50 galaxies may further contribute to the fluctuations at the high-mass end. These effects cannot be separated using the present statistical comparison.

\bibliography{sample701}{}

\begin{thebibliography}{}
\expandafter\ifx\csname natexlab\endcsname\relax\def\natexlab#1{#1}\fi
\providecommand{\url}[1]{\href{#1}{#1}}
\providecommand{\dodoi}[1]{doi:~\href{http://doi.org/#1}{\nolinkurl{#1}}}
\providecommand{\doeprint}[1]{\href{http://ascl.net/#1}{\nolinkurl{http://ascl.net/#1}}}
\providecommand{\doarXiv}[1]{\href{https://arxiv.org/abs/#1}{\nolinkurl{https://arxiv.org/abs/#1}}}

\bibitem[{ {Abdurro'uf} {et~al.}(2021){Abdurro'uf}, {Lin}, {Wu}, \& {Akiyama}}]{Abdurrouf2021}
{Abdurro'uf}, {Lin}, Y.-T., {Wu}, P.-F., \& {Akiyama}, M. 2021, \bibinfo{title}{{Introducing piXedfit: A Spectral Energy Distribution Fitting Code Designed for Resolved Sources},} \apjs, 254, 15, \dodoi{10.3847/1538-4365/abebe2}

\bibitem[{J.~A.~L. {Aguerri} {et~al.}(2009){Aguerri}, {M{\'e}ndez-Abreu}, \& {Corsini}}]{Aguerri2009}
{Aguerri}, J.~A.~L., {M{\'e}ndez-Abreu}, J., \& {Corsini}, E.~M. 2009, \bibinfo{title}{{The population of barred galaxies in the local universe. I. Detection and characterisation of bars},} \aap, 495, 491, \dodoi{10.1051/0004-6361:200810931}

\bibitem[{E. {Athanassoula}(1992){Athanassoula}}]{Athanassoula1992}
{Athanassoula}, E. 1992, \bibinfo{title}{{Morphology of bar orbits.},} \mnras, 259, 328, \dodoi{10.1093/mnras/259.2.328}

\bibitem[{E. {Athanassoula}(2003){Athanassoula}}]{Athanassoula2003}
{Athanassoula}, E. 2003, \bibinfo{title}{{What determines the strength and the slowdown rate of bars?},} \mnras, 341, 1179, \dodoi{10.1046/j.1365-8711.2003.06473.x}

\bibitem[{E. {Athanassoula} {et~al.}(2005){Athanassoula}, {Lambert}, \& {Dehnen}}]{Athanassoula2005}
{Athanassoula}, E., {Lambert}, J.~C., \& {Dehnen}, W. 2005, \bibinfo{title}{{Can bars be destroyed by a central mass concentration?- I. Simulations},} \mnras, 363, 496, \dodoi{10.1111/j.1365-2966.2005.09445.x}

\bibitem[{E. {Athanassoula} \& A. {Misiriotis}(2002){Athanassoula} \& {Misiriotis}}]{Athanassoula2002}
{Athanassoula}, E., \& {Misiriotis}, A. 2002, \bibinfo{title}{{Morphology, photometry and kinematics of N -body bars - I. Three models with different halo central concentrations},} \mnras, 330, 35, \dodoi{10.1046/j.1365-8711.2002.05028.x}

\bibitem[{E. {Athanassoula} \& J.~A. {Sellwood}(1986){Athanassoula} \& {Sellwood}}]{Athanassoula1986}
{Athanassoula}, E., \& {Sellwood}, J.~A. 1986, \bibinfo{title}{{Bi-symmetric instabilities of the Kuz'min/Toomre disc.},} \mnras, 221, 213, \dodoi{10.1093/mnras/221.2.213}

\bibitem[{M. {Baes} {et~al.}(2011){Baes}, {Verstappen}, {De Looze}, {Fritz}, {Saftly}, {Vidal P{\'e}rez}, {Stalevski}, \& {Valcke}}]{Baes2011}
{Baes}, M., {Verstappen}, J., {De Looze}, I., {et~al.} 2011, \bibinfo{title}{{Efficient Three-dimensional NLTE Dust Radiative Transfer with SKIRT},} \apjs, 196, 22, \dodoi{10.1088/0067-0049/196/2/22}

\bibitem[{F.~D. {Barazza} {et~al.}(2008){Barazza}, {Jogee}, \& {Marinova}}]{Barazza2008}
{Barazza}, F.~D., {Jogee}, S., \& {Marinova}, I. 2008, \bibinfo{title}{{Bars in Disk-dominated and Bulge-dominated Galaxies at z \raisebox{-0.5ex}\textasciitilde 0: New Insights from \raisebox{-0.5ex}\textasciitilde3600 SDSS Galaxies},} \apj, 675, 1194, \dodoi{10.1086/526510}

\bibitem[{F.~D. {Barazza} {et~al.}(2009){Barazza}, {Jablonka}, {Desai}, {Jogee}, {Arag{\'o}n-Salamanca}, {De Lucia}, {Saglia}, {Halliday}, {Poggianti}, {Dalcanton}, \& et~al.}]{Barazza2009}
{Barazza}, F.~D., {Jablonka}, P., {Desai}, V., {et~al.} 2009, \bibinfo{title}{{Frequency and properties of bars in cluster and field galaxies at intermediate redshifts},} \aap, 497, 713, \dodoi{10.1051/0004-6361/200810352}

\bibitem[{J.~E. {Barnes} \& L.~E. {Hernquist}(1991){Barnes} \& {Hernquist}}]{Barnes1991}
{Barnes}, J.~E., \& {Hernquist}, L.~E. 1991, \bibinfo{title}{{Fueling Starburst Galaxies with Gas-rich Mergers},} \apjl, 370, L65, \dodoi{10.1086/185978}

\bibitem[{I. {Berentzen} {et~al.}(2007){Berentzen}, {Shlosman}, {Martinez-Valpuesta}, \& {Heller}}]{Berentzen2007}
{Berentzen}, I., {Shlosman}, I., {Martinez-Valpuesta}, I., \& {Heller}, C.~H. 2007, \bibinfo{title}{{Gas Feedback on Stellar Bar Evolution},} \apj, 666, 189, \dodoi{10.1086/520531}

\bibitem[{E. {Bertin} \& S. {Arnouts}(1996){Bertin} \& {Arnouts}}]{Bertin1996}
{Bertin}, E., \& {Arnouts}, S. 1996, \bibinfo{title}{{SExtractor: Software for source extraction.},} \aaps, 117, 393, \dodoi{10.1051/aas:1996164}

\bibitem[{M.~R. {Blanton} {et~al.}(2001){Blanton}, {Dalcanton}, {Eisenstein}, {Loveday}, {Strauss}, {SubbaRao}, {Weinberg}, {Anderson}, {Annis}, {Bahcall}, \& et~al.}]{Blanton2001}
{Blanton}, M.~R., {Dalcanton}, J., {Eisenstein}, D., {et~al.} 2001, \bibinfo{title}{{The Luminosity Function of Galaxies in SDSS Commissioning Data},} \aj, 121, 2358, \dodoi{10.1086/320405}

\bibitem[{F. {Bournaud} \& F. {Combes}(2002){Bournaud} \& {Combes}}]{Bournaud2002}
{Bournaud}, F., \& {Combes}, F. 2002, \bibinfo{title}{{Gas accretion on spiral galaxies: Bar formation and renewal},} \aap, 392, 83, \dodoi{10.1051/0004-6361:20020920}

\bibitem[{F. {Bournaud} {et~al.}(2005){Bournaud}, {Combes}, \& {Semelin}}]{Bournaud2005}
{Bournaud}, F., {Combes}, F., \& {Semelin}, B. 2005, \bibinfo{title}{{The lifetime of galactic bars: central mass concentrations and gravity torques},} \mnras, 364, L18, \dodoi{10.1111/j.1745-3933.2005.00096.x}

\bibitem[{L. {Bradley} {et~al.}(2020){Bradley}, {Sip{\H{o}}cz}, {Robitaille}, {Tollerud}, {Vin{\'\i}cius}, {Deil}, {Barbary}, {Wilson}, {Busko}, {G{\"u}nther}, {Cara}, {Conseil}, {Bostroem}, {Droettboom}, {Bray}, {Andersen Bratholm}, {Lim}, {Barentsen}, {Craig}, {Pascual}, {Perren}, {Greco}, {Donath}, {De Val-Borro}, {Kerzendorf}, {Bach}, {Weaver}, {D'Eugenio}, {Souchereau}, \& {Ferreira}}]{Bradley2020}
{Bradley}, L., {Sip{\H{o}}cz}, B., {Robitaille}, T., {et~al.} 2020, {astropy/photutils: 1.0.0}, 1.0.0 Zenodo, \dodoi{10.5281/zenodo.4044744}

\bibitem[{G. {Bruzual} \& S. {Charlot}(2003){Bruzual} \& {Charlot}}]{Bruzual2003}
{Bruzual}, G., \& {Charlot}, S. 2003, \bibinfo{title}{{Stellar population synthesis at the resolution of 2003},} \mnras, 344, 1000, \dodoi{10.1046/j.1365-8711.2003.06897.x}

\bibitem[{M. {Bureau} \& E. {Athanassoula}(2005){Bureau} \& {Athanassoula}}]{Bureau2005}
{Bureau}, M., \& {Athanassoula}, E. 2005, \bibinfo{title}{{Bar Diagnostics in Edge-On Spiral Galaxies. III. N-Body Simulations of Disks},} \apj, 626, 159, \dodoi{10.1086/430056}

\bibitem[{E. {Cameron}(2011){Cameron}}]{Cameron2011}
{Cameron}, E. 2011, \bibinfo{title}{{On the Estimation of Confidence Intervals for Binomial Population Proportions in Astronomy: The Simplicity and Superiority of the Bayesian Approach},} \pasa, 28, 128, \dodoi{10.1071/AS10046}

\bibitem[{P. {Camps} \& M. {Baes}(2020){Camps} \& {Baes}}]{Camps2020}
{Camps}, P., \& {Baes}, M. 2020, \bibinfo{title}{{SKIRT 9: Redesigning an advanced dust radiative transfer code to allow kinematics, line transfer and polarization by aligned dust grains},} Astronomy and Computing, 31, 100381, \dodoi{10.1016/j.ascom.2020.100381}

\bibitem[{M.~K. {Cavanagh} {et~al.}(2022){Cavanagh}, {Bekki}, {Groves}, \& {Pfeffer}}]{Cavanagh2022}
{Cavanagh}, M.~K., {Bekki}, K., {Groves}, B.~A., \& {Pfeffer}, J. 2022, \bibinfo{title}{{The evolution of barred galaxies in the EAGLE simulations},} \mnras, 510, 5164, \dodoi{10.1093/mnras/stab3786}

\bibitem[{G. {Chabrier}(2003){Chabrier}}]{Chabrier2003}
{Chabrier}, G. 2003, \bibinfo{title}{{Galactic Stellar and Substellar Initial Mass Function},} \pasp, 115, 763, \dodoi{10.1086/376392}

\bibitem[{M. {Cisternas} {et~al.}(2015){Cisternas}, {Sheth}, {Salvato}, {Knapen}, {Civano}, \& {Santini}}]{Cisternas2015}
{Cisternas}, M., {Sheth}, K., {Salvato}, M., {et~al.} 2015, \bibinfo{title}{{The Role of Bars in AGN Fueling in Disk Galaxies Over the Last Seven Billion Years},} \apj, 802, 137, \dodoi{10.1088/0004-637X/802/2/137}

\bibitem[{A. {Collier} \& A.-M. {Madigan}(2021){Collier} \& {Madigan}}]{Collier2021}
{Collier}, A., \& {Madigan}, A.-M. 2021, \bibinfo{title}{{The Coupling of Galactic Dark Matter Halos with Stellar Bars},} \apj, 915, 23, \dodoi{10.3847/1538-4357/ac004d}

\bibitem[{F. {Combes} \& R.~H. {Sanders}(1981){Combes} \& {Sanders}}]{Combes1981}
{Combes}, F., \& {Sanders}, R.~H. 1981, \bibinfo{title}{{Formation and properties of persisting stellar bars.},} \aap, 96, 164

\bibitem[{L. {Costantin} {et~al.}(2023){Costantin}, {P{\'e}rez-Gonz{\'a}lez}, {Guo}, {Buttitta}, {Jogee}, {Bagley}, {Barro}, {Kartaltepe}, {Koekemoer}, {Cabello}, \& et~al.}]{Costantin2023}
{Costantin}, L., {P{\'e}rez-Gonz{\'a}lez}, P.~G., {Guo}, Y., {et~al.} 2023, \bibinfo{title}{{A Milky Way-like barred spiral galaxy at a redshift of 3},} \nat, 623, 499, \dodoi{10.1038/s41586-023-06636-x}

\bibitem[{M. {Davis} {et~al.}(1985){Davis}, {Efstathiou}, {Frenk}, \& {White}}]{Davis1985}
{Davis}, M., {Efstathiou}, G., {Frenk}, C.~S., \& {White}, S.~D.~M. 1985, \bibinfo{title}{{The evolution of large-scale structure in a universe dominated by cold dark matter},} \apj, 292, 371, \dodoi{10.1086/163168}

\bibitem[{A. {de Graaff} {et~al.}(2022){de Graaff}, {Trayford}, {Franx}, {Schaller}, {Schaye}, \& {van der Wel}}]{deGraaff2022}
{de Graaff}, A., {Trayford}, J., {Franx}, M., {et~al.} 2022, \bibinfo{title}{{Observed structural parameters of EAGLE galaxies: reconciling the mass-size relation in simulations with local observations},} \mnras, 511, 2544, \dodoi{10.1093/mnras/stab3510}

\bibitem[{V.~P. {Debattista} \& J.~A. {Sellwood}(2000){Debattista} \& {Sellwood}}]{Debattista2000}
{Debattista}, V.~P., \& {Sellwood}, J.~A. 2000, \bibinfo{title}{{Constraints from Dynamical Friction on the Dark Matter Content of Barred Galaxies},} \apj, 543, 704, \dodoi{10.1086/317148}

\bibitem[{S. {D{\'\i}az-Garc{\'\i}a} {et~al.}(2016){D{\'\i}az-Garc{\'\i}a}, {Salo}, {Laurikainen}, \& {Herrera-Endoqui}}]{DiazGarcia2016}
{D{\'\i}az-Garc{\'\i}a}, S., {Salo}, H., {Laurikainen}, E., \& {Herrera-Endoqui}, M. 2016, \bibinfo{title}{{Characterization of galactic bars from 3.6 {\ensuremath{\mu}}m S$^{4}$G imaging},} \aap, 587, A160, \dodoi{10.1051/0004-6361/201526161}

\bibitem[{H. {Dom{\'\i}nguez S{\'a}nchez} {et~al.}(2018){Dom{\'\i}nguez S{\'a}nchez}, {Huertas-Company}, {Bernardi}, {Tuccillo}, \& {Fischer}}]{Dominguez2018}
{Dom{\'\i}nguez S{\'a}nchez}, H., {Huertas-Company}, M., {Bernardi}, M., {Tuccillo}, D., \& {Fischer}, J.~L. 2018, \bibinfo{title}{{Improving galaxy morphologies for SDSS with Deep Learning},} \mnras, 476, 3661, \dodoi{10.1093/mnras/sty338}

\bibitem[{M. {Donnari} {et~al.}(2019){Donnari}, {Pillepich}, {Nelson}, {Vogelsberger}, {Genel}, {Weinberger}, {Marinacci}, {Springel}, \& {Hernquist}}]{Donnari2019}
{Donnari}, M., {Pillepich}, A., {Nelson}, D., {et~al.} 2019, \bibinfo{title}{{The star formation activity of IllustrisTNG galaxies: main sequence, UVJ diagram, quenched fractions, and systematics},} \mnras, 485, 4817, \dodoi{10.1093/mnras/stz712}

\bibitem[{H. {Du} {et~al.}(2026){Du}, {Wang}, \& {Ge}}]{Du2026}
{Du}, H., {Wang}, Y., \& {Ge}, J. 2026, \bibinfo{title}{{Revisiting the Excess of Bar-like Structures in TNG50 Early-type Galaxies: Consistency and Tension with Observations},} Research in Astronomy and Astrophysics, 26, 075021, \dodoi{10.1088/1674-4527/ae5a5a}

\bibitem[{M. {Du} {et~al.}(2017){Du}, {Debattista}, {Shen}, {Ho}, \& {Erwin}}]{Du2017}
{Du}, M., {Debattista}, V.~P., {Shen}, J., {Ho}, L.~C., \& {Erwin}, P. 2017, \bibinfo{title}{{Black Hole Growth in Disk Galaxies Mediated by the Secular Evolution of Short Bars},} \apjl, 844, L15, \dodoi{10.3847/2041-8213/aa7ecb}

\bibitem[{P. {Erwin}(2005){Erwin}}]{Erwin2005}
{Erwin}, P. 2005, \bibinfo{title}{{How large are the bars in barred galaxies?},} \mnras, 364, 283, \dodoi{10.1111/j.1365-2966.2005.09560.x}

\bibitem[{P. {Erwin}(2018){Erwin}}]{Erwin2018}
{Erwin}, P. 2018, \bibinfo{title}{{The dependence of bar frequency on galaxy mass, colour, and gas content - and angular resolution - in the local universe},} \mnras, 474, 5372, \dodoi{10.1093/mnras/stx3117}

\bibitem[{P. {Erwin}(2019){Erwin}}]{Erwin2019}
{Erwin}, P. 2019, \bibinfo{title}{{What determines the sizes of bars in spiral galaxies?},} \mnras, 489, 3553, \dodoi{10.1093/mnras/stz2363}

\bibitem[{P.~B. {Eskridge} {et~al.}(2000){Eskridge}, {Frogel}, {Pogge}, {Quillen}, {Davies}, {DePoy}, {Houdashelt}, {Kuchinski}, {Ram{\'\i}rez}, {Sellgren}, \& et~al.}]{Eskridge2000}
{Eskridge}, P.~B., {Frogel}, J.~A., {Pogge}, R.~W., {et~al.} 2000, \bibinfo{title}{{The Frequency of Barred Spiral Galaxies in the Near-Infrared},} \aj, 119, 536, \dodoi{10.1086/301203}

\bibitem[{F. {Fragkoudi} {et~al.}(2020){Fragkoudi}, {Grand}, {Pakmor}, {Bl{\'a}zquez-Calero}, {Gargiulo}, {Gomez}, {Marinacci}, {Monachesi}, {Ness}, {Perez}, \& et~al.}]{Fragkoudi2020}
{Fragkoudi}, F., {Grand}, R.~J.~J., {Pakmor}, R., {et~al.} 2020, \bibinfo{title}{{Chemodynamics of barred galaxies in cosmological simulations: On the Milky Way's quiescent merger history and in-situ bulge},} \mnras, 494, 5936, \dodoi{10.1093/mnras/staa1104}

\bibitem[{N. {Frankel} {et~al.}(2022){Frankel}, {Pillepich}, {Rix}, {Rodriguez-Gomez}, {Sanders}, {Bovy}, {Kollmeier}, {Murray}, \& {Mackereth}}]{Frankel2022}
{Frankel}, N., {Pillepich}, A., {Rix}, H.-W., {et~al.} 2022, \bibinfo{title}{{Simulated Bars May Be Shorter but Are Not Slower Than Those Observed: TNG50 versus MaNGA},} \apj, 940, 61, \dodoi{10.3847/1538-4357/ac9972}

\bibitem[{D.~A. {Gadotti}(2008){Gadotti}}]{Gadotti2008}
{Gadotti}, D.~A. 2008, \bibinfo{title}{{Image decomposition of barred galaxies and AGN hosts},} \mnras, 384, 420, \dodoi{10.1111/j.1365-2966.2007.12723.x}

\bibitem[{D.~A. {Gadotti}(2011){Gadotti}}]{Gadotti2011}
{Gadotti}, D.~A. 2011, \bibinfo{title}{{Secular evolution and structural properties of stellar bars in galaxies},} \mnras, 415, 3308, \dodoi{10.1111/j.1365-2966.2011.18945.x}

\bibitem[{D.~A. {Gadotti} \& R.~E. {de Souza}(2003){Gadotti} \& {de Souza}}]{Gadotti2003}
{Gadotti}, D.~A., \& {de Souza}, R.~E. 2003, \bibinfo{title}{{NGC 4608 and NGC 5701: Barred Galaxies without Disks?},} \apjl, 583, L75, \dodoi{10.1086/368159}

\bibitem[{D.~A. {Gadotti} {et~al.}(2020){Gadotti}, {Bittner}, {Falc{\'o}n-Barroso}, {M{\'e}ndez-Abreu}, {Kim}, {Fragkoudi}, {de Lorenzo-C{\'a}ceres}, {Leaman}, {Neumann}, {Querejeta}, \& et~al.}]{Gadotti2020}
{Gadotti}, D.~A., {Bittner}, A., {Falc{\'o}n-Barroso}, J., {et~al.} 2020, \bibinfo{title}{{Kinematic signatures of nuclear discs and bar-driven secular evolution in nearby galaxies of the MUSE TIMER project},} \aap, 643, A14, \dodoi{10.1051/0004-6361/202038448}

\bibitem[{G. {Gavazzi} {et~al.}(2015){Gavazzi}, {Consolandi}, {Dotti}, {Fanali}, {Fossati}, {Fumagalli}, {Viscardi}, {Savorgnan}, {Boselli}, {Guti{\'e}rrez}, \& et~al.}]{Gavazzi2015}
{Gavazzi}, G., {Consolandi}, G., {Dotti}, M., {et~al.} 2015, \bibinfo{title}{{H{\ensuremath{\alpha}}3: an H{\ensuremath{\alpha}} imaging survey of HI selected galaxies from ALFALFA. VI. The role of bars in quenching star formation from z = 3 to the present epoch},} \aap, 580, A116, \dodoi{10.1051/0004-6361/201425351}

\bibitem[{S. {Genel} {et~al.}(2018){Genel}, {Nelson}, {Pillepich}, {Springel}, {Pakmor}, {Weinberger}, {Hernquist}, {Naiman}, {Vogelsberger}, {Marinacci}, \& et~al.}]{Genel2018}
{Genel}, S., {Nelson}, D., {Pillepich}, A., {et~al.} 2018, \bibinfo{title}{{The size evolution of star-forming and quenched galaxies in the IllustrisTNG simulation},} \mnras, 474, 3976, \dodoi{10.1093/mnras/stx3078}

\bibitem[{M. {Gerin} {et~al.}(1990){Gerin}, {Combes}, \& {Athanassoula}}]{Gerin1990}
{Gerin}, M., {Combes}, F., \& {Athanassoula}, E. 1990, \bibinfo{title}{{The influence of galaxy interactions on stellar bars.},} \aap, 230, 37

\bibitem[{G.~F. {Gon{\c{c}}alves} \& R.~E.~G. {Machado}(2024){Gon{\c{c}}alves} \& {Machado}}]{Goncalves2024}
{Gon{\c{c}}alves}, G.~F., \& {Machado}, R. E.~G. 2024, \bibinfo{title}{{Ellipse-fitting in Mock Images of TNG50 Barred Galaxies},} Research Notes of the American Astronomical Society, 8, 217, \dodoi{10.3847/2515-5172/ad758f}

\bibitem[{G.~F. {Gon{\c{c}}alves} {et~al.}(2025){Gon{\c{c}}alves}, {Machado}, {Men{\'e}ndez-Delmestre}, \& {Bueno-Dalpiaz}}]{Gonccalves2025}
{Gon{\c{c}}alves}, G.~F., {Machado}, R. E.~G., {Men{\'e}ndez-Delmestre}, K., \& {Bueno-Dalpiaz}, T. 2025, \bibinfo{title}{{Bar properties as a function of wavelength in TNG50: analysis of mock images},} \mnras, 542, 3154, \dodoi{10.1093/mnras/staf1384}

\bibitem[{G.~F. {Gon{\c{c}}alves} {et~al.}(2026){Gon{\c{c}}alves}, {Machado}, {Valen{\c{c}}a}, {Athanassoula}, {Men{\'e}ndez-Delmestre}, \& {Bueno-Dalpiaz}}]{Goncalves2026}
{Gon{\c{c}}alves}, G.~F., {Machado}, R. E.~G., {Valen{\c{c}}a}, R.~R., {et~al.} 2026, \bibinfo{title}{{The detectability of bars at high redshift: a case study using Euclid-like mock observations of TNG50 simulated galaxies},} \mnras, 546, stag263, \dodoi{10.1093/mnras/stag263}

\bibitem[{J.-Y. {Gong} {et~al.}(2025){Gong}, {Lin}, {Tang}, \& {Lan}}]{Gong2025}
{Gong}, J.-Y., {Lin}, W., {Tang}, L., \& {Lan}, Y. 2025, \bibinfo{title}{{Mock Observations: Morphological Analysis of Galaxies in TNG100 Simulations},} \apjs, 279, 16, \dodoi{10.3847/1538-4365/add5e4}

\bibitem[{B. {Groves} {et~al.}(2008){Groves}, {Dopita}, {Sutherland}, {Kewley}, {Fischera}, {Leitherer}, {Brandl}, \& {van Breugel}}]{Groves2008}
{Groves}, B., {Dopita}, M.~A., {Sutherland}, R.~S., {et~al.} 2008, \bibinfo{title}{{Modeling the Pan-Spectral Energy Distribution of Starburst Galaxies. IV. The Controlling Parameters of the Starburst SED},} \apjs, 176, 438, \dodoi{10.1086/528711}

\bibitem[{M. {Guo} {et~al.}(2020){Guo}, {Du}, {Ho}, {Debattista}, \& {Zhao}}]{Guo2020}
{Guo}, M., {Du}, M., {Ho}, L.~C., {Debattista}, V.~P., \& {Zhao}, D. 2020, \bibinfo{title}{{A New Channel of Bulge Formation via the Destruction of Short Bars},} \apj, 888, 65, \dodoi{10.3847/1538-4357/ab584a}

\bibitem[{Y. {Guo} {et~al.}(2023){Guo}, {Jogee}, {Finkelstein}, {Chen}, {Wise}, {Bagley}, {Barro}, {Wuyts}, {Kocevski}, {Kartaltepe}, {McGrath}, {Ferguson}, {Mobasher}, {Giavalisco}, {Lucas}, {Zavala}, {Lotz}, {Grogin}, {Huertas-Company}, {Vega-Ferrero}, {Hathi}, {Arrabal Haro}, {Dickinson}, {Koekemoer}, {Papovich}, {Pirzkal}, {Yung}, {Backhaus}, {Bell}, {Calabr{\`o}}, {Cleri}, {Coogan}, {Cooper}, {Costantin}, {Croton}, {Davis}, {Dekel}, {Franco}, {Gardner}, {Holwerda}, {Hutchison}, {Pandya}, {P{\'e}rez-Gonz{\'a}lez}, {Ravindranath}, {Rose}, {Trump}, {de la Vega}, \& {Wang}}]{Guo2023}
{Guo}, Y., {Jogee}, S., {Finkelstein}, S.~L., {et~al.} 2023, \bibinfo{title}{{First Look at z > 1 Bars in the Rest-frame Near-infrared with JWST Early CEERS Imaging},} \apjl, 945, L10, \dodoi{10.3847/2041-8213/acacfb}

\bibitem[{A. {Habibi} {et~al.}(2024){Habibi}, {Roshan}, {Hosseinirad}, {Khosroshahi}, {Aguerri}, {Cuomo}, \& {Abbassi}}]{Habibi2024}
{Habibi}, A., {Roshan}, M., {Hosseinirad}, M., {et~al.} 2024, \bibinfo{title}{{The redshift evolution of galactic bar pattern speed in TNG50},} \aap, 691, A122, \dodoi{10.1051/0004-6361/202451028}

\bibitem[{L. {Hao} {et~al.}(2009){Hao}, {Jogee}, {Barazza}, {Marinova}, \& {Shen}}]{Hao2009}
{Hao}, L., {Jogee}, S., {Barazza}, F.~D., {Marinova}, I., \& {Shen}, J. 2009, \bibinfo{title}{{Bars in Starbursts and AGNs - A Quantitative Reexamination},} in Astronomical Society of the Pacific Conference Series, Vol. 419, Galaxy Evolution: Emerging Insights and Future Challenges, ed. S.~{Jogee}, I.~{Marinova}, L.~{Hao}, \& G.~A. {Blanc}, 402, \dodoi{10.48550/arXiv.0910.3960}

\bibitem[{T.~G. {Hawarden} {et~al.}(1986){Hawarden}, {Mountain}, {Leggett}, \& {Puxley}}]{Hawarden1986}
{Hawarden}, T.~G., {Mountain}, C.~M., {Leggett}, S.~K., \& {Puxley}, P.~J. 1986, \bibinfo{title}{{Enhanced star formation - the importance of bars in spiral galaxies.},} \mnras, 221, 41P, \dodoi{10.1093/mnras/221.1.41P}

\bibitem[{C.~H. {Heller} {et~al.}(2007){Heller}, {Shlosman}, \& {Athanassoula}}]{Heller2007}
{Heller}, C.~H., {Shlosman}, I., \& {Athanassoula}, E. 2007, \bibinfo{title}{{Structure Formation inside Triaxial Dark Matter Halos: Galactic Disks, Bulges, and Bars},} \apj, 671, 226, \dodoi{10.1086/523260}

\bibitem[{L.~C. {Ho} {et~al.}(1997){Ho}, {Filippenko}, \& {Sargent}}]{Ho1997}
{Ho}, L.~C., {Filippenko}, A.~V., \& {Sargent}, W. L.~W. 1997, \bibinfo{title}{{The Influence of Bars on Nuclear Activity},} \apj, 487, 591, \dodoi{10.1086/304643}

\bibitem[{M. {Huertas-Company} {et~al.}(2019){Huertas-Company}, {Rodriguez-Gomez}, {Nelson}, {Pillepich}, {Bottrell}, {Bernardi}, {Dom{\'\i}nguez-S{\'a}nchez}, {Genel}, {Pakmor}, {Snyder}, \& et~al.}]{Huertas-Company2019}
{Huertas-Company}, M., {Rodriguez-Gomez}, V., {Nelson}, D., {et~al.} 2019, \bibinfo{title}{{The Hubble Sequence at z {\ensuremath{\sim}} 0 in the IllustrisTNG simulation with deep learning},} \mnras, 489, 1859, \dodoi{10.1093/mnras/stz2191}

\bibitem[{S. {Jogee} {et~al.}(2005){Jogee}, {Scoville}, \& {Kenney}}]{Jogee2005}
{Jogee}, S., {Scoville}, N., \& {Kenney}, J. D.~P. 2005, \bibinfo{title}{{The Central Region of Barred Galaxies: Molecular Environment, Starbursts, and Secular Evolution},} \apj, 630, 837, \dodoi{10.1086/432106}

\bibitem[{S. {Jogee} {et~al.}(2004){Jogee}, {Barazza}, {Rix}, {Shlosman}, {Barden}, {Wolf}, {Davies}, {Heyer}, {Beckwith}, {Bell}, \& et~al.}]{Jogee2004}
{Jogee}, S., {Barazza}, F.~D., {Rix}, H.-W., {et~al.} 2004, \bibinfo{title}{{Bar Evolution over the Last 8 Billion Years: A Constant Fraction of Strong Bars in the GEMS Survey},} \apjl, 615, L105, \dodoi{10.1086/426138}

\bibitem[{T. {Kim} {et~al.}(2016){Kim}, {Gadotti}, {Athanassoula}, {Bosma}, {Sheth}, \& {Lee}}]{Kim2016}
{Kim}, T., {Gadotti}, D.~A., {Athanassoula}, E., {et~al.} 2016, \bibinfo{title}{{Evidence of bar-induced secular evolution in the inner regions of stellar discs in galaxies: what shapes disc galaxies?},} \mnras, 462, 3430, \dodoi{10.1093/mnras/stw1899}

\bibitem[{T. {Kimm} \& S.~K. {Yi}(2007){Kimm} \& {Yi}}]{Kimm2007}
{Kimm}, T., \& {Yi}, S.~K. 2007, \bibinfo{title}{{Intrinsic Axis Ratio Distribution of Early-Type Galaxies from the Sloan Digital Sky Survey},} \apj, 670, 1048, \dodoi{10.1086/522573}

\bibitem[{J.~H. {Knapen} {et~al.}(1995){Knapen}, {Beckman}, {Heller}, {Shlosman}, \& {de Jong}}]{Knapen1995}
{Knapen}, J.~H., {Beckman}, J.~E., {Heller}, C.~H., {Shlosman}, I., \& {de Jong}, R.~S. 1995, \bibinfo{title}{{The Central Region in M100: Observations and Modeling},} \apj, 454, 623, \dodoi{10.1086/176516}

\bibitem[{J. {Kormendy}(1993){Kormendy}}]{Kormendy1993}
{Kormendy}, J. 1993, \bibinfo{title}{{Kinematics of extragalactic bulges: evidence that some bulges are really disks},} in IAU Symposium, Vol. 153, Galactic Bulges, ed. H.~{Dejonghe} \& H.~J. {Habing}, 209

\bibitem[{J. {Kormendy}(2013){Kormendy}}]{Kormendy2013}
{Kormendy}, J. 2013, \bibinfo{title}{{Secular Evolution in Disk Galaxies},} in {Secular Evolution of Galaxies}, ed. J.~{Falc{\'o}n-Barroso} \& J.~H. {Knapen} (Cambridge: Cambridge University Press), 1--154, \dodoi{10.1017/CBO9781139547420.003}

\bibitem[{J. {Kormendy} \& R.~C. {Kennicutt}(2004){Kormendy} \& {Kennicutt}}]{Kormendy2004}
{Kormendy}, J., \& {Kennicutt}, Jr., R.~C. 2004, \bibinfo{title}{{Secular Evolution and the Formation of Pseudobulges in Disk Galaxies},} \araa, 42, 603, \dodoi{10.1146/annurev.astro.42.053102.134024}

\bibitem[{D. {Krajnovi{\'c}} {et~al.}(2018){Krajnovi{\'c}}, {Emsellem}, {den Brok}, {Marino}, {Schmidt}, {Steinmetz}, \& {Weilbacher}}]{Krajnovic2018}
{Krajnovi{\'c}}, D., {Emsellem}, E., {den Brok}, M., {et~al.} 2018, \bibinfo{title}{{Climbing to the top of the galactic mass ladder: evidence for frequent prolate-like rotation among the most massive galaxies},} \mnras, 477, 5327, \dodoi{10.1093/mnras/sty1031}

\bibitem[{K. {Kuijken} \& M.~R. {Merrifield}(1995){Kuijken} \& {Merrifield}}]{Kuijken1995}
{Kuijken}, K., \& {Merrifield}, M.~R. 1995, \bibinfo{title}{{Establishing the Connection between Peanut-shaped Bulges and Galactic Bars},} \apjl, 443, L13, \dodoi{10.1086/187824}

\bibitem[{Y. {Lan} {et~al.}(2024){Lan}, {Tang}, {Lin}, \& {Gong}}]{Lan2024}
{Lan}, Y., {Tang}, L., {Lin}, W., \& {Gong}, J. 2024, \bibinfo{title}{{Mock Observations: Three Different Types of Galaxy Alignment in TNG100 Simulations},} \apj, 974, 40, \dodoi{10.3847/1538-4357/ad6982}

\bibitem[{R. {Lange} {et~al.}(2016){Lange}, {Moffett}, {Driver}, {Robotham}, {Lagos}, {Kelvin}, {Conselice}, {Margalef-Bentabol}, {Alpaslan}, {Baldry}, \& et~al.}]{Lange2016}
{Lange}, R., {Moffett}, A.~J., {Driver}, S.~P., {et~al.} 2016, \bibinfo{title}{{Galaxy And Mass Assembly (GAMA): M\_star - R\_e relations of z = 0 bulges, discs and spheroids},} \mnras, 462, 1470, \dodoi{10.1093/mnras/stw1495}

\bibitem[{Z.~A. {Le Conte} {et~al.}(2024){Le Conte}, {Gadotti}, {Ferreira}, {Conselice}, {de S{\'a}-Freitas}, {Kim}, {Neumann}, {Fragkoudi}, {Athanassoula}, \& {Adams}}]{Le2024}
{Le Conte}, Z.~A., {Gadotti}, D.~A., {Ferreira}, L., {et~al.} 2024, \bibinfo{title}{{A JWST investigation into the bar fraction at redshifts 1 {\ensuremath{\leq}} z {\ensuremath{\leq}} 3},} \mnras, 530, 1984, \dodoi{10.1093/mnras/stae921}

\bibitem[{Z.~A. {Le Conte} {et~al.}(2026){Le Conte}, {Gadotti}, {Ferreira}, {Conselice}, {de S{\'a}-Freitas}, {Kim}, {Neumann}, {Fragkoudi}, {Athanassoula}, \& {Adams}}]{LeConte2026}
{Le Conte}, Z.~A., {Gadotti}, D.~A., {Ferreira}, L., {et~al.} 2026, \bibinfo{title}{{The evolution of the bar fraction and bar lengths in the last 12 billion years},} \mnras, 545, staf2010, \dodoi{10.1093/mnras/staf2010}

\bibitem[{G.-H. {Lee} {et~al.}(2012){Lee}, {Park}, {Lee}, \& {Choi}}]{Lee2012}
{Lee}, G.-H., {Park}, C., {Lee}, M.~G., \& {Choi}, Y.-Y. 2012, \bibinfo{title}{{Dependence of Barred Galaxy Fraction on Galaxy Properties and Environment},} \apj, 745, 125, \dodoi{10.1088/0004-637X/745/2/125}

\bibitem[{Z.-Z. {Liang} {et~al.}(2025){Liang}, {Wang}, {Gao}, {Ho}, \& {Athanassoula}}]{Liang2025}
{Liang}, Z.-Z., {Wang}, J., {Gao}, H., {Ho}, L.~C., \& {Athanassoula}, E. 2025, \bibinfo{title}{{Connection between Nonaxisymmetric Structures and Neutral Gas Distribution in Disk Galaxies},} \apj, 983, 61, \dodoi{10.3847/1538-4357/ad87f1}

\bibitem[{E.~L. {{\L}okas}(2021){{\L}okas}}]{Lokas2021}
{{\L}okas}, E.~L. 2021, \bibinfo{title}{{Bar-like galaxies in IllustrisTNG},} \aap, 647, A143, \dodoi{10.1051/0004-6361/202040056}

\bibitem[{S. {Lu} {et~al.}(2025){Lu}, {Du}, \& {Debattista}}]{Lu2025}
{Lu}, S., {Du}, M., \& {Debattista}, V.~P. 2025, \bibinfo{title}{{IllustrisTNG insights: Factors affecting the presence of bars in disk galaxies},} \aap, 697, A236, \dodoi{10.1051/0004-6361/202453143}

\bibitem[{D. {Lynden-Bell} \& A.~J. {Kalnajs}(1972){Lynden-Bell} \& {Kalnajs}}]{Lynden-Bell1972}
{Lynden-Bell}, D., \& {Kalnajs}, A.~J. 1972, \bibinfo{title}{{On the generating mechanism of spiral structure},} \mnras, 157, 1, \dodoi{10.1093/mnras/157.1.1}

\bibitem[{F. {Marinacci} {et~al.}(2018){Marinacci}, {Vogelsberger}, {Pakmor}, {Torrey}, {Springel}, {Hernquist}, {Nelson}, {Weinberger}, {Pillepich}, {Naiman}, \& et~al.}]{Marinacci2018}
{Marinacci}, F., {Vogelsberger}, M., {Pakmor}, R., {et~al.} 2018, \bibinfo{title}{{First results from the IllustrisTNG simulations: radio haloes and magnetic fields},} \mnras, 480, 5113, \dodoi{10.1093/mnras/sty2206}

\bibitem[{I. {Marinova} \& S. {Jogee}(2007){Marinova} \& {Jogee}}]{Marinova2007}
{Marinova}, I., \& {Jogee}, S. 2007, \bibinfo{title}{{Characterizing Bars at z \raisebox{-0.5ex}\textasciitilde 0 in the Optical and NIR: Implications for the Evolution of Barred Disks with Redshift},} \apj, 659, 1176, \dodoi{10.1086/512355}

\bibitem[{K.~L. {Masters} {et~al.}(2011){Masters}, {Nichol}, {Hoyle}, {Lintott}, {Bamford}, {Edmondson}, {Fortson}, {Keel}, {Schawinski}, {Smith}, \& et~al.}]{Masters2011}
{Masters}, K.~L., {Nichol}, R.~C., {Hoyle}, B., {et~al.} 2011, \bibinfo{title}{{Galaxy Zoo: bars in disc galaxies},} \mnras, 411, 2026, \dodoi{10.1111/j.1365-2966.2010.17834.x}

\bibitem[{K.~L. {Masters} {et~al.}(2012){Masters}, {Nichol}, {Haynes}, {Keel}, {Lintott}, {Simmons}, {Skibba}, {Bamford}, {Giovanelli}, \& {Schawinski}}]{Masters2012}
{Masters}, K.~L., {Nichol}, R.~C., {Haynes}, M.~P., {et~al.} 2012, \bibinfo{title}{{Galaxy Zoo and ALFALFA: atomic gas and the regulation of star formation in barred disc galaxies},} \mnras, 424, 2180, \dodoi{10.1111/j.1365-2966.2012.21377.x}

\bibitem[{A. {Meert} {et~al.}(2015){Meert}, {Vikram}, \& {Bernardi}}]{Meert2015}
{Meert}, A., {Vikram}, V., \& {Bernardi}, M. 2015, \bibinfo{title}{{A catalogue of 2D photometric decompositions in the SDSS-DR7 spectroscopic main galaxy sample: preferred models and systematics},} \mnras, 446, 3943, \dodoi{10.1093/mnras/stu2333}

\bibitem[{T. {Melvin} {et~al.}(2014){Melvin}, {Masters}, {Lintott}, {Nichol}, {Simmons}, {Bamford}, {Casteels}, {Cheung}, {Edmondson}, {Fortson}, \& et~al.}]{Melvin2014}
{Melvin}, T., {Masters}, K., {Lintott}, C., {et~al.} 2014, \bibinfo{title}{{Galaxy Zoo: an independent look at the evolution of the bar fraction over the last eight billion years from HST-COSMOS},} \mnras, 438, 2882, \dodoi{10.1093/mnras/stt2397}

\bibitem[{K. {Men{\'e}ndez-Delmestre} {et~al.}(2007){Men{\'e}ndez-Delmestre}, {Sheth}, {Schinnerer}, {Jarrett}, \& {Scoville}}]{Menendez2007}
{Men{\'e}ndez-Delmestre}, K., {Sheth}, K., {Schinnerer}, E., {Jarrett}, T.~H., \& {Scoville}, N.~Z. 2007, \bibinfo{title}{{A Near-Infrared Study of 2MASS Bars in Local Galaxies: An Anchor for High-Redshift Studies},} \apj, 657, 790, \dodoi{10.1086/511025}

\bibitem[{R. {Michard}(2002){Michard}}]{Michard2002}
{Michard}, R. 2002, \bibinfo{title}{{PSF far wings and ``red halo'' in the photometry of galaxies. One more source of errors},} \aap, 384, 763, \dodoi{10.1051/0004-6361:20011813}

\bibitem[{J.~P. {Naiman} {et~al.}(2018){Naiman}, {Pillepich}, {Springel}, {Ramirez-Ruiz}, {Torrey}, {Vogelsberger}, {Pakmor}, {Nelson}, {Marinacci}, {Hernquist}, \& et~al.}]{Naiman2018}
{Naiman}, J.~P., {Pillepich}, A., {Springel}, V., {et~al.} 2018, \bibinfo{title}{{First results from the IllustrisTNG simulations: a tale of two elements - chemical evolution of magnesium and europium},} \mnras, 477, 1206, \dodoi{10.1093/mnras/sty618}

\bibitem[{P.~B. {Nair} \& R.~G. {Abraham}(2010){Nair} \& {Abraham}}]{Nair2010}
{Nair}, P.~B., \& {Abraham}, R.~G. 2010, \bibinfo{title}{{A Catalog of Detailed Visual Morphological Classifications for 14,034 Galaxies in the Sloan Digital Sky Survey},} \apjs, 186, 427, \dodoi{10.1088/0067-0049/186/2/427}

\bibitem[{L. {Nanni} {et~al.}(2022){Nanni}, {Thomas}, {Trayford}, {Maraston}, {Neumann}, {Law}, {Hill}, {Pillepich}, {Yan}, {Chen}, \& et~al.}]{Nanni2022}
{Nanni}, L., {Thomas}, D., {Trayford}, J., {et~al.} 2022, \bibinfo{title}{{iMaNGA: mock MaNGA galaxies based on IllustrisTNG and MaStar SSPs - I. Construction and analysis of the mock data cubes},} \mnras, 515, 320, \dodoi{10.1093/mnras/stac1531}

\bibitem[{D. {Nelson} {et~al.}(2018){Nelson}, {Pillepich}, {Springel}, {Weinberger}, {Hernquist}, {Pakmor}, {Genel}, {Torrey}, {Vogelsberger}, {Kauffmann}, {Marinacci}, \& {Naiman}}]{Nelson2018}
{Nelson}, D., {Pillepich}, A., {Springel}, V., {et~al.} 2018, \bibinfo{title}{{First results from the IllustrisTNG simulations: the galaxy colour bimodality},} \mnras, 475, 624, \dodoi{10.1093/mnras/stx3040}

\bibitem[{D. {Nelson} {et~al.}(2019){Nelson}, {Pillepich}, {Springel}, {Pakmor}, {Weinberger}, {Genel}, {Torrey}, {Vogelsberger}, {Marinacci}, \& {Hernquist}}]{Nelson2019}
{Nelson}, D., {Pillepich}, A., {Springel}, V., {et~al.} 2019, \bibinfo{title}{{First results from the TNG50 simulation: galactic outflows driven by supernovae and black hole feedback},} \mnras, 490, 3234, \dodoi{10.1093/mnras/stz2306}

\bibitem[{J. {Neumann} {et~al.}(2024){Neumann}, {Thomas}, {Maraston}, {Gleis}, {Mao}, {Schinnerer}, \& {Stuber}}]{Neumann2024}
{Neumann}, J., {Thomas}, D., {Maraston}, C., {et~al.} 2024, \bibinfo{title}{{Azimuthal variations of stellar populations in barred galaxies},} \mnras, 534, 2438, \dodoi{10.1093/mnras/stae2252}

\bibitem[{M. {Noguchi}(1987){Noguchi}}]{Noguchi1987}
{Noguchi}, M. 1987, \bibinfo{title}{{Close encounter between galaxies - II. Tidal deformation of a disc galaxy stabilized by massive halo.},} \mnras, 228, 635, \dodoi{10.1093/mnras/228.3.635}

\bibitem[{S. {Oh} {et~al.}(2012){Oh}, {Oh}, \& {Yi}}]{Oh2012}
{Oh}, S., {Oh}, K., \& {Yi}, S.~K. 2012, \bibinfo{title}{{Bar Effects on Central Star Formation and Active Galactic Nucleus Activity},} \apjs, 198, 4, \dodoi{10.1088/0067-0049/198/1/4}

\bibitem[{N.~D. {Padilla} \& M.~A. {Strauss}(2008){Padilla} \& {Strauss}}]{Padilla2008}
{Padilla}, N.~D., \& {Strauss}, M.~A. 2008, \bibinfo{title}{{The shapes of galaxies in the Sloan Digital Sky Survey},} \mnras, 388, 1321, \dodoi{10.1111/j.1365-2966.2008.13480.x}

\bibitem[{C.~Y. {Peng} {et~al.}(2002){Peng}, {Ho}, {Impey}, \& {Rix}}]{Peng2002}
{Peng}, C.~Y., {Ho}, L.~C., {Impey}, C.~D., \& {Rix}, H.-W. 2002, \bibinfo{title}{{Detailed Structural Decomposition of Galaxy Images},} \aj, 124, 266, \dodoi{10.1086/340952}

\bibitem[{A. {Pillepich} {et~al.}(2018{\natexlab{a}}){Pillepich}, {Nelson}, {Hernquist}, {Springel}, {Pakmor}, {Torrey}, {Weinberger}, {Genel}, {Naiman}, {Marinacci}, \& {Vogelsberger}}]{Pillepich2018_firstresult}
{Pillepich}, A., {Nelson}, D., {Hernquist}, L., {et~al.} 2018{\natexlab{a}}, \bibinfo{title}{{First results from the IllustrisTNG simulations: the stellar mass content of groups and clusters of galaxies},} \mnras, 475, 648, \dodoi{10.1093/mnras/stx3112}

\bibitem[{A. {Pillepich} {et~al.}(2018{\natexlab{b}}){Pillepich}, {Springel}, {Nelson}, {Genel}, {Naiman}, {Pakmor}, {Hernquist}, {Torrey}, {Vogelsberger}, {Weinberger}, \& et~al.}]{Pillepich2018}
{Pillepich}, A., {Springel}, V., {Nelson}, D., {et~al.} 2018{\natexlab{b}}, \bibinfo{title}{{Simulating galaxy formation with the IllustrisTNG model},} \mnras, 473, 4077, \dodoi{10.1093/mnras/stx2656}

\bibitem[{A. {Pillepich} {et~al.}(2019){Pillepich}, {Nelson}, {Springel}, {Pakmor}, {Torrey}, {Weinberger}, {Vogelsberger}, {Marinacci}, {Genel}, {van der Wel}, \& {Hernquist}}]{Pillepich2019}
{Pillepich}, A., {Nelson}, D., {Springel}, V., {et~al.} 2019, \bibinfo{title}{{First results from the TNG50 simulation: the evolution of stellar and gaseous discs across cosmic time},} \mnras, 490, 3196, \dodoi{10.1093/mnras/stz2338}

\bibitem[{ {Planck Collaboration} {et~al.}(2016){Planck Collaboration}, {Ade}, {Aghanim}, {Arnaud}, {Ashdown}, {Aumont}, {Baccigalupi}, {Banday}, {Barreiro}, {Bartlett}, \& et~al.}]{Planck2016}
{Planck Collaboration}, {Ade}, P.~A.~R., {Aghanim}, N., {et~al.} 2016, \bibinfo{title}{{Planck 2015 results. XIII. Cosmological parameters},} \aap, 594, A13, \dodoi{10.1051/0004-6361/201525830}

\bibitem[{J. {Reddish} {et~al.}(2022){Reddish}, {Kraljic}, {Petersen}, {Tep}, {Dubois}, {Pichon}, {Peirani}, {Bournaud}, {Choi}, {Devriendt}, \& et~al.}]{Reddish2022}
{Reddish}, J., {Kraljic}, K., {Petersen}, M.~S., {et~al.} 2022, \bibinfo{title}{{The NewHorizon simulation - to bar or not to bar},} \mnras, 512, 160, \dodoi{10.1093/mnras/stac494}

\bibitem[{H.-W. {Rix} {et~al.}(2004){Rix}, {Barden}, {Beckwith}, {Bell}, {Borch}, {Caldwell}, {H{\"a}ussler}, {Jahnke}, {Jogee}, {McIntosh}, \& et~al.}]{Rix2004}
{Rix}, H.-W., {Barden}, M., {Beckwith}, S. V.~W., {et~al.} 2004, \bibinfo{title}{{GEMS: Galaxy Evolution from Morphologies and SEDs},} \apjs, 152, 163, \dodoi{10.1086/420885}

\bibitem[{V. {Rodriguez-Gomez} {et~al.}(2019){Rodriguez-Gomez}, {Snyder}, {Lotz}, {Nelson}, {Pillepich}, {Springel}, {Genel}, {Weinberger}, {Tacchella}, {Pakmor}, \& et~al.}]{Rodriguez-Gomez2019}
{Rodriguez-Gomez}, V., {Snyder}, G.~F., {Lotz}, J.~M., {et~al.} 2019, \bibinfo{title}{{The optical morphologies of galaxies in the IllustrisTNG simulation: a comparison to Pan-STARRS observations},} \mnras, 483, 4140, \dodoi{10.1093/mnras/sty3345}

\bibitem[{A.~B. {Romeo} {et~al.}(2023){Romeo}, {Agertz}, \& {Renaud}}]{Romeo2023}
{Romeo}, A.~B., {Agertz}, O., \& {Renaud}, F. 2023, \bibinfo{title}{{The specific angular momentum of disc galaxies and its connection with galaxy morphology, bar structure, and disc gravitational instability},} \mnras, 518, 1002, \dodoi{10.1093/mnras/stac3074}

\bibitem[{Y. {Rosas-Guevara} {et~al.}(2020){Rosas-Guevara}, {Bonoli}, {Dotti}, {Zana}, {Nelson}, {Pillepich}, {Ho}, {Izquierdo-Villalba}, {Hernquist}, \& {Pakmor}}]{Rosas2020}
{Rosas-Guevara}, Y., {Bonoli}, S., {Dotti}, M., {et~al.} 2020, \bibinfo{title}{{The buildup of strongly barred galaxies in the TNG100 simulation},} \mnras, 491, 2547, \dodoi{10.1093/mnras/stz3180}

\bibitem[{Y. {Rosas-Guevara} {et~al.}(2022){Rosas-Guevara}, {Bonoli}, {Dotti}, {Izquierdo-Villalba}, {Lupi}, {Zana}, {Bonetti}, {Nelson}, {Springel}, {Hernquist}, \& et~al.}]{Rosas-Guevara2022}
{Rosas-Guevara}, Y., {Bonoli}, S., {Dotti}, M., {et~al.} 2022, \bibinfo{title}{{The evolution of the barred galaxy population in the TNG50 simulation},} \mnras, 512, 5339, \dodoi{10.1093/mnras/stac816}

\bibitem[{M. {Roshan} {et~al.}(2021){Roshan}, {Ghafourian}, {Kashfi}, {Banik}, {Haslbauer}, {Cuomo}, {Famaey}, \& {Kroupa}}]{Roshan2021}
{Roshan}, M., {Ghafourian}, N., {Kashfi}, T., {et~al.} 2021, \bibinfo{title}{{Fast galaxy bars continue to challenge standard cosmology},} \mnras, 508, 926, \dodoi{10.1093/mnras/stab2553}

\bibitem[{K. {Saha} \& B. {Elmegreen}(2018){Saha} \& {Elmegreen}}]{Saha2018}
{Saha}, K., \& {Elmegreen}, B. 2018, \bibinfo{title}{{Why Are Some Galaxies Not Barred?},} \apj, 858, 24, \dodoi{10.3847/1538-4357/aabacd}

\bibitem[{K. {Saha} \& T. {Naab}(2013){Saha} \& {Naab}}]{Saha2013}
{Saha}, K., \& {Naab}, T. 2013, \bibinfo{title}{{Spinning dark matter haloes promote bar formation},} \mnras, 434, 1287, \dodoi{10.1093/mnras/stt1088}

\bibitem[{S. {Salim} {et~al.}(2016){Salim}, {Lee}, {Janowiecki}, {da Cunha}, {Dickinson}, {Boquien}, {Burgarella}, {Salzer}, \& {Charlot}}]{Salim2016}
{Salim}, S., {Lee}, J.~C., {Janowiecki}, S., {et~al.} 2016, \bibinfo{title}{{GALEX-SDSS-WISE Legacy Catalog (GSWLC): Star Formation Rates, Stellar Masses, and Dust Attenuations of 700,000 Low-redshift Galaxies},} \apjs, 227, 2, \dodoi{10.3847/0067-0049/227/1/2}

\bibitem[{R.~H. {Sanders} \& A.~D. {Tubbs}(1980){Sanders} \& {Tubbs}}]{Sanders1980}
{Sanders}, R.~H., \& {Tubbs}, A.~D. 1980, \bibinfo{title}{{Gas as a tracer of barred spiral dynamics},} \apj, 235, 803, \dodoi{10.1086/157683}

\bibitem[{J.~A. {Sellwood} \& A. {Wilkinson}(1993){Sellwood} \& {Wilkinson}}]{Sellwood1993}
{Sellwood}, J.~A., \& {Wilkinson}, A. 1993, \bibinfo{title}{{Dynamics of barred galaxies},} Reports on Progress in Physics, 56, 173, \dodoi{10.1088/0034-4885/56/2/001}

\bibitem[{J. {Shen} {et~al.}(2010){Shen}, {Rich}, {Kormendy}, {Howard}, {De Propris}, \& {Kunder}}]{Shen2010}
{Shen}, J., {Rich}, R.~M., {Kormendy}, J., {et~al.} 2010, \bibinfo{title}{{Our Milky Way as a Pure-disk Galaxy{\textemdash}A Challenge for Galaxy Formation},} \apjl, 720, L72, \dodoi{10.1088/2041-8205/720/1/L72}

\bibitem[{J. {Shen} \& J.~A. {Sellwood}(2004){Shen} \& {Sellwood}}]{Shen2004}
{Shen}, J., \& {Sellwood}, J.~A. 2004, \bibinfo{title}{{The Destruction of Bars by Central Mass Concentrations},} \apj, 604, 614, \dodoi{10.1086/382124}

\bibitem[{V. {Springel}(2010){Springel}}]{Springel2010}
{Springel}, V. 2010, \bibinfo{title}{{E pur si muove: Galilean-invariant cosmological hydrodynamical simulations on a moving mesh},} \mnras, 401, 791, \dodoi{10.1111/j.1365-2966.2009.15715.x}

\bibitem[{V. {Springel} {et~al.}(2018){Springel}, {Pakmor}, {Pillepich}, {Weinberger}, {Nelson}, {Hernquist}, {Vogelsberger}, {Genel}, {Torrey}, {Marinacci}, \& et~al.}]{Springel2018}
{Springel}, V., {Pakmor}, R., {Pillepich}, A., {et~al.} 2018, \bibinfo{title}{{First results from the IllustrisTNG simulations: matter and galaxy clustering},} \mnras, 475, 676, \dodoi{10.1093/mnras/stx3304}

\bibitem[{S. {Tacchella} {et~al.}(2019){Tacchella}, {Diemer}, {Hernquist}, {Genel}, {Marinacci}, {Nelson}, {Pillepich}, {Rodriguez-Gomez}, {Sales}, {Springel}, \& et~al.}]{Tacchella2019}
{Tacchella}, S., {Diemer}, B., {Hernquist}, L., {et~al.} 2019, \bibinfo{title}{{Morphology and star formation in IllustrisTNG: the build-up of spheroids and discs},} \mnras, 487, 5416, \dodoi{10.1093/mnras/stz1657}

\bibitem[{L. {Tang} {et~al.}(2023){Tang}, {Lin}, {Wang}, {Li}, \& {Lan}}]{Tang2023}
{Tang}, L., {Lin}, W., {Wang}, Y., {Li}, J., \& {Lan}, Y. 2023, \bibinfo{title}{{Mock Observations: Formation and Evolution of Diffuse Light in Galaxy Groups and Clusters in the IllustrisTNG Simulations},} \apj, 959, 104, \dodoi{10.3847/1538-4357/ad05ca}

\bibitem[{L. {Tang} {et~al.}(2021){Tang}, {Lin}, {Wang}, \& {Napolitano}}]{Tang2021}
{Tang}, L., {Lin}, W., {Wang}, Y., \& {Napolitano}, N.~R. 2021, \bibinfo{title}{{The importance of mock observations in validating galaxy properties for cosmological simulations},} \mnras, 508, 3321, \dodoi{10.1093/mnras/stab2722}

\bibitem[{B.~A. {Terrazas} {et~al.}(2020){Terrazas}, {Bell}, {Pillepich}, {Nelson}, {Somerville}, {Genel}, {Weinberger}, {Habouzit}, {Li}, {Hernquist}, \& et~al.}]{Terrazas2020}
{Terrazas}, B.~A., {Bell}, E.~F., {Pillepich}, A., {et~al.} 2020, \bibinfo{title}{{The relationship between black hole mass and galaxy properties: examining the black hole feedback model in IllustrisTNG},} \mnras, 493, 1888, \dodoi{10.1093/mnras/staa374}

\bibitem[{A. {Toomre}(1964){Toomre}}]{Toomre1964}
{Toomre}, A. 1964, \bibinfo{title}{{On the gravitational stability of a disk of stars.},} \apj, 139, 1217, \dodoi{10.1086/147861}

\bibitem[{V. {Vikram} {et~al.}(2010){Vikram}, {Wadadekar}, {Kembhavi}, \& {Vijayagovindan}}]{Vikram2010}
{Vikram}, V., {Wadadekar}, Y., {Kembhavi}, A.~K., \& {Vijayagovindan}, G.~V. 2010, \bibinfo{title}{{PYMORPH: automated galaxy structural parameter estimation using PYTHON},} \mnras, 409, 1379, \dodoi{10.1111/j.1365-2966.2010.17426.x}

\bibitem[{M.~D. {Weinberg}(1985){Weinberg}}]{Weinberg1985}
{Weinberg}, M.~D. 1985, \bibinfo{title}{{Evolution of barred galaxies by dynamical friction.},} \mnras, 213, 451, \dodoi{10.1093/mnras/213.3.451}

\bibitem[{H. {Wu} {et~al.}(2005){Wu}, {Shao}, {Mo}, {Xia}, \& {Deng}}]{Wu2005}
{Wu}, H., {Shao}, Z., {Mo}, H.~J., {Xia}, X., \& {Deng}, Z. 2005, \bibinfo{title}{{Optical and Near-Infrared Color Profiles in Nearby Early-Type Galaxies and the Implied Age and Metallicity Gradients},} \apj, 622, 244, \dodoi{10.1086/427821}

\bibitem[{T. {Zana} {et~al.}(2022){Zana}, {Lupi}, {Bonetti}, {Dotti}, {Rosas-Guevara}, {Izquierdo-Villalba}, {Bonoli}, {Hernquist}, \& {Nelson}}]{Zana2022}
{Zana}, T., {Lupi}, A., {Bonetti}, M., {et~al.} 2022, \bibinfo{title}{{Morphological decomposition of TNG50 galaxies: methodology and catalogue},} \mnras, 515, 1524, \dodoi{10.1093/mnras/stac1708}

\bibitem[{D. {Zhao} {et~al.}(2020){Zhao}, {Du}, {Ho}, {Debattista}, \& {Shi}}]{Zhao2020}
{Zhao}, D., {Du}, M., {Ho}, L.~C., {Debattista}, V.~P., \& {Shi}, J. 2020, \bibinfo{title}{{Barred Galaxies in the IllustrisTNG Simulation},} \apj, 904, 170, \dodoi{10.3847/1538-4357/abbe1b}

\bibitem[{X. {Zhou} {et~al.}(2025){Zhou}, {Yang}, {Li}, {Xiong}, {Deng}, {Meng}, {Ye}, {Shen}, {Wei}, {Cui}, \& et~al.}]{Zhou2025}
{Zhou}, X., {Yang}, H., {Li}, N., {et~al.} 2025, \bibinfo{title}{{GalaxyGenius: Mock galaxy image generator for various telescopes from hydrodynamical simulations},} \aap, 700, A120, \dodoi{10.1051/0004-6361/202554287}

\bibitem[{Z.-B. {Zhou} {et~al.}(2020){Zhou}, {Zhu}, {Wang}, \& {Feng}}]{Zhou2020}
{Zhou}, Z.-B., {Zhu}, W., {Wang}, Y., \& {Feng}, L.-L. 2020, \bibinfo{title}{{Barred Galaxies in the Illustris-1 and TNG100 Simulations: A Comparison Study},} \apj, 895, 92, \dodoi{10.3847/1538-4357/ab8d32}

\bibitem[{S. {Zibetti} {et~al.}(2009){Zibetti}, {Charlot}, \& {Rix}}]{Zibetti2009}
{Zibetti}, S., {Charlot}, S., \& {Rix}, H.-W. 2009, \bibinfo{title}{{Resolved stellar mass maps of galaxies - I. Method and implications for global mass estimates},} \mnras, 400, 1181, \dodoi{10.1111/j.1365-2966.2009.15528.x}

\bibitem[{E. {Zinger} {et~al.}(2020){Zinger}, {Pillepich}, {Nelson}, {Weinberger}, {Pakmor}, {Springel}, {Hernquist}, {Marinacci}, \& {Vogelsberger}}]{Zinger2020}
{Zinger}, E., {Pillepich}, A., {Nelson}, D., {et~al.} 2020, \bibinfo{title}{{Ejective and preventative: the IllustrisTNG black hole feedback and its effects on the thermodynamics of the gas within and around galaxies},} \mnras, 499, 768, \dodoi{10.1093/mnras/staa2607}

\bibitem[{V. {Zubko} {et~al.}(2004){Zubko}, {Dwek}, \& {Arendt}}]{Zubko2004}
{Zubko}, V., {Dwek}, E., \& {Arendt}, R.~G. 2004, \bibinfo{title}{{Interstellar Dust Models Consistent with Extinction, Emission, and Abundance Constraints},} \apjs, 152, 211, \dodoi{10.1086/382351}

\end{thebibliography}
\bibliographystyle{aasjournalv7}

\end{document}